\documentclass[useAMS,usenatbib]{mn2e}
\usepackage{epsfig}
\usepackage{graphicx}
\usepackage{subfigure}

\newcommand{\Zs}{\mathrm{Z}_\odot}
\newcommand{\Ms}{\mathrm{M}_\odot}
\newcommand{\Ls}{\mathrm{L}_\odot}

\newcommand{\mnras}{MNRAS}
\newcommand{\aap}{A\&A}
\newcommand{\araa}{ARA\&A}
\newcommand{\aj}{AJ}
\newcommand{\apj}{ApJ}
\newcommand{\apjl}{ApJ}
\newcommand{\apjs}{ApJS}
\newcommand{\nat}{Nature}
\newcommand{\sci}{Science}

\title[Formation of Isolated Dwarf Galaxies with Feedback]{Formation
  of Isolated Dwarf Galaxies with Feedback}

\author[Sawala, Scannapieco, Maio and White] { Till
  Sawala$^{1}$\thanks{E-Mail: till@mpa-garching.mpg.de}, Cecilia
  Scannapieco$^{1}$, Umberto Maio$^{2}$ and Simon White$^{1}$
  \\ $^{1}$Max-Planck Institute for Astrophysics,
  Karl-Schwarzschild-Strasse 1, 85748 Garching, Germany
  \\ $^{2}$Max-Planck Institute for Extraterrestrial Physics,
  Giessenbachstrasse 1, 85748 Garching, Germany}

\begin{document}

\date{Accepted 2009 January 1. Received 2009 January 1; in original
  form 2009 January 1}

\pagerange{\pageref{firstpage}--\pageref{lastpage}} \pubyear{2009}

\maketitle

\label{firstpage}

\begin{abstract}
We present results of high resolution hydrodynamical simulations of
the formation and evolution of dwarf galaxies. Our simulations start
from cosmological initial conditions at high redshift. They include
metal-dependent cooling, star formation, feedback from type II and
type Ia supernovae and UV background radiation, with physical recipes
identical to those applied in a previous study of Milky Way type
galaxies. We find that a combination of feedback and the cosmic UV
background results in the formation of galaxies with properties
similar to the Local Group dwarf spheroidals, and that their effect is
strongly moderated by the depth of the gravitational potential. Taking
this into account, our models naturally reproduce the observed
luminosities and metallicities. The final objects have halo masses
between 2.3~$\times~10^8$ and 1.1~$\times~10^9~\Ms$, mean velocity
dispersions between 6.5 and 9.7~kms$^{-1}$, stellar masses ranging
from 5~$\times~10^5$ to 1.2~$\times 10^7~\Ms$, median metallicities
between [Fe/H]~$=-1.8$ and $-1.1$, and half-light radii of the order
of 200 to 300 pc, all comparable with Local Group dwarf
spheroidals. Our simulations also indicate that the dwarf spheroidal
galaxies observed today lie near a halo mass threshold around
$10^9~\Ms$, in agreement with stellar kinematic data, where supernova
feedback not only suffices to completely expel the interstellar medium
and leave the residual gas-free, but where the combination of
feedback, UV radiation and self-shielding establishes a dichotomy of
age distributions similar to that observed in the Milky Way and M31
satellites.
\end{abstract}


\section{Introduction}

\begin{table*} 

  \centering

  \begin{minipage}{\textwidth}

    \renewcommand{\footnoterule}{}
    \renewcommand{\thefootnote}{\alph{footnote}}
  \end{minipage}
  
  \begin{minipage}{\textwidth}
    \caption{Overview of Numerical Simulation Results\label{tab:params}}

    \begin{tabular}{@{}llrlrrrrrrllll@{}}


      \hline 

    Label & f$_s$ & M$_\star$ & M$_\mathrm{g}$ & M$_\mathrm{tot}$ &
    M$_{0.3}$ & M$_{1.8}$ & M$/$L$_{0.3}$ & M$/$L$_{1.8}$ & r$_{1/2}$
    & $\Delta$ age & $[$Fe$/$H$]$ & $\sigma$ \\
    & & [$10^6 \Ms$] & [$10^6 \Ms$] & [$10^6 \Ms$] & [$10^6 \Ms$] &
    [$10^6 \Ms$] &[$[\Ms / \Ls]_V$]& [$[\Ms / \Ls]_V$]& [pc] & [Gyrs] & & [kms$^{-1}$]\\
     
    \hline

    \multicolumn{12}{l}{Simulations including feedback, UV radiation
      and self-shielding \footnotemark[1]} \\

  1 & 0.368 &
  0.55 & 0.099 & 233.8 &
  8.48 & 79.0 & 
  129 & 555 &
  244 & 1.13 & -1.78 & 6.55\\
  
  2 & 0.422&
  0.96 & 0.005 & 348.8 &
  9.71 & 103.6 &
  100 & 409 &
  303 & 1.15 & -1.76 & 7.26 \\

  3 & 0.464&
  2.10 & 0.009 & 466.1 &
  11.5 & 125.7 &
  39 & 223 &
  226 & 2.51 & -1.52 & 7.54 \\

  4 & 0.500 &
  2.68 & 0.006 & 585.3 &
  13.2 & 147.3 &
  39 & 209 &
  246 & 1.78 & -1.54 & 8.12 \\

  5 & 0.531 &
  3.94 & 0.14 & 701.8 &
  14.8 & 165.9 &
  26 & 156 &
  212 & 1.21 & -1.46 & 8.47 \\

  6 & 0.559 &
  9.02 & 0.010 & 809.3 & 
  16.3 & 185.7 & 
  13 & 81 &
  164 & 3.10 & -1.12 & 8.62 \\

  7 & 0.585 &
  10.02 & 0.005 & 922.3 &
  17.9 & 203.5 &
  13 & 80 &
  171 & 3.23 & -1.17 & 9.08\\

  8 & 0.608 &
  12.20 & 0.011 & 1042 &
  17.9 & 218.5 &
  12 & 71 &
  190 & 3.30 & -1.15 & 9.62 \\

  9 & 0.630 &
  12.26 & 0.012 & 1162 &
  21.0 & 235.1 &
  14 & 76 &
  181 & 3.21 & -1.14 & 9.71 \\

    \hline

  \multicolumn{12}{l}{Simulations including UV radiation but no
    supernova feedback \footnotemark[1]~\footnotemark[3] } \\

  10 & 0.422 &
  18.77 & 0.82 & 318.4 &
  14.5 & 113.3 &
  8 & 28 &
  118 & 3.75 & -0.71 & - \\

  11 & 0.531 &
  99.50 & 9.64 & 645.6 &
  27.0 & 216.5 &
  6 & 14 &
  160 & 6.46 & -0.21 & -\\

    \hline

 \multicolumn{12}{l}{Simulations including feedback, UV radiation but
   no shielding\footnotemark[1]} \\

  12 & 0.368 &
  0.53 & 0.002 & 233.2 &
  8.40 & 79.2 & 
  129 & 571 &
  271 & 1.09 & -1.78 & 6.39\\
  
  13 & 0.422 &
  0.99 & 0.005 & 351.7 &
  10.3 & 103.8 &
  98 & 407 &
  289 & 1.11 & -1.70 & 7.43\\

  14 & 0.464 &
  1.95 & 0.015 & 463.8 &
  11.6 & 126.0 &
  48 & 240 &
  254 & 1.11 & -1.52 & 7.67\\

  15 & 0.500 &
  2.75 &  0.19 & 582.7 &
  13.5 & 148.1 &
  38 & 201 &
  248 & 1.16 & -1.48 & 8.31\\

  16 & 0.531 &
  4.19 & 0.20 & 697.8 &
  15.5 & 168.6 &
  27 & 152 &
  208 & 1.17 & -1.38 & 8.40\\

  17 & 0.559 &
  6.91 & 0.55 & 811.2 & 
  16.6 & 187.6 & 
  17 & 105 &
  186 & 1.14 & -1.27 & 8.45\\

  18 & 0.585 &
  6.69 & 0.38 & 939.4 &
  18.0 & 202.6 &
  18 & 113 &
  180 & 1.21 & -1.29 & 9.12\\

  19 & 0.608 &
  9.31 & 0.58 & 1051 &
  18.1 & 217.9 &
  15 & 89 &
  197 & 1.18 & -1.32 & 9.38\\

  20 & 0.630 &
  9.65 & 0.65 & 1179 &
  18.9 & 235.2 &
  16 & 93 &
  196 & 1.12 & -1.31 & 9.74\\

    \hline

 \multicolumn{12}{l}{Simulations including feedback but no UV
   radiation\footnotemark[1]}\\

  21 & 0.368 &
  1.38 & 0.077 & 229.6 &
  9.08 &  79.24 &
  36 & 202 &
  180 & 7.96 & -1.09 & 6.28\\
  
  22 & 0.422 &
  2.58 & 0.10 & 344.2 &
  11.8 &  105.3 &
  27 & 143 &
  180 & 7.60 & -1.03 & 7.11\\
   
  23 & 0.464 &
  4.58 & 0.08 & 458.4 &
  13.4 &  128.6 &
  19 & 105 &
  178 & 5.49 & -1.08 & 7.39\\
   
  24 & 0.500 &
  5.72 & 0.11 & 576.2 &
  15.4 &  150.3 &
  18 & 99 &
  165 & 5.25 & -1.06 & 8.17 \\
   
  25 & 0.531 &
  7.14 & 0.098 & 685.9 &
  16.3 &  158.5 &
  16 & 90 &
  171 & 4.53 & -1.08 & 8.33\\
  
  26 & 0.559 &
  9.26 & 0.086 & 798.5 &
  17.4 & 186.9 &
  14 & 81 &
  157 & 2.59 & -1.12 & 8.75\\
   
  27 & 0.585 &
  10.95 & 0.064 & 909.0 &
  17.5 & 204.3 &
  13 & 76 &
  169 & 3.66 & -1.05 & 9.17\\
   
  28 & 0.630 &
  13.70 & 0.003 & 1147 &
  20.1 & 236.9 &
  12 & 69 &
  181 & 3.71 & -1.09 & 10.0\\

    \hline

 \multicolumn{12}{l}{Simulations including feedback, UV radiation and
   self-shielding\footnotemark[2]}\\

  29 & 0.422 &
  0.59 & 0.008 & 338.7 &
  6.88 & 104.1 &
  157 & 737 &
  375 & 1.21 & -1.78 & 7.20\\

  30 & 0.630 &
  7.35 & 0.066 & 1141 &
  15.5 & 235.4 &
  18 & 133 &
  245 & 2.62 & -1.11 & 9.50\\

\hline
\end{tabular}

\small{Notes: Col.~2: Scale factor (length) of the initial conditions
  relative to \cite{Hayashi-2003}, Col.~3: Stellar mass, Col.~4: Gas
  mass within 1.8 kpc, Col.~5: Halo mass (M$_{200}$), Col.~6: Dark
  matter mass within 0.3 kpc, Col.~7: Dark matter mass within 1.8 kpc,
  Col.~8: Mass-to-light ratio (V-Band) within 0.3 kpc, Col.~9:
  Mass-to-light ratio (V-Band) within 1.8 kpc, Col.~10: Half light
  radius (projected), Col.~11: Formation time interval containing 90
  \% of M$_\star$, Col.~12: Median stellar iron abundance, Col.~13:
  RMS stellar velocity dispersion}

\footnotetext[1]{Initial number of particles: $1.7 \times 10^5$ gas,
  $8.7 \times 10^5$ dark matter.}

\footnotetext[2]{Initial Number of particles: $1.21 \times 10^6$ gas,
  $2.83 \times 10^6$ dark matter.}

\footnotetext[3]{Simulations terminated at $z=0.68$}

\end{minipage}

\end{table*}

Dwarf spheroidal galaxies are amongst the smallest and faintest known
galactic systems, and at first sight, should be easy to understand.
Their name indicates a simple morphology, they possess low rotation,
little or no interstellar gas and no active star formation. Their
stellar masses range from less than $10^4$ to a few times $10^7 \Ms$,
which even at the more luminous end, makes them comparable to the
brightest globular clusters. However, whilst all observed dwarf
spheroidal galaxies contain at least a fraction of very old stars
\citep{Grebel-1997}, this is where the similarities with globular
clusters end. Spectroscopic surveys of individual stars in several
dwarf spheroidal galaxies of the Local Group
\citep[e.g.][]{DART-Fornax-2006} have revealed surprisingly complex
star formation histories, sometimes over several Gyrs, and at least in
one case in multiple bursts \citep{Koch-Carina-2006, Koch-Carina-2008,
  Orban-2008}

About two dozen dwarf spheroidal galaxies have so far been discovered
as satellites of the Milky Way, while estimates using luminosity
functions corrected for completeness and bias predict the total number
of faint satellites to be an order of magnitude higher
\citep{Tollerud-2008}. The known dwarf spheroidal galaxies in the
Local Group reside in a variety of environments. There are a few near
both M31 and the Milky Way, with distances of $\sim 30$ kpc and
clearly within their hosts' dark matter haloes, as well as some remote
objects like Cetus \citep{Lewis-2007} and Tucana
\citep{Castellani-1996, Fraternali-2009}, which can be considered to
have evolved in isolation.

It has been proposed for a long time \citep[e.g.][]{Faber-Lin-1983}
and is now widely believed that the luminous component of dwarf
spheroidal galaxies is not all there is to them. Their dynamics appear
to be largely dark matter dominated, and measurements of stellar
velocity dispersions \citep[e.g.][]{Koch-2007, Walker-2007,
  Mateo-2008, Walker-2009} indicate that they possess the highest
mass-to-light ratios of any known galactic systems. Recent studies
further suggest that despite the spread in luminosities, the total
mass within the central 300 pc of each galaxy lies within a small
range of around $10^7 \Ms$ \citep{Strigari-2008}.

It is also worth pointing out that in galaxy formation, small size can
breed complexity. Shallow potential wells make these systems
susceptible to both internal and external effects, such as violent
supernova feedback, photoionization and heating from the cosmic UV
background, tidal interactions and ram-pressure stripping. All of
these processes have the potential to shape the evolution of dwarf
galaxies, and to leave their mark on the star formation history and
the chemical abundances, as well as on the morphology and dynamics of
the final objects. They may explain some of the peculiar properties of
dwarf spheroidals, including their very high mass-to-light ratios, and
may also be responsible for the observed scaling laws
\citep[e.g.][]{Woo-2008}. In this sense, the evolution of dwarf
spheroidal galaxies can be considered an extreme case, but at the same
time, an extremely good laboratory for astrophysical and cosmological
processes \citep{Marlowe-1995}. While the sensitivity to many
parameters represents a considerable challenge for simulations, the
large number of dwarf galaxies in the Local Group, together with the
availability of high quality observational data also provides an
unusually high number of constraints. Revaz et al. (2009 in prep.)
exploit this fact by studying a large number of idealised models with
non-cosmological initial conditions, which they can tune to reproduce
the observed relations.

The number of dwarf galaxies observed in the Local Group continues to
grow as new, `ultra-faint' satellite galaxies are discovered
\citep[e.g.][]{Martin-2006, Chapman-2008}. Nevertheless, it is still
much smaller than the total number of dark matter subhaloes found in
high-resolution simulations of spiral galaxy haloes in the standard
$\Lambda$CDM cosmology \citep[e.g.][]{Klypin-1999, Moore-1999,
  Diemand-2007, Springel-2008}. This has become known as the `missing
satellites problem'. However, this is only an apparent discrepancy. It
is removed when one accounts for the fact that not all subhaloes
contain stars. Two possible mechanisms that can produce a number of
visible satellite galaxies similar to that observed are the following.
Perhaps many haloes were able to form a few stars initially, but the
baryonic components of all haloes below some critical mass were
subsequently destroyed by supernova feedback
\citep[e.g.][]{Dekel-1986, Ferrara-2000}. Alternatively (or perhaps
additionally) photoionization may have prevented star formation in the
smallest haloes \citep[e.g.][]{Efstathiou-1992, Somerville-2002,
  Simon-2007}. As dwarf spheroidals are the faintest known galaxies, a
detailed understanding of their evolution should eventually reveal the
influences of these two effects.

Examples of earlier numerical studies of the formation of dwarf
galaxies include simulations by \cite{Read-2006},
\cite{Mashchenko-2008}, \cite{Stinson-2007, Stinson-2009} and
\cite{Valcke-2008}. The latter two have investigated the collapse of
gas clouds in dark matter haloes of constant mass. Both find evidence
of prolonged and self-regulated star formation. However, while they do
observe significant supernova-driven outflows, at a halo mass of $10^9
\Ms$, \citeauthor{Stinson-2007} find better agreement with dwarf
irregular galaxies. \citeauthor{Read-2006} performed simulations of
the formation of the first baryonic building blocks in a cosmological
volume at high redshift. They confirm the importance of supernova
feedback and UV heating (assumed to begin at $z=13$) for removing the
gas from the smallest haloes. However, they do not follow the
evolution of the surviving objects to the present day, terminating
their simulations at $z=10$. \citeauthor{Mashchenko-2008} have also
performed cosmological simulations, albeit of noticeably more massive
haloes, which they follow up to $z = 5$. They do not include UV
radiation, and would require an additional mechanism to remove
the gas from the galaxy, in order to form a system comparable to
  observed dwarf spheroidals. However, they find stellar properties
in good agreement with the Fornax dwarf spheroidal, including globular
clusters. They also predict that supernova feedback induces the
formation of extended dark matter cores via gravitational resonance
heating.

In this work, we model the formation and evolution of dwarf galaxies
in fully cosmological, smoothed particle hydrodynamical (SPH)
simulations. We simulate a cosmological volume with periodic boundary
conditions, in which the haloes grow from small density perturbations
imposed at high redshift. Our initial conditions are chosen to
reproduce galaxies of halo masses similar to the ones inferred for the
Local Group dwarf spheroidals.  Because we follow the evolution to $z
= 0$, and because we include the environment in a consistent way, our
mass-resolution is somewhat lower compared with simulations of
isolated haloes, or simulations which end at high redshift. However,
the cosmological nature of our simulation allows us to simultaneously
follow the growth of the dark matter halo, and the evolution of the
dissipative component. The full time-evolution also reveals the effect
of the UV background, and lets us directly compare present-day
properties to the observations. Our numerical model includes cooling,
star formation, chemical enrichment and feedback, and we allow for
exchange of material with the intergalactic medium. We also include
cosmological effects such as reionization. We use the same code, and
with a few notable exceptions owing to the different physical effects
that play a role in the two regimes, we use the same physics model and
basic parameters as those employed by \cite{Scannapieco-2008d} in
their study of the formation of Milky Way type disk galaxies, some
$10^4$ times larger in terms of stellar mass than the dwarf galaxies
we consider here. We do not study local environmental effects, which
may play a role for the closest companion satellite galaxies to the
Milky Way. The main questions that we will address are how it is
possible that systems of such low luminosity and seemingly similar
total masses undergo such complex and diverse star formation
histories, why dwarf spheroidal galaxies have such high mass-to-light
ratios, why they appear to follow certain scaling laws, and if their
formation and evolution in a cosmological constant dominated Cold Dark
Matter ($\Lambda$CDM) universe can be explained by a consistent
physical model. In Section~\ref{sec:methods}, we present the
computational methods which we have used and our choice of initial
conditions. Section~\ref{sec:evolution} follows with a presentation of
the results of our simulations in broad terms, while we pay closer
attention to the relevance of individual physical processes,
particularly supernova feedback and UV radiation, in
Section~\ref{sec:processes}. In Section~\ref{sec:parameters}, we focus
on the observed scaling laws, and present the dependence of our
results on model parameters. We conclude with a summary where we
discuss the achievements and shortcomings of the simulations in
Section~\ref{sec:conclusion}, and look forward to our future work.

\section{Methods} \label{sec:methods}
The simulations presented here have been performed using the Tree-PM
code \textsc{GADGET-3} \citep{Springel-2005, Springel-2008}, which
includes gravity and smoothed particle hydrodynamics. As an extension,
metal-dependent cooling, star formation, chemical enrichment and
energy injection from type II and type Ia supernovae have been
implemented in the multiphase gas model of \cite{Scannapieco-2005,
  Scannapieco-2006}. This model has previously been used to study the
effect of feedback on galaxy evolution in general terms
\citep{Scannapieco-2006} and the formation of disk galaxies in
particular \citep{Scannapieco-2008d, Scannapieco-2009}. In addition,
some of our simulations contain an approximative treatment of
self-shielding and low temperature cooling, which were not included in
the previous model. In this section, we explain the most important
characteristics of our model.

\subsection{Cooling}\label{sec:methods:cooling}

\begin{figure}
  \includegraphics[scale = .5]{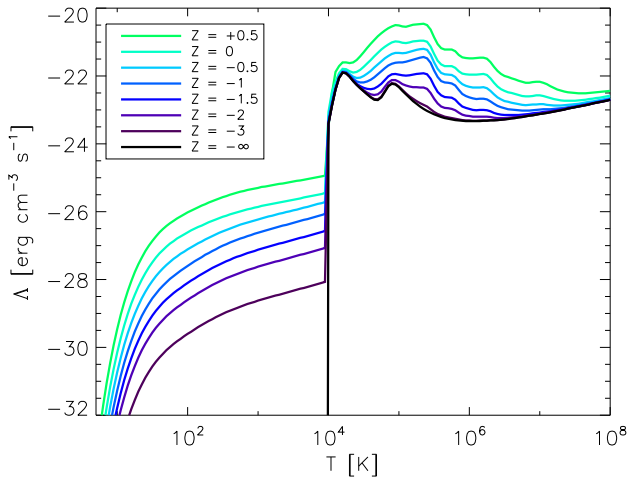}
  \caption{Normalized net cooling rate $\left[\lambda / n_H^2\right]$,
    adopted from Sutherland \& Dopita for temperatures above $10^4$~K,
    and Maio et al. for lower temperatures. The metal-dependency is
    expressed in solar units, assuming $\Zs = 0.02$.}
\label{fig:temp-cooling}
\end{figure}

Above the hydrogen ionization temperature of $10^4$~K, our gas cooling
model is based on metal-dependent cooling functions of
\cite{Sutherland-1993}. When we include cooling below $10^4$~K, we use
the extension of \cite{Maio-2007}. The adopted cooling function
$\lambda(T)$, normalized to $n_H=$1 cm$^{-3}$, is shown in Figure
\ref{fig:temp-cooling} for eight different metallicities. In our
models, the gas density is typically below $n_H=3\times
10^3$~cm$^{-3}$. Consequentially, we do not consider the effect of
rotational excitations, and the resulting linear density dependence in
the high density limit \citep{Dalgarno-1972}.

The ionization states of H, He, and the electron number density are
computed analytically, following the model of
\cite{Katz-1996}. Compton cooling, which is not included in
Figure~\ref{fig:temp-cooling}, is the main coolant at high redshift,
and implemented as a function of the free electron density as well as
the temperature difference between the gas and the evolving CMB
temperature.

The effect of the UV background on the net cooling function, and on
the temperature evolution of the gas, is described in
Section~\ref{sec:model-UV}, and depends on the abundance of the
different species of H and He. We note that the metal-dependent
cooling functions of \cite{Sutherland-1993} assume collisional
ionization equilibrium. As \cite{Wiersma-2009} have recently shown,
the presence of an ionizing radiation background can significantly
decrease the cooling efficiently compared to collisional equilibrium
in optically thin gas. This effect is in addition to the direct UV
heating, and is neglected in our model. However, because the galaxies
in this study form before reionization (see
Section~\ref{sec:evolution}), and because the current implementation
already predicts inefficient cooling of the optically thin gas after
the UV background is switched on (see Section \ref{sec:UV}), we expect
that it would not qualitatively alter our results.

Below $10^4 K$, the cooling rate is a strong function of metallicity,
which \cite{Maio-2007} have calibrated to different iron
abundances. In section \ref{sec:LowTCooling}, we investigate the
significance of low temperature cooling in our simulations. In
general, the difference is small, except for objects with virial
temperatures significantly below $10^4$~K.

\subsection{Star Formation} \label{sec:sf}
Star formation is implemented so that gas particles can spawn, or be
converted into, star particles, subject to certain conditions. We
require the gas particle to be in a region of convergent flow. In
addition, we impose a physical density threshold $\rho_c$ on the local
gas density. The existence of a threshold for star formation is
motivated by observations \citep[e.g.][]{Kennicutt-1989,
  Kennicutt-1998}. Calculations by \cite{Quirk-1972} as well as
numerical simulations, e.g. by \cite{Katz-1996, Springel-2003,
  Bush-2008} and others have shown that the observed Kennicutt-Schmidt
relation can be reproduced imposing a volume density threshold, even
though slightly different values are derived. Recently,
\cite{Koyama-2009} demonstrated with high-resolution simulations of
the turbulent interstellar medium that the star formation rate depends
only weakly on the choice of $\rho_c$, and we adopt the value of
$\rho_c = 7 \times 10^{-26}\mathrm{gcm}^{-3}$, that was used in
\cite{Scannapieco-2006}. We also impose a threshold of
$\rho_\mathrm{g} / \overline{\rho_\mathrm{g}} \ge 57.7 $ on the local
gas overdensity (where $\overline{\rho_\mathrm{g}}$ is the global mean
density). This corresponds to a mean enclosed overdensity which is the
minimum overdensity of a spherical, $r^{-2}$ perturbation for
gravitational collapse, and ensures that even at high redshift, star
formation takes place only in virialised regions
\citep{Katz-1996}. Subject to these constraints, the local star
formation efficiency is regulated by a single efficiency parameter
$c_\star$, so that the star formation rate is given by $$\frac{d
  \rho_\star}{dt} = c_\star \frac{\rho_g}{t_\mathrm{dyn}}$$ where
$t_\mathrm{dyn}$ is the local gas dynamical time. The creation of an
individual stellar particle of mass $m_\star$ from a gas particle of
mass $m_g$ during the time interval $\Delta t$ is stochastic, with the
probability given by $$ p_\star = \frac{m_g}{m_\star} \left( 1 - exp
\left( - c_\star \frac{\Delta t}{t_\mathrm{dyn}} \right) \right). $$
In most of our simulations, we adopt a choice of $c_\star=0.05$. We
study the effect of different values in Section \ref{sec:sfe}. Each
star particle thus produced contains a single stellar population,
whose metallicity is inherited from the parent gas particle. For
simplicity, we assume a Salpeter initial mass function
\citep{Salpeter-1955}. We calculate the luminosities at any given time
using the stellar evolution model of \cite{BC-2003}.

\subsection{Multiphase Interstellar Medium and Feedback} \label{sec:ism}
For each star particle, we determine the rate as well as the yields of
supernovae type II and type Ia. Chemical yields are calculated
separately for both types, following \cite{Woosley-1995} and
\cite{Thielemann-1993} respectively. Supernovae type II are assumed to
be instantaneous, while for supernovae type Ia, we assume a uniform
delay time distribution with given minimum and maximum delay times, as
discussed in Section \ref{sec:lifetimes}. We assume a constant energy
production of $10^{51}$ ergs per supernova, which is released into the
interstellar medium (ISM) as purely thermal energy.

The multiphase characteristic of the ISM, in which components of a
wide range in temperature and density coexist, is lost in simple SPH
models, where the smoothing kernel is a function of position
only. This leads to an overestimation of the density in diffuse clouds
neighbouring high density regions, and results in an underestimation
of their cooling times, artificially increasing the star formation
rate. It also means that feedback from supernovae is released
primarily to the gas in star forming regions, where the densities are
normally so high that the energy is lost immediately via radiative
cooling. As a result, outflows and self-regulation of star formation
are severely suppressed, and metals remain confined \citep{Katz-1992,
  Marri-2003}.
 
Most simulators fix the second problem by switching off cooling in the
reheated particles for some time \citep[e.g.][]{Thacker-2000,
  Governato-2007}, or by giving them a kick of arbitrarily specified
amplitudes \citep[e.g.][]{Navarro-1994, DallaVecchia-2008}. The
multiphase scheme for the interstellar medium of
\cite{Scannapieco-2006}, addresses the problems at a fundamental
level. It allows an overlap of diffuse and dense gaseous components by
considering as neighbours in the smoothing kernel only gas particles
with similar thermodynamic properties. Specifically, particles $i$ and
$j$ are mutually excluded as neighbours if the ratio of their entropic
functions $A(s)_{ij}$ exceeds a certain threshold and their
pairwise-averaged velocity divergence multiplied by their mutual
separationfalls below the local sound speed, which avoids the
decoupling of shock-waves.

However, this approach introduces some additional freedom in
determining how the energy and metals released by supernovae are
shared between the gas particles of the multiphase medium, which in
our simulations each receive half of the total energy. In the dwarf
galaxies we have simulated, most of the ejecta given to gas particles
in the hot and diffuse phase eventually escape from the system,
leaving mostly those that go to the cold phase to be included in
subsequent generations of stars. Thus, increasing the fraction of
metals given to the cold phase increases the final metallicity for a
given stellar mass, whereas a high fraction of metals given to the hot
phase creates strongly metal-enhanced winds. To some extent, we can
use the observed metallicity-luminosity relation of dwarf spheroidals,
shown in Figure~\ref{fig:lum-met}, in order to calibrate this
parameter. Because it effects all elements in the same way, the
remaining degeneracy with the supernova Ia lifetimes can be partially
broken by also considering the [Ca/Fe] ratios. We find relatively good
agreement if $25\%$ of the metals and energy are injected to the cold
phase, and we use this value for all the simulations presented in this
work.

\subsection{UV background} \label{sec:model-UV}
Quasar spectra indicate that the universe has been fully ionised from
about redshift $z=6$ \citep{Fan-2002}. This has prompted us to include
UV background radiation in our models, and we discuss its influence in
Section \ref{sec:UV}. The question of whether dwarf galaxies survive
the cosmic reionization epoch has been an intense area of study
\cite[e.g.][]{ Kitayama-2000, Susa-2004, Hoeft-2006, Hoeft-2008}. In
hydrodynamical simulations, \cite{Hoeft-2008} find that UV heating
reduces the baryonic fraction in galaxies below a characteristic total
mass, $6\times10^9\Ms$. However, \cite{Grebel-2004} found no clear
signature of a widespread impact from reionization in their analysis
of age distributions of nearby dwarf galaxies. In those simulations
where the UV background is included, we have modified the cooling
function for partially ionized gas by a heating term. Apart from tests
where we have decreased the UV intensity, the intensity evolution of
the UV background follows that of \cite{Haardt-1996}.

\subsection{Initial Conditions} \label{sec:ICs}
All simulations are performed in the context of a $\Lambda$CDM
cosmology, with $\Omega_\Lambda$ = 0.7 and $\Omega_\mathrm{m}$ =
0.3. We use a set of initial conditions based on pure dark matter
simulations of isolated haloes by \cite{Hayashi-2003}. The halo on
which our simulations are based (labeled D1 by
\citeauthor{Hayashi-2003}) was selected from their set of dwarf haloes
in order to yield an object whose high redshift progenitors fill a
compact region in space, enabling us to limit the high resolution
region to a small fraction of the total volume, whose (unscaled) side
length is 35.25~h$^{-1}$~Mpc. The resimulations start at redshift $z_i
= 74$ with density fluctuations corresponding to a present value of
$\sigma_8 = 0.9$ in the unscaled initial conditions. To the dark
matter, gas particles were added at a rate of $\Omega_\mathrm{b} =
0.04$ and $\Omega_\mathrm{DM} = 0.26$. As indicated in Table~1, we
have scaled the initial conditions at constant density, to give final
halo masses between $2.33 \times 10^8$ and $1.18 \times 10^9$ $h^{-1}
\Ms$, but identical formation redshift and (scaled) assembly histories
for all our objects. This causes an effective change of the
normalisation of the power spectrum between the simulations, but as
\cite{Colin-2004} have shown, due to the early formation of dwarf
haloes and the flatness of the linear fluctuation amplitude in this
mass regime, the influence on the evolution of individual haloes is
expected to be insignificant compared to the scatter between objects.

As described in Section~\ref{sec:resolution}, we have also performed
simulations of varying particle numbers (up to 2.83 $\times 10^6$ for
dark matter and $1.21 \times 10^6$ for gas). The gravitational
softenings for each particle type were fixed to 1/10th of the
respective mean interparticle separation in comoving coordinates in
the initial conditions and limited, in physical coordinates, to $\sim$
1/5th of the mean separation within the collapsed haloes. This allowed
a spatial resolution typically below 100 pc (depending on the scale
and the number of particles). Haloes were identified using a
friends-of-friends method with a linking length of 0.2. In each case,
over a hundred small haloes with 32 particles or more were formed in
the simulated volume, and depending on the choice of parameters of the
baryonic physics model, several of them formed stars. However, in each
case we limit our analysis to the most massive one, for which the
effective resolution is highest. We have made tests to confirm the
scale-free behaviour of the pure dark matter simulations. We find that
in all cases, the dark matter profiles are well-fitted by a Navarro,
Frenk and White (NFW) model, down to the resolution limit. Note that
all our simulations have the same assembly history, apart from
resolution effects. This means that we cannot say anything about the
scatter in properties expected among similar mass haloes. On the other
hand, differences between our various simulations must therefore be
due entirely to differences in the assumed physics or the numerical
parameters. Cosmic variance plays no role.

\subsection{Effects of Resolution} \label{sec:resolution}
The hydrodynamical model, and the recipies for feedback and star
formation may also be influenced by resolution
effects. \cite{Scannapieco-2006} have tested the model for numerical
convergence. Since we extend their model to a new mass range, we have
performed additional tests. Simulations~1-28, summarized in Table~1,
are run with a constant number of $8.7 \times 10^5$ dark matter
particles, $1.7 \times 10^5$ of which are in the high resolution
Lagrangian volume that also contains $1.7 \times 10^5$ gas
particles. The corresponding particle masses range from $6.6 \times
10^3 \Ms$ and $1.2 \times 10^3 \Ms$ in simulation~1, to $3.3 \times
10^4 \Ms$ and $5.8 \times 10^3 \Ms$ in simulation~9, for dark matter
and gas particles, respectively. The number of stellar particles
varies, depending on star formation rate, and the stellar particle
masses range between $5.4 \times 10^2 \Ms$ in simulation~1 and $2.7
\times 10^3 \Ms$ in simulation~9. Wherever we have changed the other
parameters of the model, we have kept the resolution fixed. We have
also performed two simulations, 29 and 30, with a mass resolution
increased by a factor of eight compared to simulations~2 and 9,
respectively, while all other parameters were kept constant. The
results are shown in the bottom rows of Table~1. While the respective
total masses of the systems are constant to within a few percent, a
statement of the fact that the gravitational part of the force
calculations is largely resolution-independent, and that the coupling
of the dark matter to the baryons in our simulations is small, the
total stellar mass produced in both cases decreases by $\sim 40
\%$. With increased resolution, star formation begins slightly earlier
and at lower halo masses, resulting in quicker heating and outflows of
the gas. However, this difference is small compared to the influence
of physical parameters, such as total mass. The simulations at
different resolutions also show similar properties with respect to the
response to the UV background, the self-shielding threshold and the
metal-enrichment. The results remain consistent with global scaling
relations, as illustrated in Figures \ref{fig:lum-met} and
\ref{fig:l-ml}, where we have included the high resolution results of
simulations~29 and 30, together with the results of simulations~1-9.

\section{Formation and Evolution} \label{sec:evolution}
We find that the evolution of the dwarf galaxies that we simulate is
strongly affected both by supernova feedback and by the UV background
radiation. It is the combination of these two effects that shapes the
evolution of the galaxy. We begin this section by showing the
evolution of a typical dwarf galaxy up to the present time, including
all the different effects that play a role, but focusing on the global
picture. We then try to disentangle the effects of feedback and UV
radiation, and look in more detail at how they each influence the
evolution in Section \ref{sec:processes}.

\subsection{Time Evolution} \label{sec:time-evolution}

\begin{figure*}
  \begin{center}
    \begin{tabular}{ccccc}
      \hspace{-.2in} \includegraphics*[width = .25\textwidth]{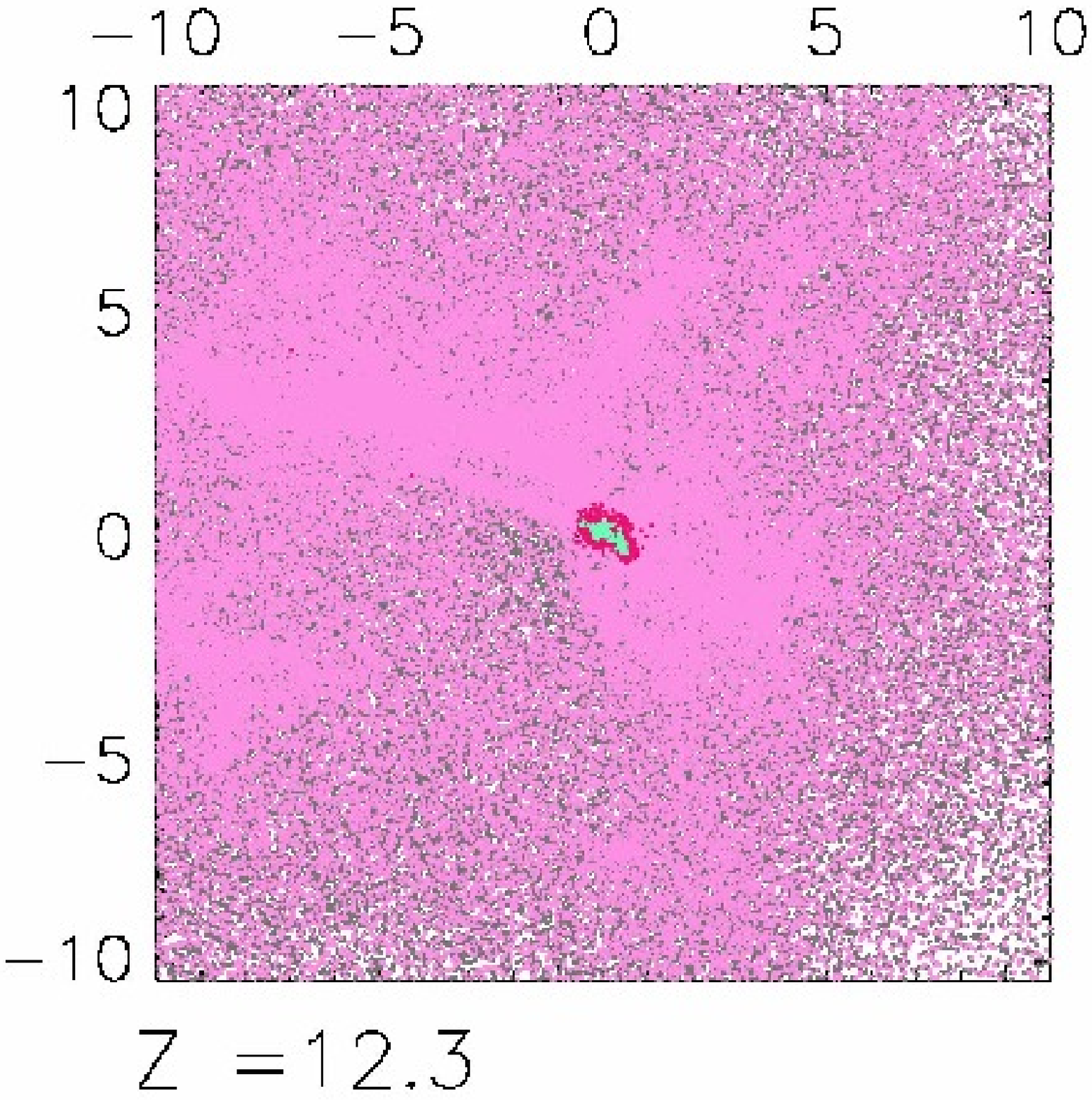} &
      \hspace{-.2in} \includegraphics*[width = .2\textwidth]{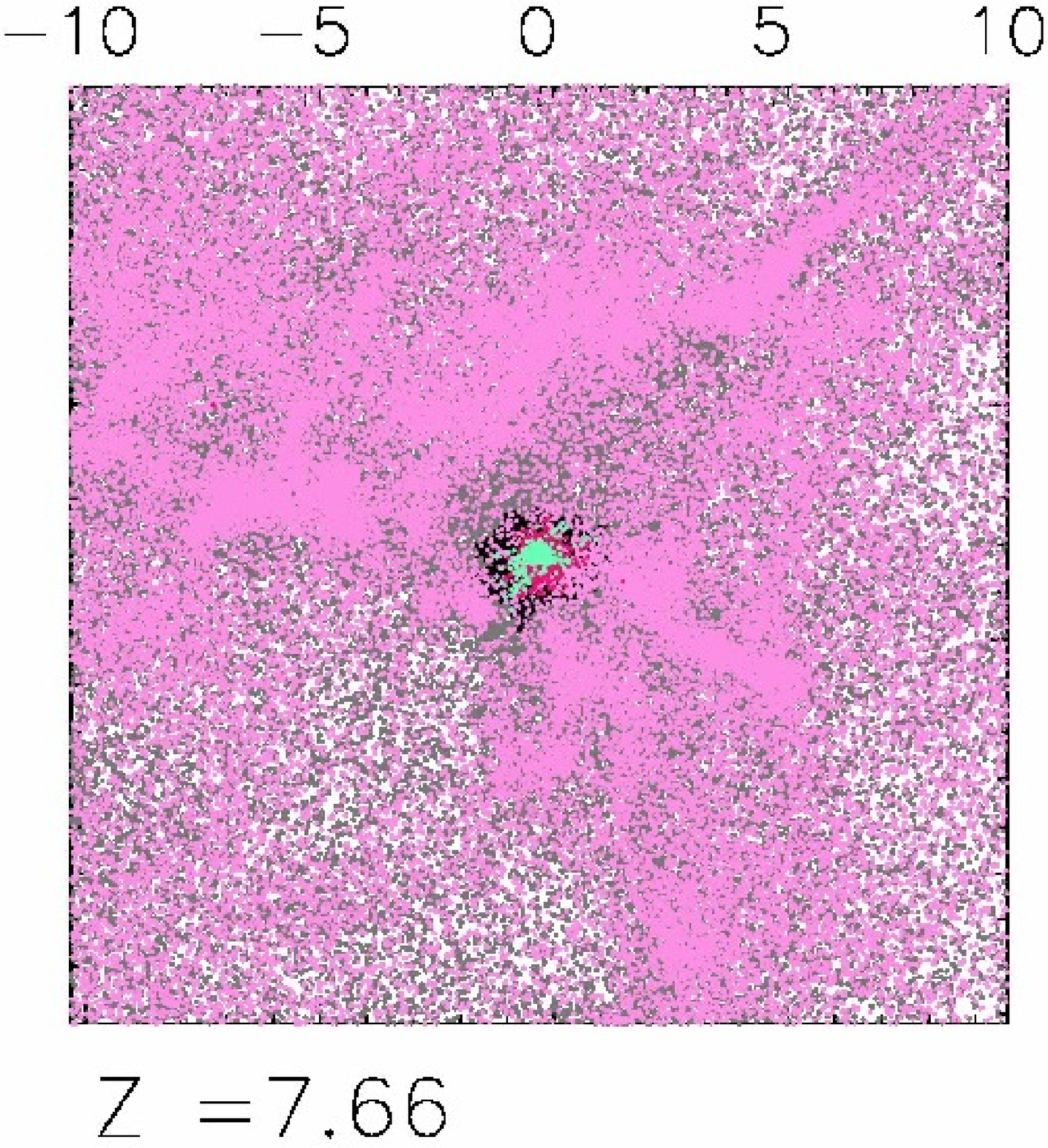} &
      \hspace{-.2in} \includegraphics*[width = .2\textwidth]{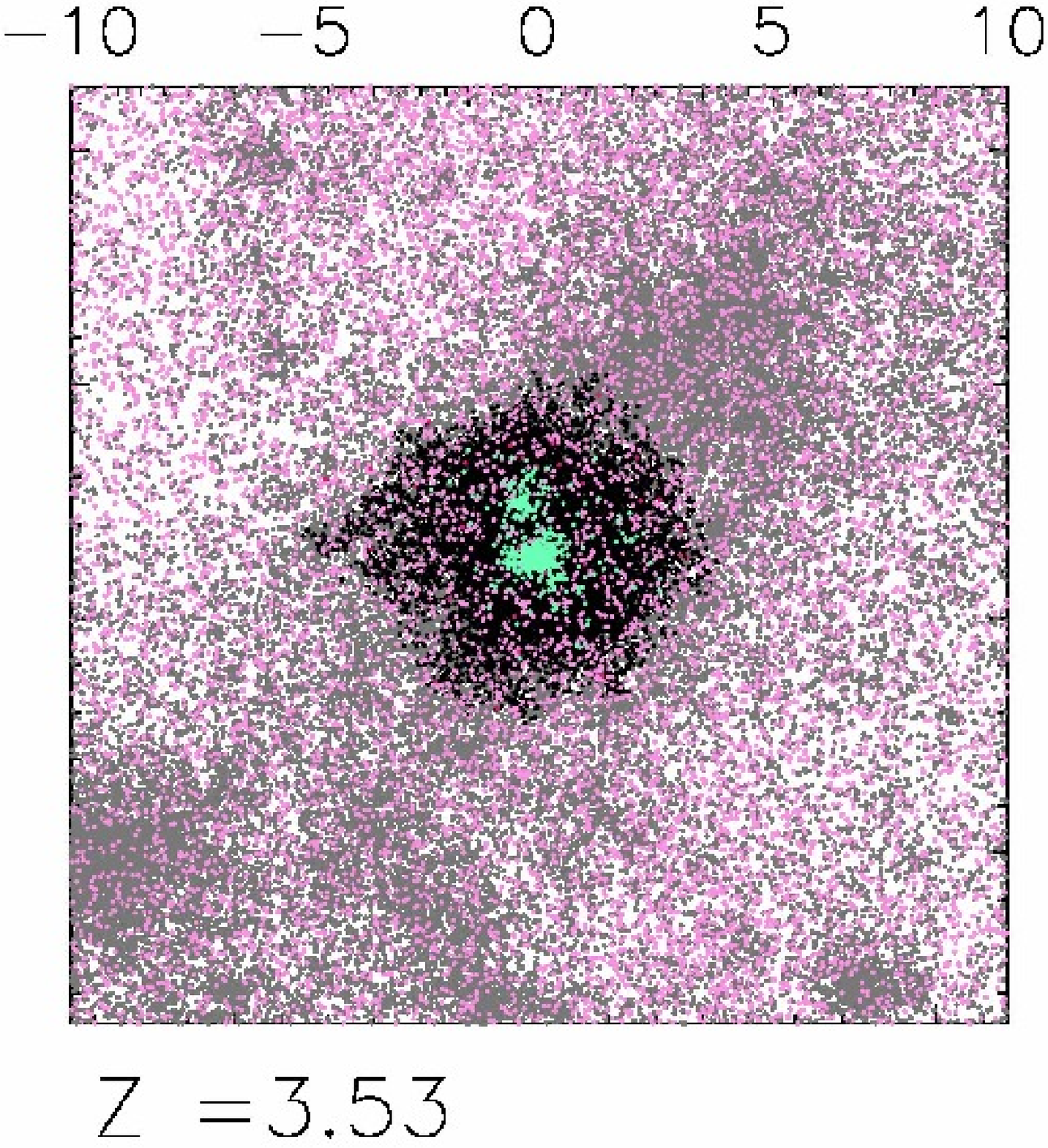} &
      \hspace{-.2in} \includegraphics*[width = .2\textwidth]{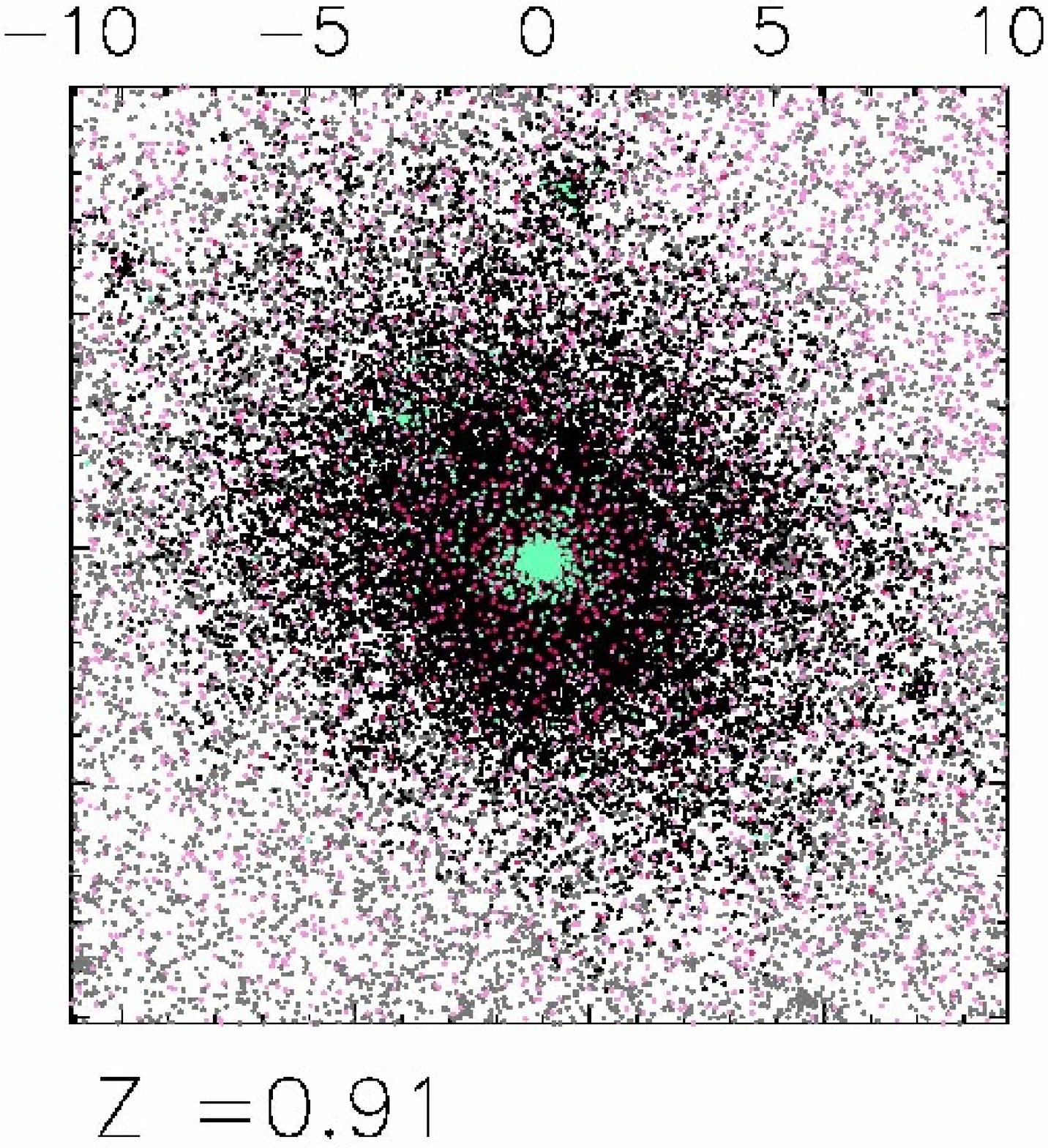} &
      \hspace{-.2in} \includegraphics*[width = .2\textwidth]{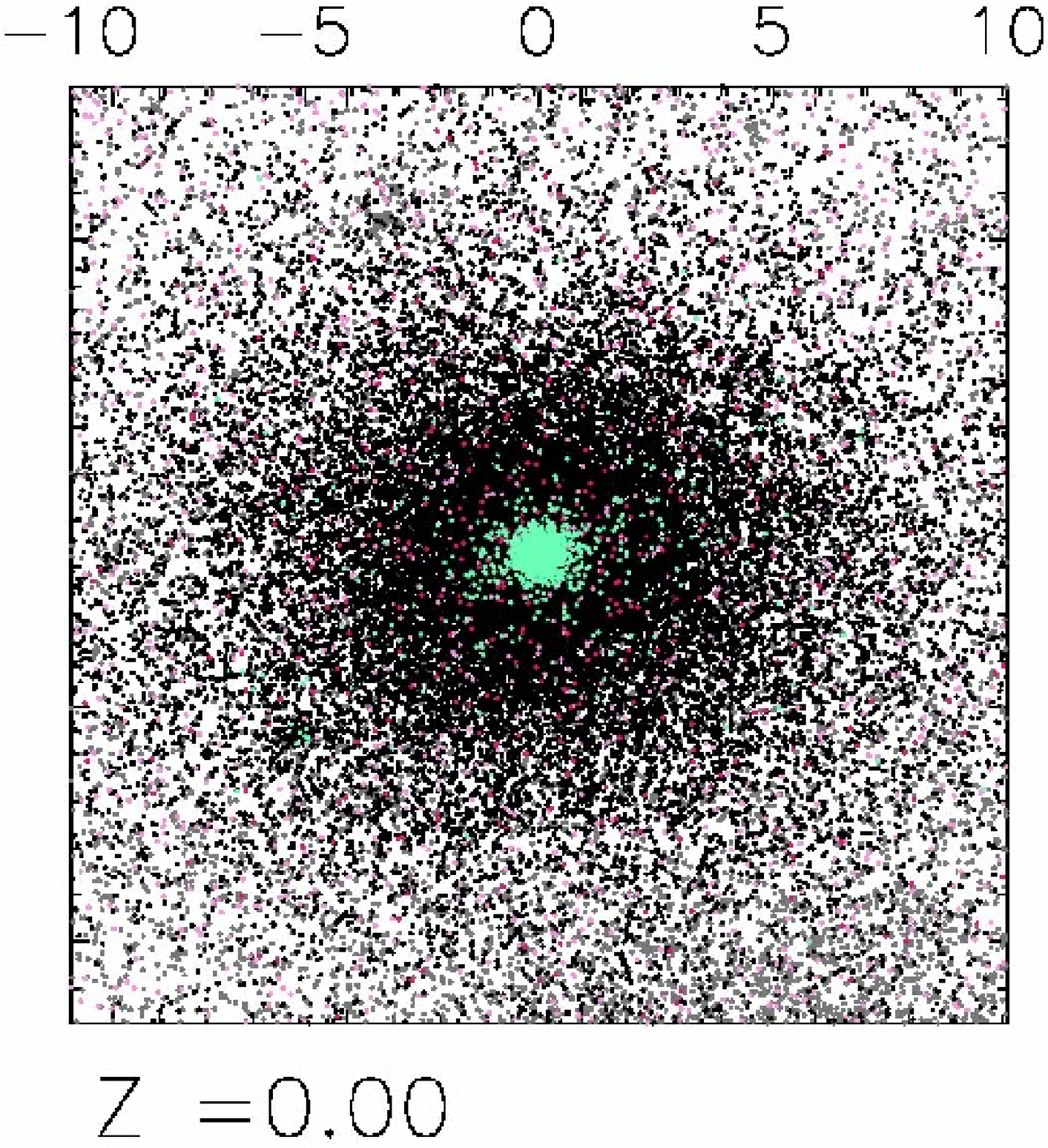} \\ 
      \hspace{-.2in} \includegraphics*[width = .25\textwidth]{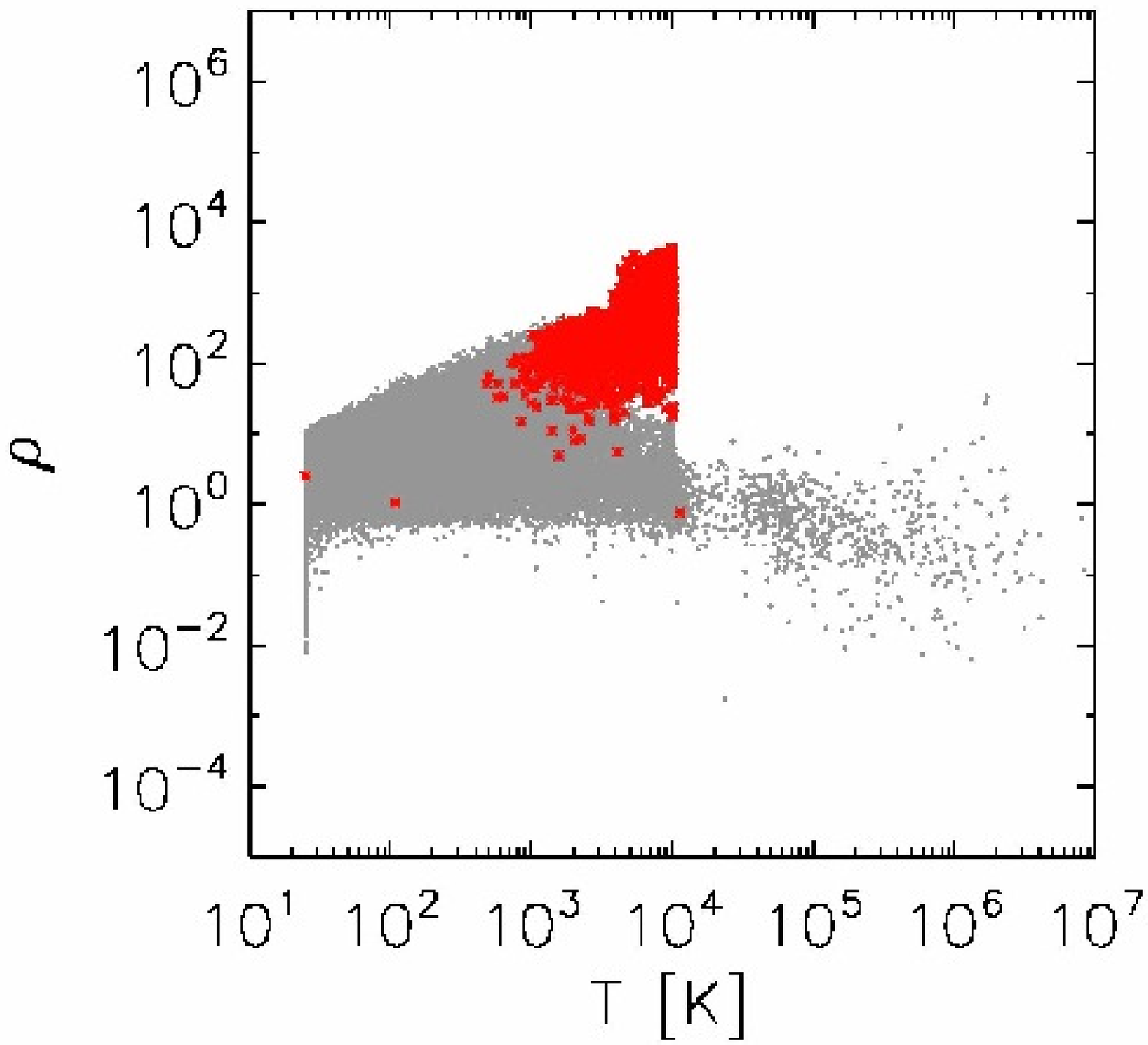} &
      \hspace{-.2in} \includegraphics*[width = .2\textwidth]{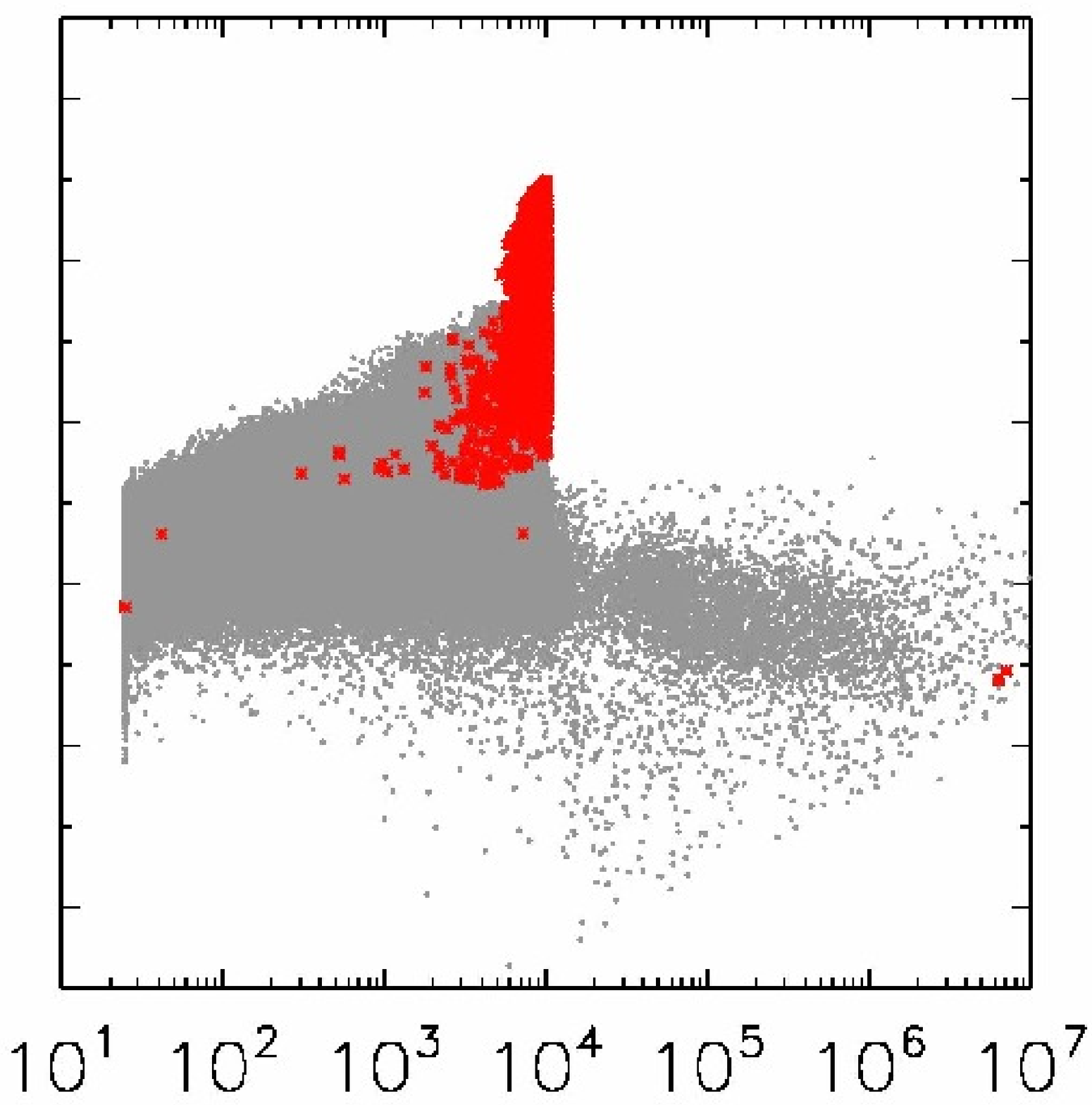} &
      \hspace{-.2in} \includegraphics*[width = .2\textwidth]{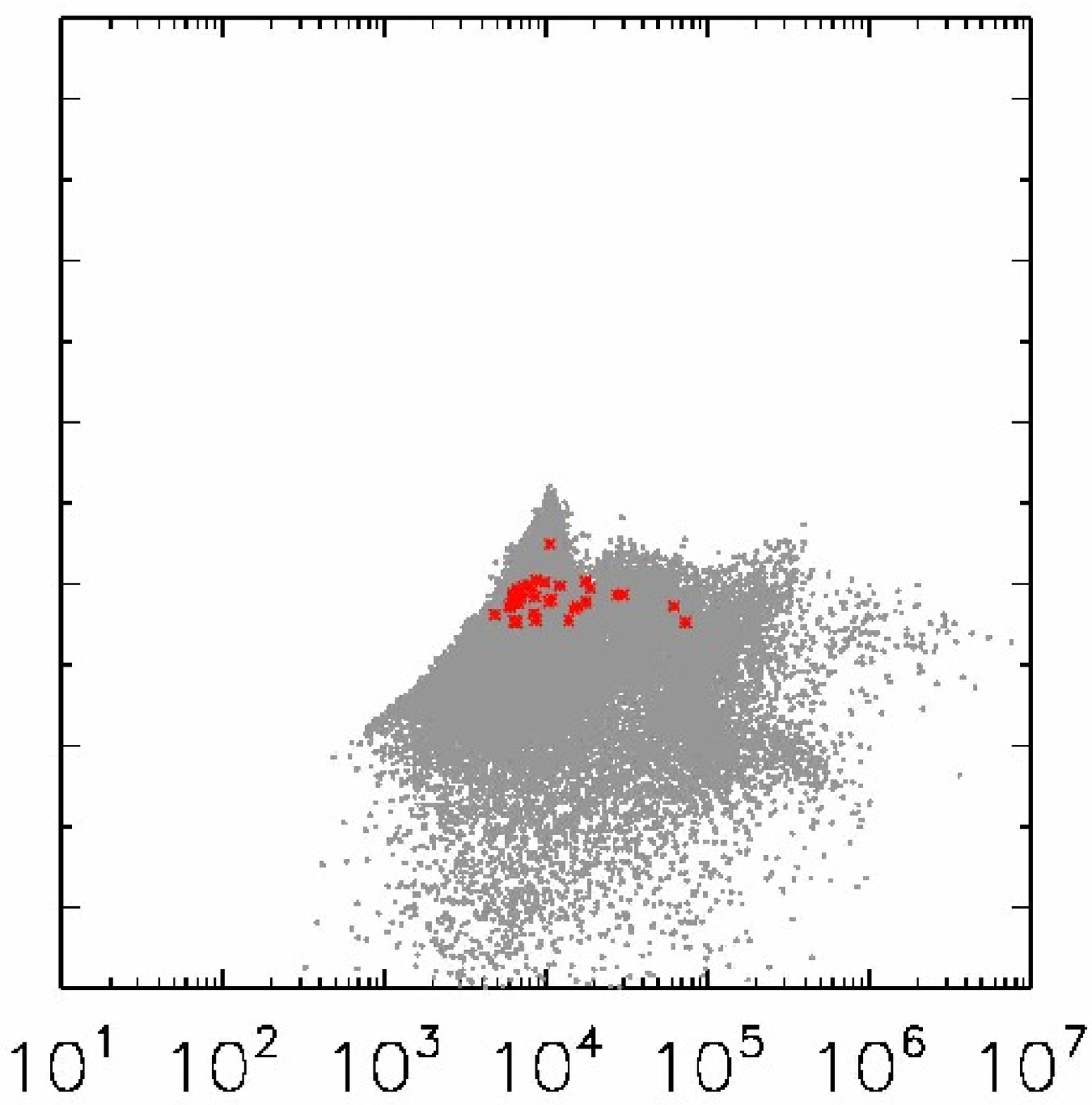} &
      \hspace{-.2in} \includegraphics*[width = .2\textwidth]{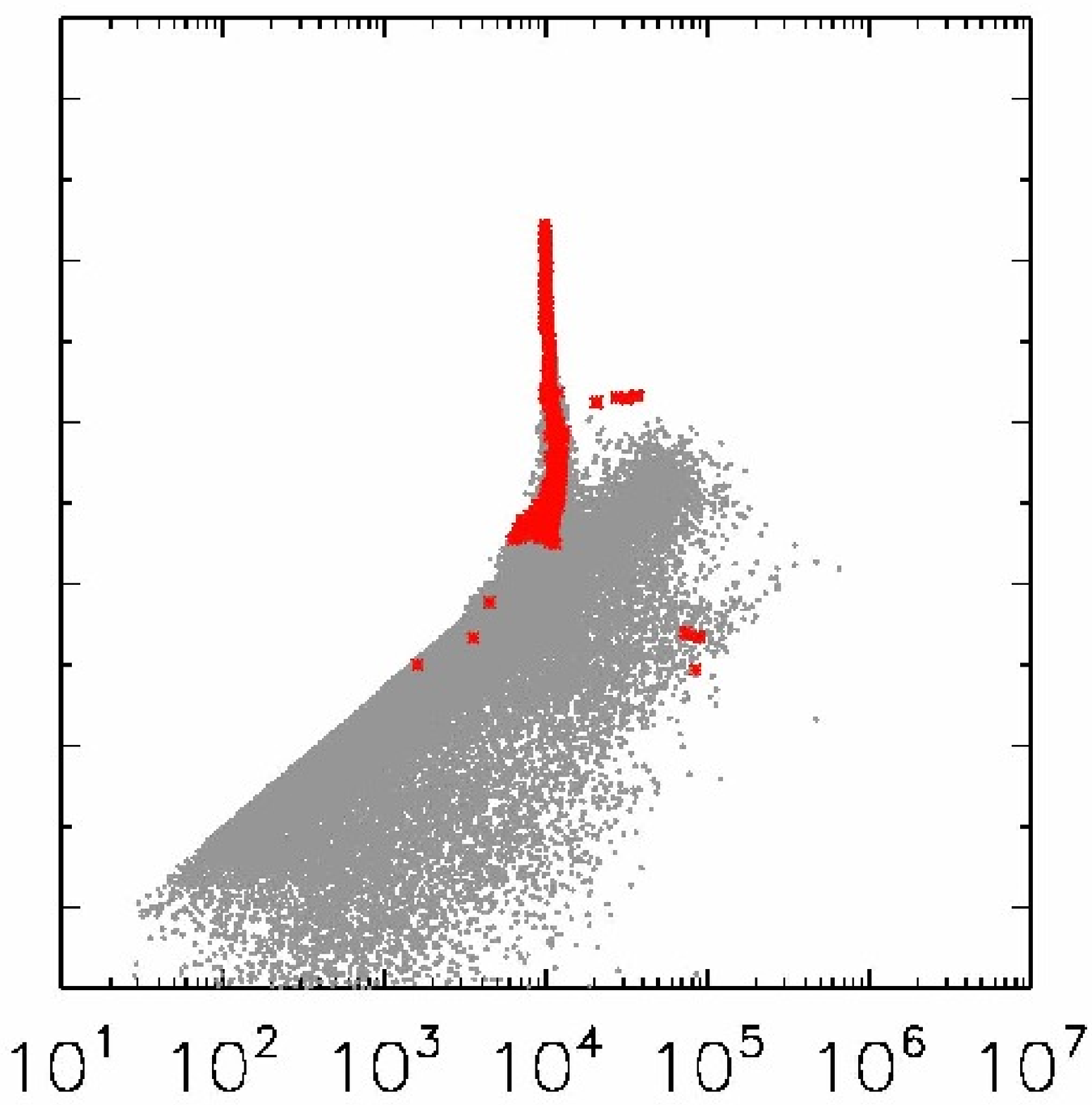} &
      \hspace{-.2in} \includegraphics*[width = .2\textwidth]{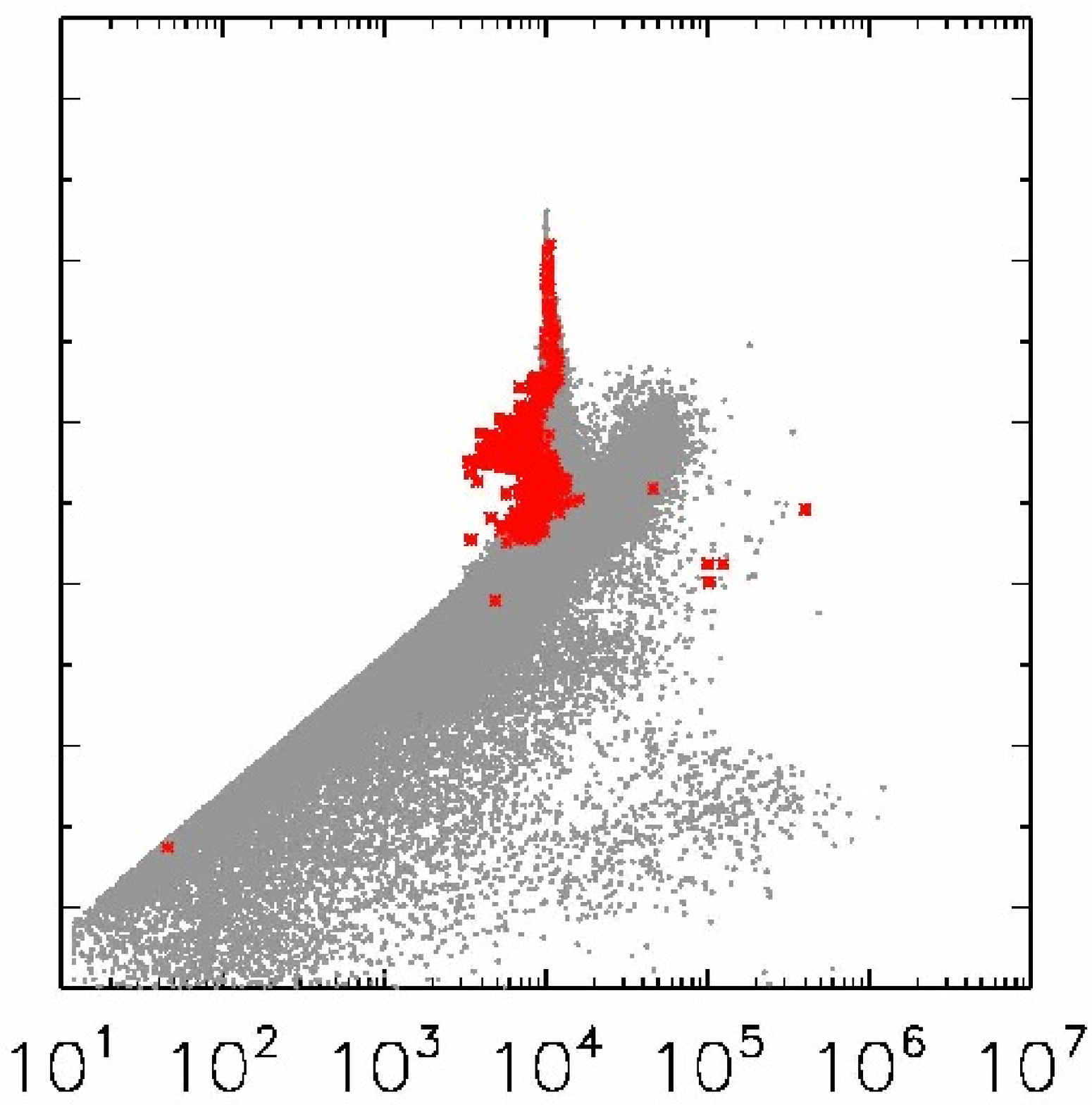}
    \end{tabular}
  \end{center}
  \caption{Top row: Spatial distribution of particles of different
    types at different redshifts of simulation~16 in Table~1. Dark
    matter particles are shown in black (or grey), gas particles in
    red (or purple), depending on whether they are bound to the object
    in the centre, or whether they are part of other haloes or the
    intergalactic medium. Star particles are shown in green. Bottom
    row: Temperature and density of gas particles. Red dots indicate
    gas that is bound to the central halo, while grey dots are for
    particles in all other parts of the simulated volume. Both
    supernova feedback and UV radiation are included in this
    simulation, which has a final halo mass of
    $\sim~7\times10^8\Ms$. It can be seen that the central halo is
    almost gas-free at redshift $z=3.5$, due to the combined effect of
    feedback and the UV background. Feedback heats the gas and blows
    some of it out during the early stages of the evolution. After
    redshift $z=6$, UV radiation heats the remaining gas above the
    haloes virial temperature, quickly removing it from the halo, and
    it also heats the intergalactic medium. Some gas falls back to the
    main halo at later times, but does not lead to significant amounts
    of star formation. Smaller haloes without star formation, and
    hence not subject to feedback, also lose their gas due to the UV
    radiation.}
  \label{fig:xy-begin}
\end{figure*}

Figure~\ref{fig:xy-begin} illustrates different stages in the
evolution of a proto-galaxy (labeled simulation~16 in Table~1)
together with its environment. The top row shows the position of dark
matter, gas and star particles. The scale of the panels is kept
constant in physical coordinates with a side length of 20 kpc, hence
the volume displayed shrinks in terms of comoving coordinates and the
view zooms in on the central galaxy as the redshift decreases from
left to right. In the first two columns, the filamentary structure of
the environment is still recognisable, together with a number of
smaller haloes that have accumulated gas, but not yet begun star
formation. The bottom row shows the distribution of gas particles on
the density-temperature plane, both within and outside of the most
massive halo.

As the halo forms, gas begins to fall in, contracts and gets
heated. At a temperature of 10$^4$ K, radiative cooling becomes so
efficient that the gas can contract essentially isothermally, until
the central density reaches the threshold for star formation, as
described in Section~\ref{sec:sf}. At $z=12.3$, which
corresponds to the leftmost column of Figure~\ref{fig:xy-begin}, the
first stars have already formed in the central object, and supernovae
of type II have started heating the gas, already pushing some of it
out. This is visible also in the bottom row of
Figure~\ref{fig:xy-begin}, where the gas particles that start
appearing to the right of $10^4$ K, which indicates that they have
been heated by supernovae, are no longer bound to the halo.

\begin{figure}
  \includegraphics*[scale = .5]{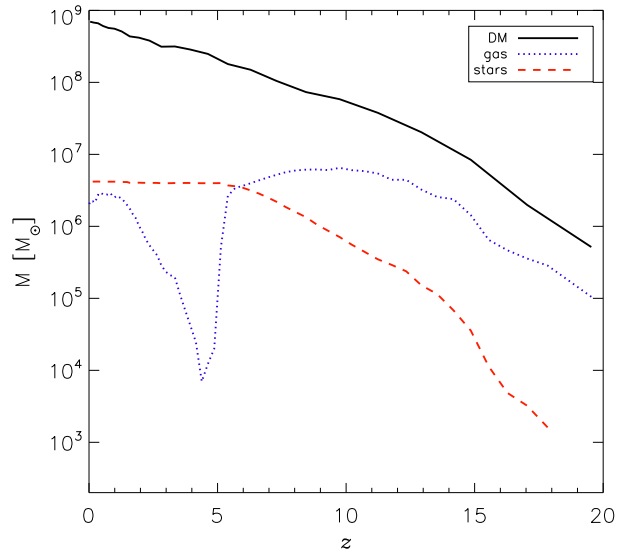}
  \caption{Evolution of the bound dark matter mass (solid black), gas
    mass (dotted blue) and stellar mass (dashed red) as function of
    redshift for simulation~16, the same simulation that is shown in
    Figure~\ref{fig:xy-begin} in several snapshots. The simulation
    includes cooling, star formation, supernova feedback and a cosmic
    UV background, but no self-shielding. It reaches a virial mass of
    $\sim 7\times10^8~\Ms$, and a stellar mass of $\sim4\times10^6
    \Ms$ at $z=0$. Other properties are summarised in Table~1.}
  \label{fig:components}
\end{figure}

The total masses of the three components; dark matter, gas and stars
identified as belonging to the halo by a friends-of-friends algorithm,
are shown as a function of time in Figure~\ref{fig:components}. Star
formation in the galaxy continues for about one Gyr, as more gas gets
accreted and cools, whilst supernovae of both type II and type Ia
continue to expel the interstellar medium. Ejection and heating
balance accretion and cooling at $z=9$, and the star formation rate
peaks at $z=8$. By redshift $z=6$, the star formation rate has
already decreased by a factor of two from its peak value of $3 \times
10^{-1} \Ms$ yr$^{-1}$ due to feedback.

At $z=6$, the UV background suddenly switches on. In this particular
model, it is sufficient to heat the remaining gas above the virial
temperature of the halo in a very short time, resulting in its
expulsion, and a sharp end to star formation. Some gas falls back at a
later stage, but does not reach sufficient density for significant
star formation.

It can also be seen in Figure~\ref{fig:components} that the dark
matter halo in this simulation continues to grow over time through
accretion and minor mergers. It is worth noting that throughout the
period of star formation, from the onset around redshift $z=16$ to the
end shortly after redshift $z=6$, the halo mass is several times
smaller than the final value, which might be observed today. This
behaviour is common to all of our simulations, independent of the
baryonic physics. It contributes to the high efficiency of the winds
in our models. It also suggests that the impact that supernova
feedback might have had during the history of a particular dwarf
galaxy not only depends on its `mass' as it is presently observed, but
also on the co-evolution of its star formation and the assembly of its
halo at earlier times. This fact is taken into account explicitly in
semi-analytic models like that of \cite{Ferrara-2000}, but it is
overlooked in non-cosmological simulations that assume collapse in a
static potential.

In our simulations, the halo continues to grow unperturbed up to
$z=0$. This is not necessarily true for haloes of satellite galaxies,
which may have experienced truncation upon infall
\citep{Nagai-2005}. However, at least according to our models, it is
likely that star formation would have finished before a typical infall
redshift of $z\sim1$ or below \citep{Li-2008}. While we therefore do
not expect environmental effects to significantly alter the stellar
population, they may further skew the correspondence between observed
halo masses today, and gravitational potential in place at the epoch
of star formation. We also have to assume that the late infall of gas,
leaving in some cases a small amount of gas at $z=0$, which would be
in disagreement with observations, is prevented in the Local Group
environment.

\section{The Role of Physical Processes}\label{sec:processes}

As we have seen, feedback from supernovae is sufficient to expel gas
from the shallow potential wells of forming dwarf-galaxies, and it is
responsible for regulating star formation at least up to redshift
$z=6$. In order to investigate the relative importance of feedback and
UV heating, and to disentangle their respective contributions over
time, we have performed test simulations where only one of the two
processes is included.

\begin{figure}
  \includegraphics*[scale = .5]{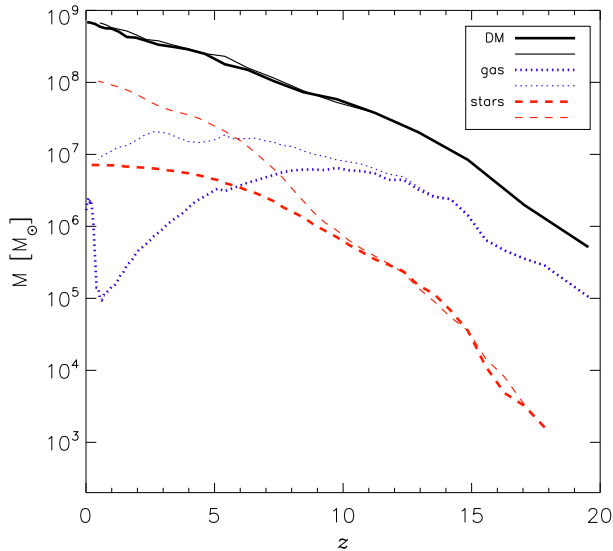}
  \caption{Evolution of the dark matter mass (solid black), gas mass
    (dotted blue) and stellar mass (dashed red) as functions of
    redshift, for simulations~25 and~11. Initial conditions and the
    final dark-matter mass of $\sim 7 \times 10^8 \Ms$ are identical
    to those of simulation~16, shown in Figure~\ref{fig:components},
    but the evolution is different. The thick lines show the evolution
    of simulation~25, where feedback is the only source of heating,
    whereas the thin lines are for simulation~11, that includes UV
    radiation from redshift $z=6$ but no feedback. In the first case,
    feedback alone is sufficient to remove most of the gas, but more
    slowly compared to Figure~\ref{fig:components}. In the case of
    simulation~11, in the absence of feedback, the decline in the gas
    mass is solely due to consumption and conversion to
    stars. Furthermore, without feedback, the UV background present
    from $z=6$ has almost no effect on the gas mass or the star
    formation rate. The resulting stellar masses are vastly different:
    $7\times10^6\Ms$ without UV for simulation~25, and $\sim 10^8 \Ms$
    for simulation~11 without feedback.}
  \label{fig:components-feedback}
\end{figure}

Figure~\ref{fig:components-feedback} illustrates two such `incomplete'
scenarios. We show the evolution of dark matter, gas and stellar mass,
for simulation~25, where feedback is the {\it only} source of thermal
energy, and for simulation~11, where UV radiation is included, but
stellar feedback is ignored. They can be compared with our reference
simulation~16 in Figure~\ref{fig:components}, where the combined
effect of supernova feedback and UV background radiation are
shown. All three simulations have identical initial conditions and
numerical resolution. While the growth of the dark matter mass appears
unaffected by the baryonic physics, the dashed and dotted lines, which
indicate the stellar and gas mass, respectively, show large
differences. The outflow induced by feedback in simulation~25 causes
the thick dotted line representing the gas mass in
Figure~\ref{fig:components-feedback} to peak at about $z=9$ and
decline thereafter, similar to Figure~\ref{fig:components}. The star
formation rate (not shown) also declines and the thick dashed line,
representing the total stellar mass, increases ever more slowly,
reaching 7 $\times 10^6 \Ms$ at redshift $z=0$. In contrast, the thin
dotted line in Figure~\ref{fig:components-feedback}, which represents
the gas mass without feedback, shows no decline at high redshift. The
total baryon fraction of the halo stays constant at around $1/6$th,
indicating that the late decline of the gas mass is due solely to
consumption by star formation. It is worth reiterating that this
simulation includes the full UV background (see Section~\ref{sec:UV}),
without self-shielding (see
Section~\ref{sec:self-shielding}). However, contrary to the results of
Figure~\ref{fig:components}, we find that when thermal feedback is
ignored, the UV radiation has no effect either. The gas density is so
high that the gas can cool fast enough to balance any heating due to
the cosmic UV background.

\subsection{The Importance of Feedback} \label{sec:feedback}
In summary, we find that feedback alone can blow out all the remaining
gas before redshift $z=0$ even in the absence of photoelectric
heating, albeit at a much slower rate, resulting in a larger number of
intermediate age stars. Even in this case, only between $3\%$ and $
6\%$ of the total amount of gas ever bound to the halo gets turned
into stars, depending on the mass of the object. Most of the gas still
escapes to the intergalactic medium, enriching it with metals.

In simulations without thermal feedback (simulations 10 and 11 in
Table~1, thin lines in Figure~\ref{fig:components-feedback}), the
  picture is drastically different. Not only is the star formation
more efficient during the early stages, the interstellar gas also
becomes so dense that all effects of the UV background radiation
discussed in Section~\ref{sec:UV} are eliminated due to very efficient
cooling. The result is a system of large stellar mass (up to $10^8
\Ms$ in the case of simulation~11, compared to $4 \times 10^6 \Ms$ for
the same initial conditions run with feedback), low mass-to-light
ratio, high metallicity, an abundance of young stars, and a high gas
content. All these properties are incompatible with observations of
Local Group dwarf spheroidals.

We conclude that feedback is necessary to shut down star formation in
those haloes massive and dense enough to cool and begin forming
stars. Under the assumption, supported by observations
\citep[e.g.][]{Lewis-2007, Fraternali-2009}, that at least some of the
local Group dwarf spheroidals have evolved in isolation, these results
suggest that supernova feedback is the key factor in determining their
stellar evolution.

\subsection{The Influence of the UV Background}\label{sec:UV}
\begin{figure*}
  \begin{center}
    \begin{tabular}{ccccc}
      \hspace{-.2in} \includegraphics*[width =
        .25\textwidth]{box-2mass_gas_4_res600/box-xy_left-040.eps} &
      \hspace{-.2in} \includegraphics*[width = .2\textwidth]{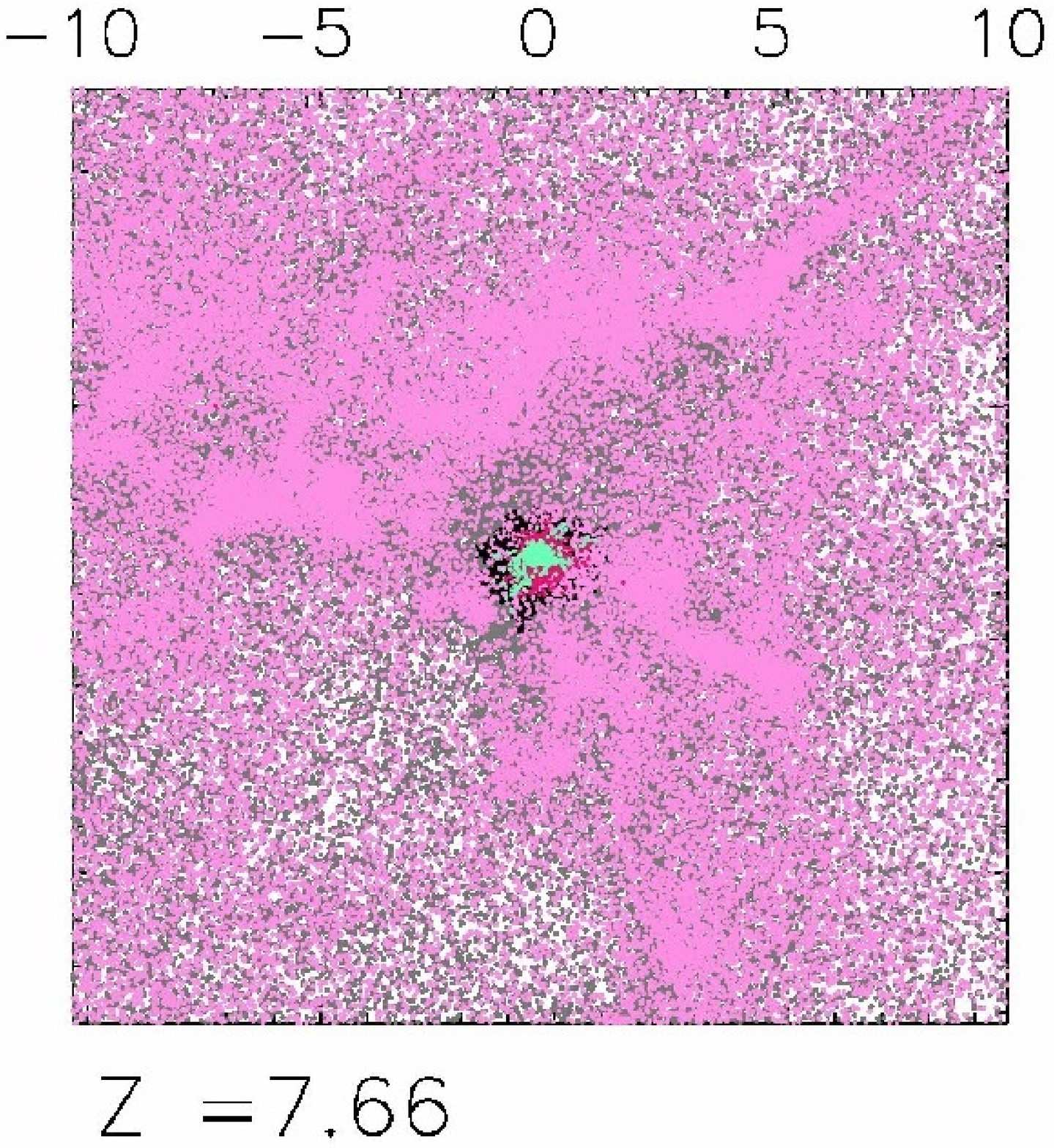} &
      \hspace{-.2in} \includegraphics*[width = .2\textwidth]{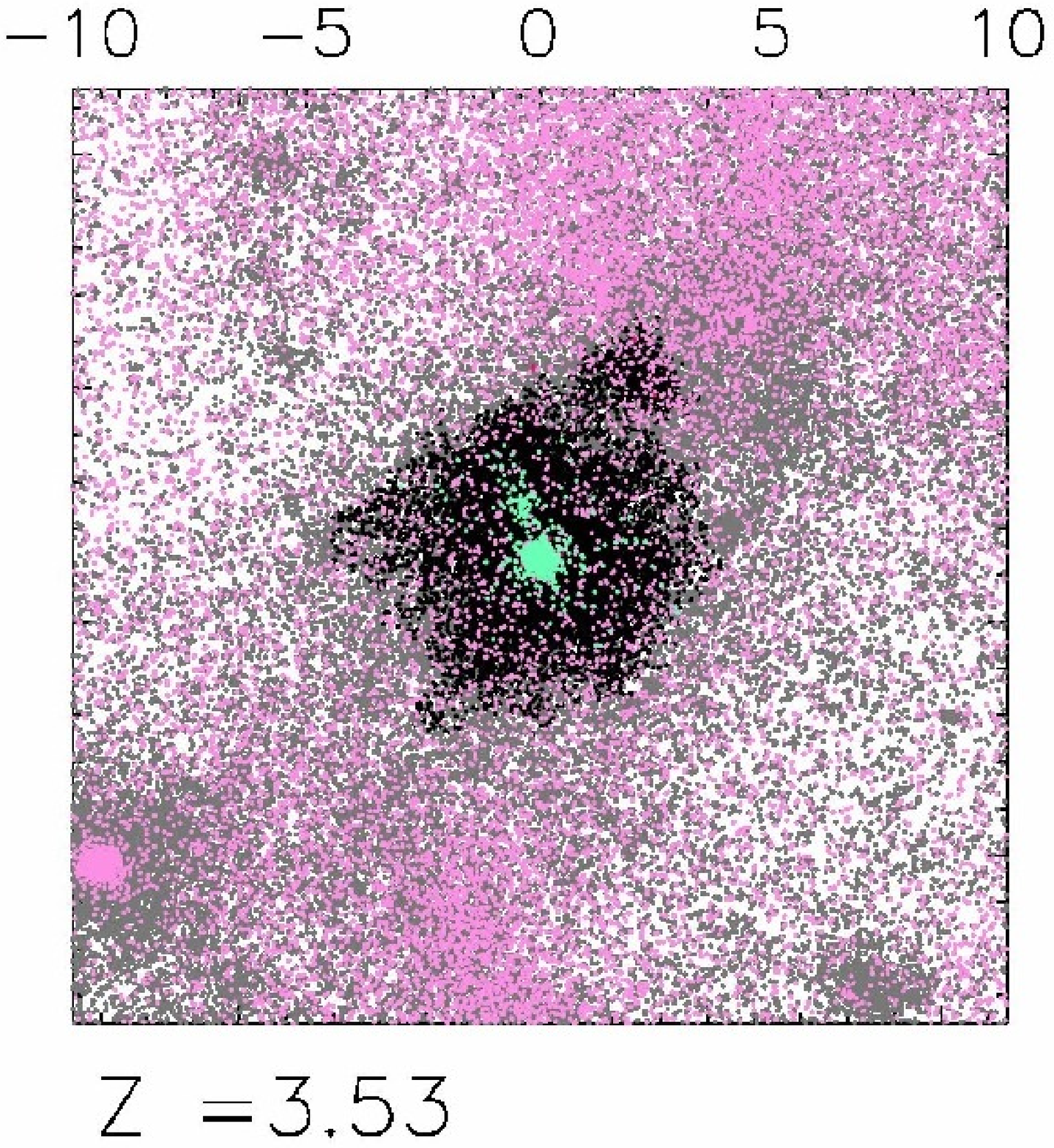} &
      \hspace{-.2in} \includegraphics*[width = .2\textwidth]{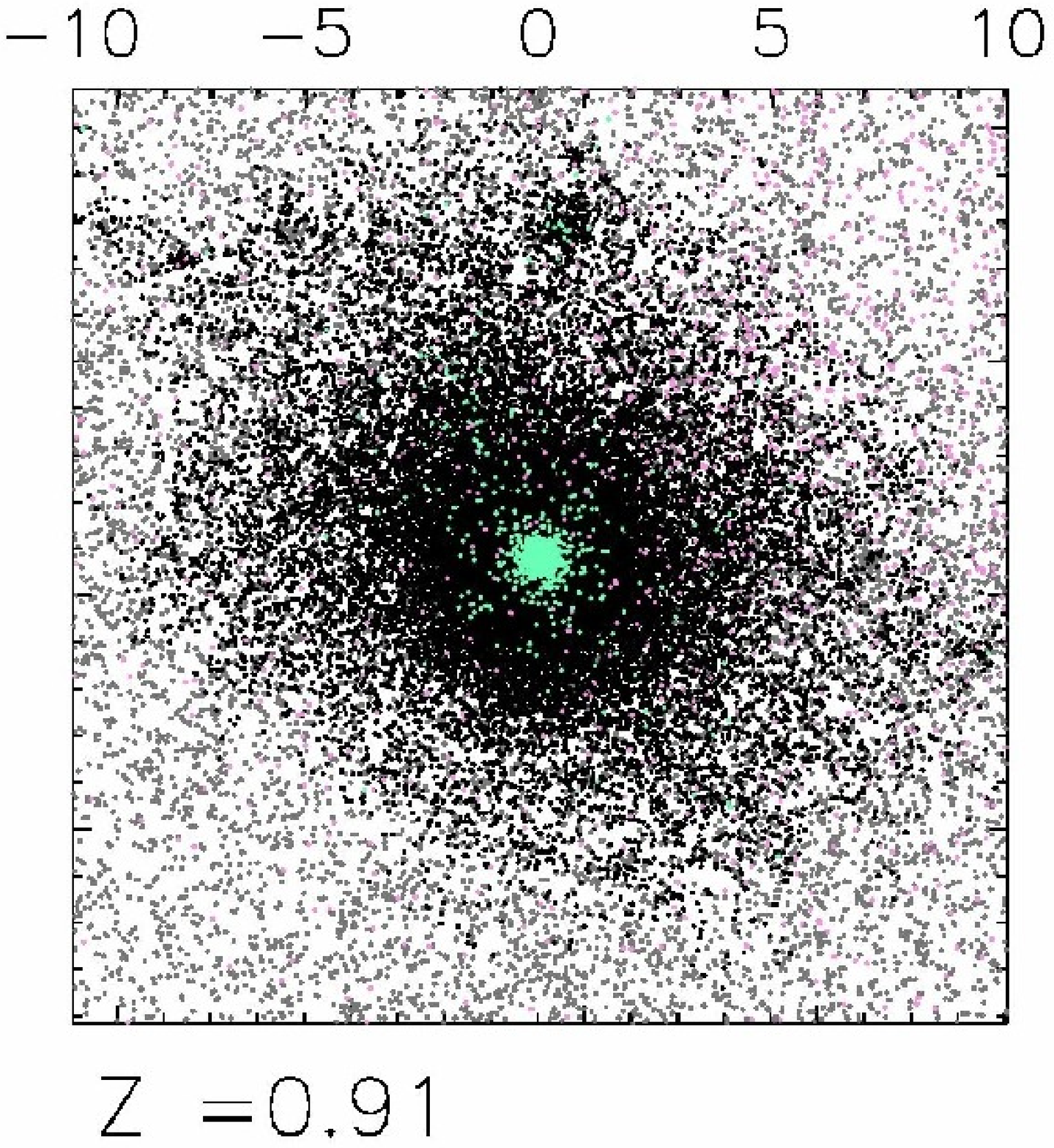} &
      \hspace{-.2in} \includegraphics*[width = .2\textwidth]{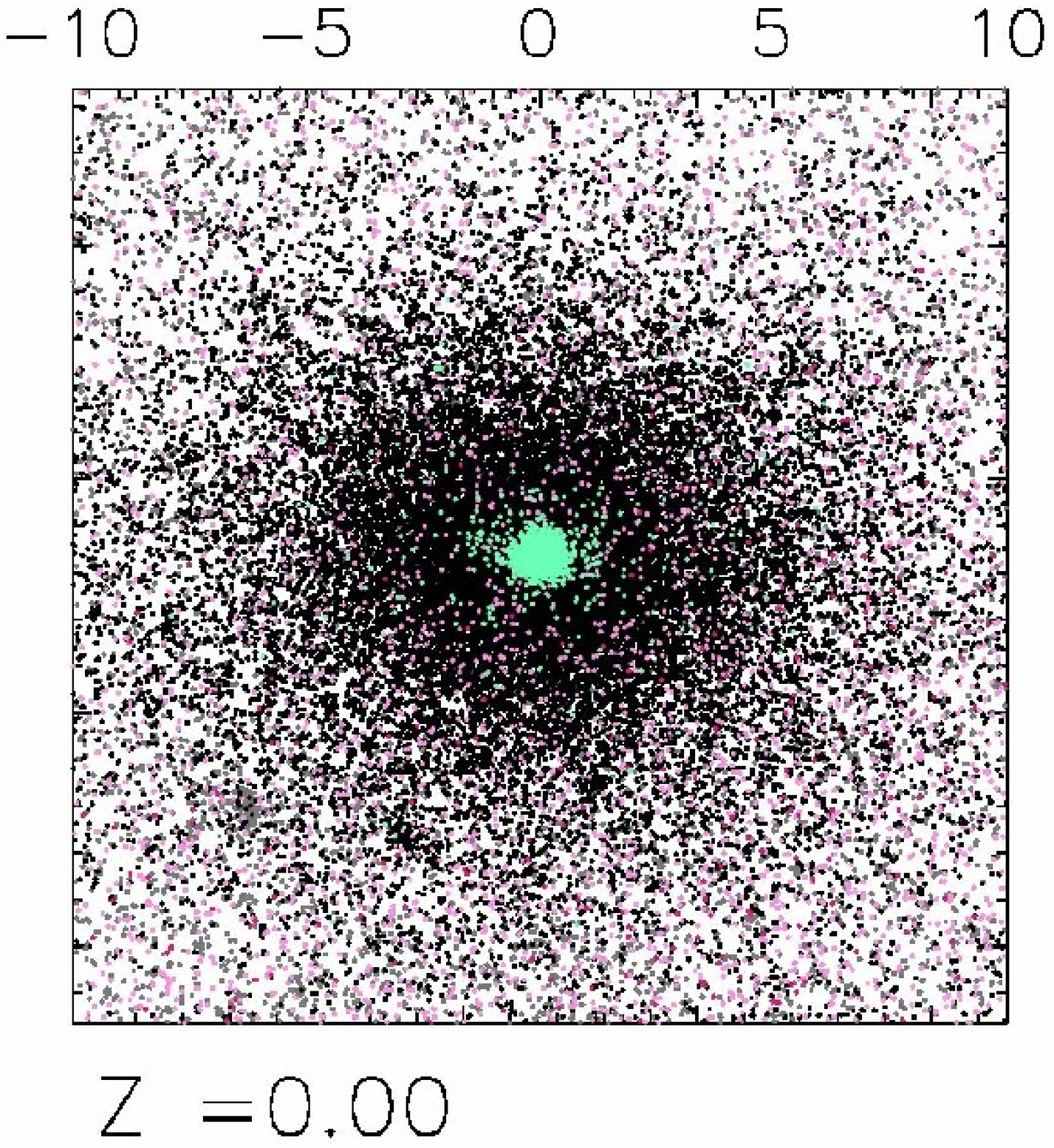} \\ 
      \hspace{-.2in} \includegraphics*[width = .25\textwidth]{box-2mass_gas_4_res600/rho-temp_left-040.eps} &
      \hspace{-.2in} \includegraphics*[width = .2\textwidth]{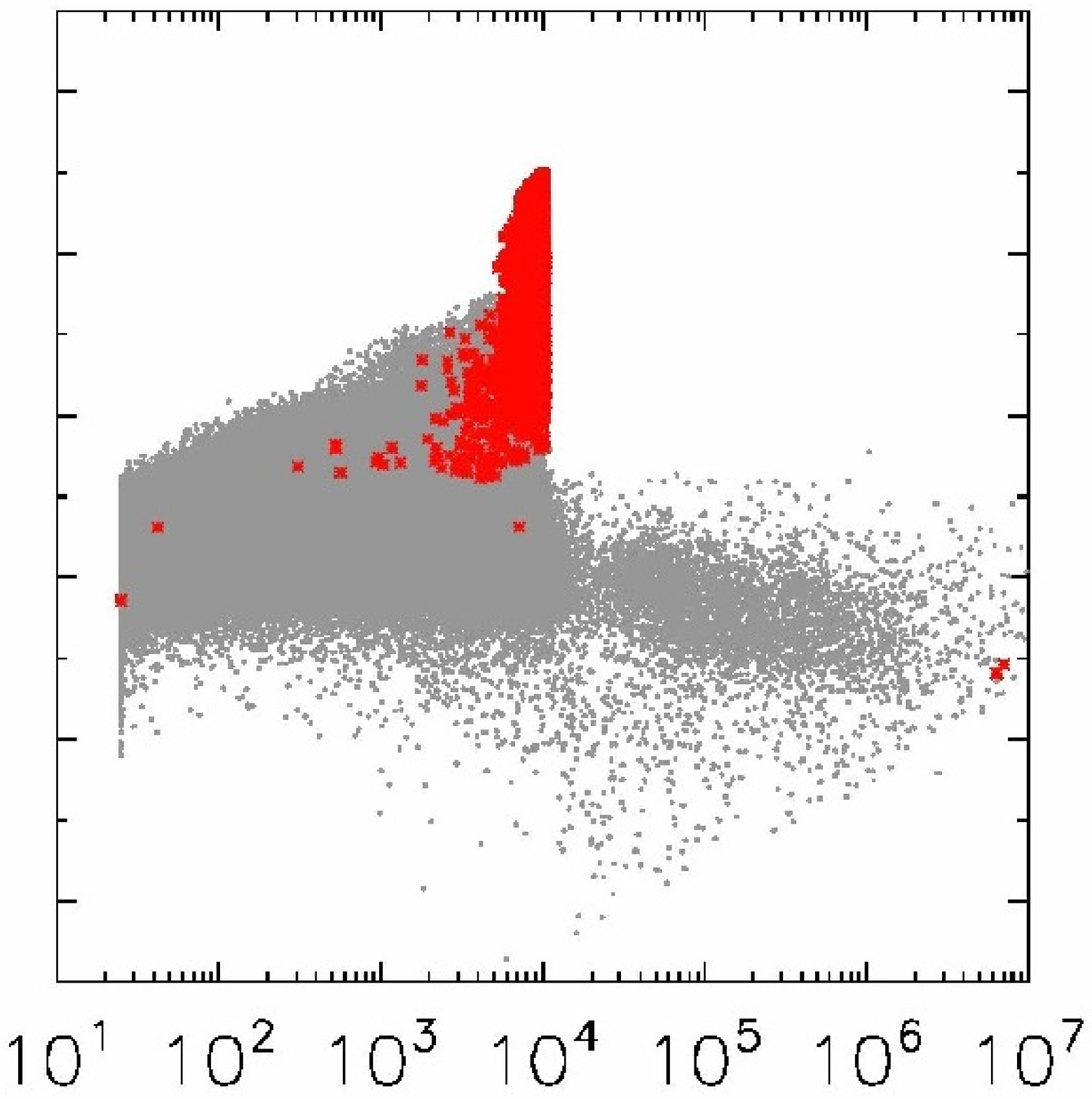} &
      \hspace{-.2in} \includegraphics*[width = .2\textwidth]{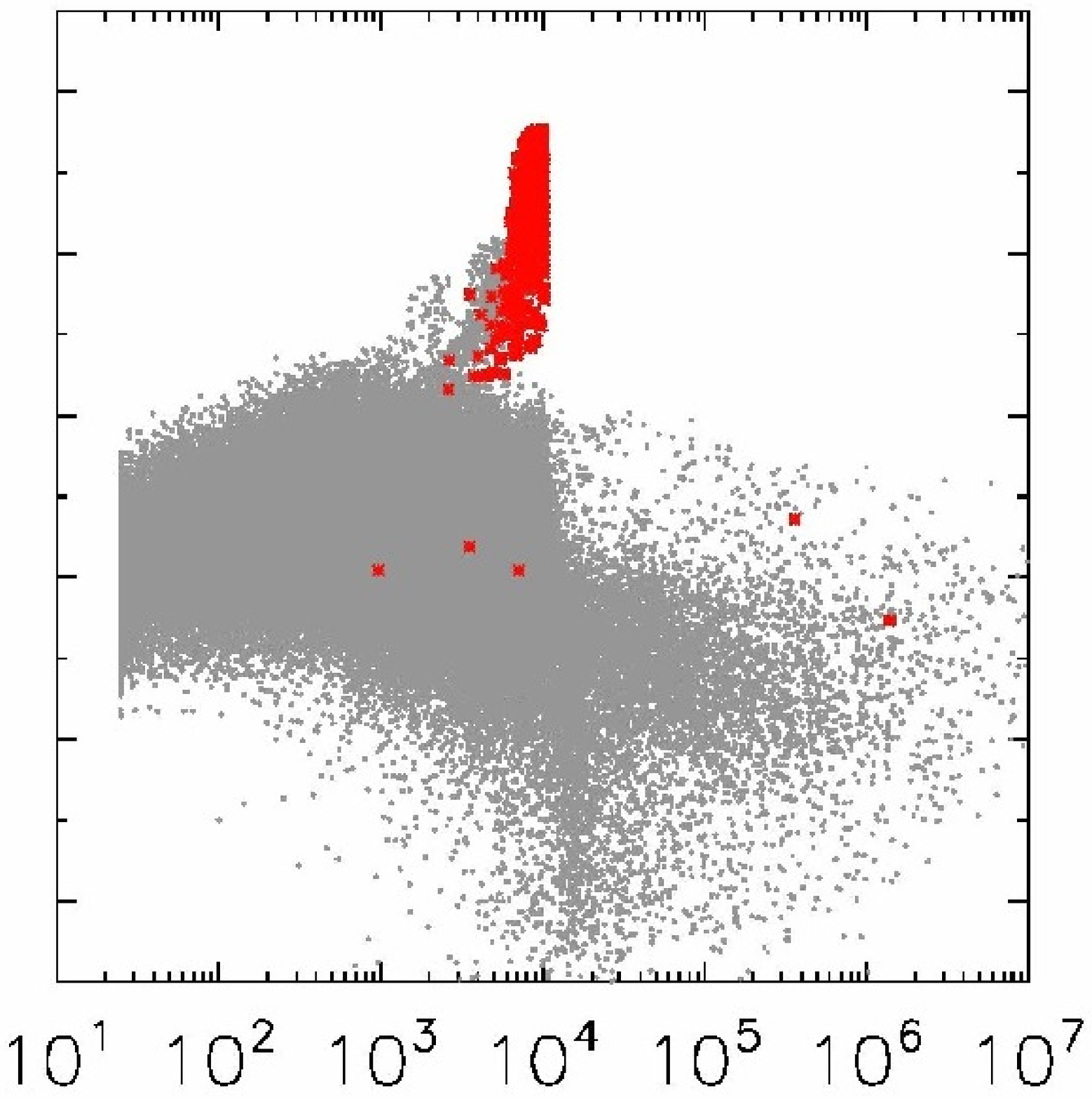} &
      \hspace{-.2in} \includegraphics*[width = .2\textwidth]{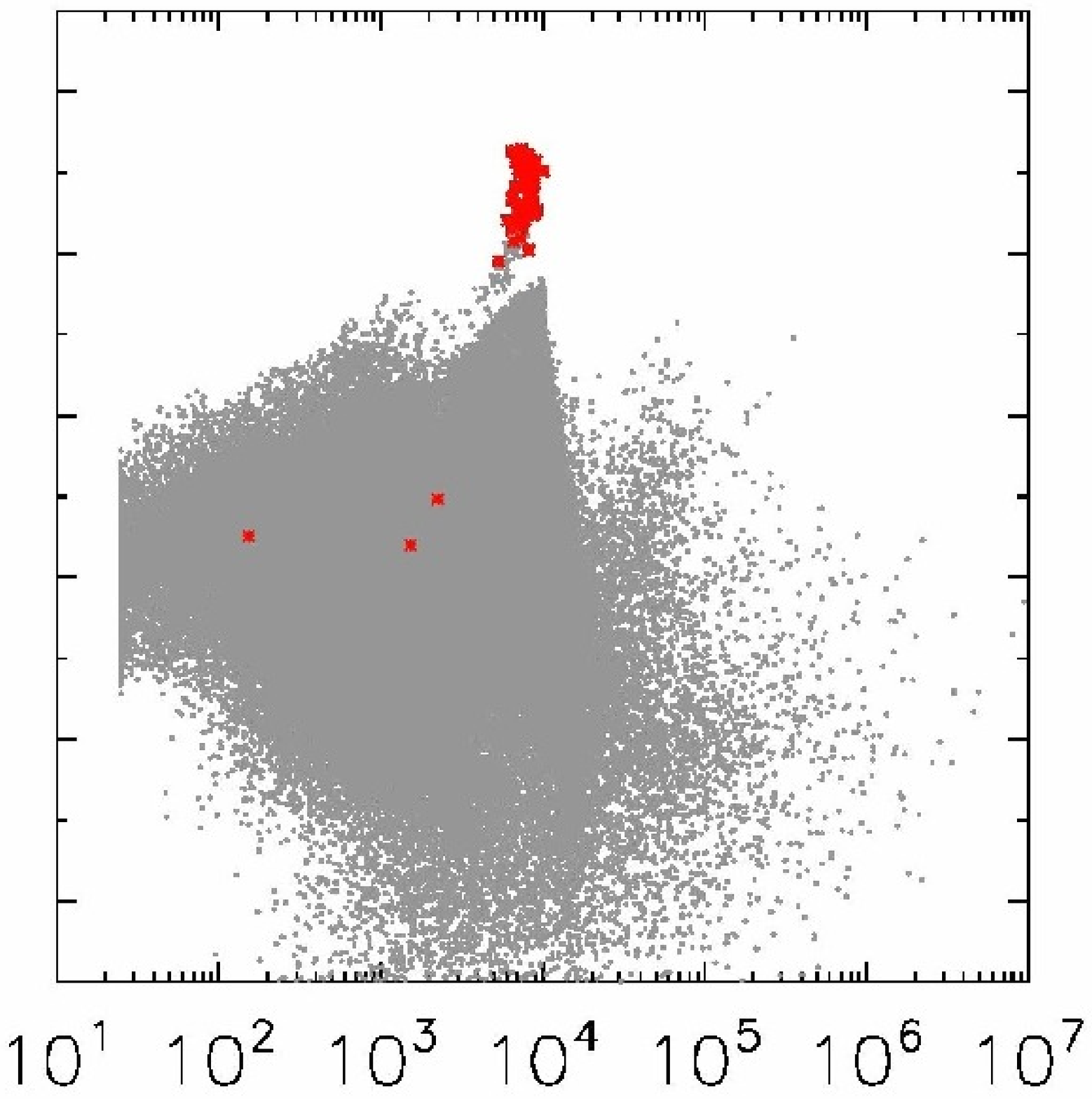} &
      \hspace{-.2in} \includegraphics*[width = .2\textwidth]{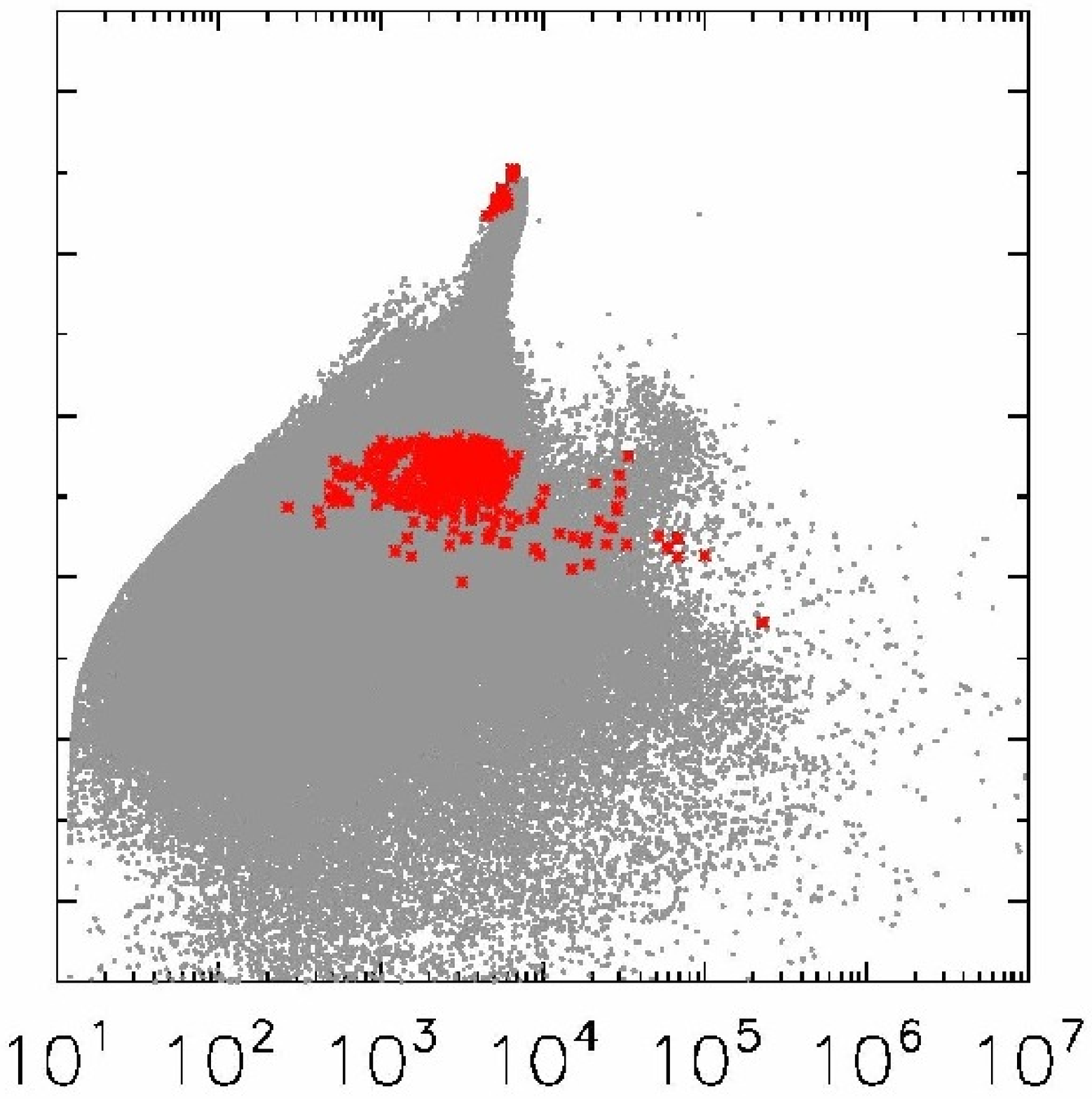}
    \end{tabular}
  \end{center}

  \caption{Top row: Spatial distribution of particles of different
    types at different redshifts of simulation~25. Dark matter
    particles are shown in black (or grey), gas particles in red (or
    purple), depending on whether they are bound to the object in the
    centre, or whether they are part of other haloes or the
    intergalactic medium. Star particles are shown in green. Bottom
    row: Temperature and density of gas particles. Red dots indicate
    gas that is bound to the central halo, while grey dots are for
    particles in all other parts of the simulated
    volume. Simulation~25, also shown in
    Figure~\ref{fig:components-feedback} (thick lines), has initial
    conditions identical to the one shown in
    Figure~\ref{fig:xy-begin}, but contains no UV radiation. It
    reaches a final halo mass of $\sim7\times10^8$. While the system
    still looses almost all its gas, this happens more slowly compared
    to the case with UV radiation, and the stellar mass continues to
    grow beyond redshift $z=6$. The most noticeable difference,
    however, is in the smaller haloes ($10^6~\Ms$ or less), which did
    not form stars early on, and which would now be able to retain
    their gas.}

  \label{fig:xy-no_uv}

\end{figure*}
To further elucidate the influence of the UV radiation, in
Figure~\ref{fig:xy-no_uv}, we show the evolution of a system that
includes feedback but no UV radiation (simulation~25 in Table~1), a
simulation otherwise identical to our reference simulation 16
described in Section~\ref{sec:time-evolution}, which is shown in
Figures~\ref{fig:xy-begin} and~\ref{fig:components}. The most obvious
difference to note when comparing the two sets of figures is in the
low density regions of the intergalactic medium not part of our main
halo, where heating by the UV background is most effective.

We find that all haloes that are not massive enough to accrete
sufficient gas to form stars before $z=6$ subsequently lose their gas
when the effects of photoelectric heating are included. One
consequence of this is that while the main halo grows through
accretion of smaller haloes, these minor mergers are essentially
gas-free and do not trigger renewed star formation. It also supports
the idea that reionization establishes a lower mass threshold for
dwarf galaxy haloes, and so provides part of the solution to the
`missing satellites problem'. However, we do not observe star
formation in these haloes, even when UV radiation is ignored, most
likely due to inefficient cooling as a result of insufficient
resolution.

\begin{figure}
 \subfigure{
     \includegraphics*[scale = .5]{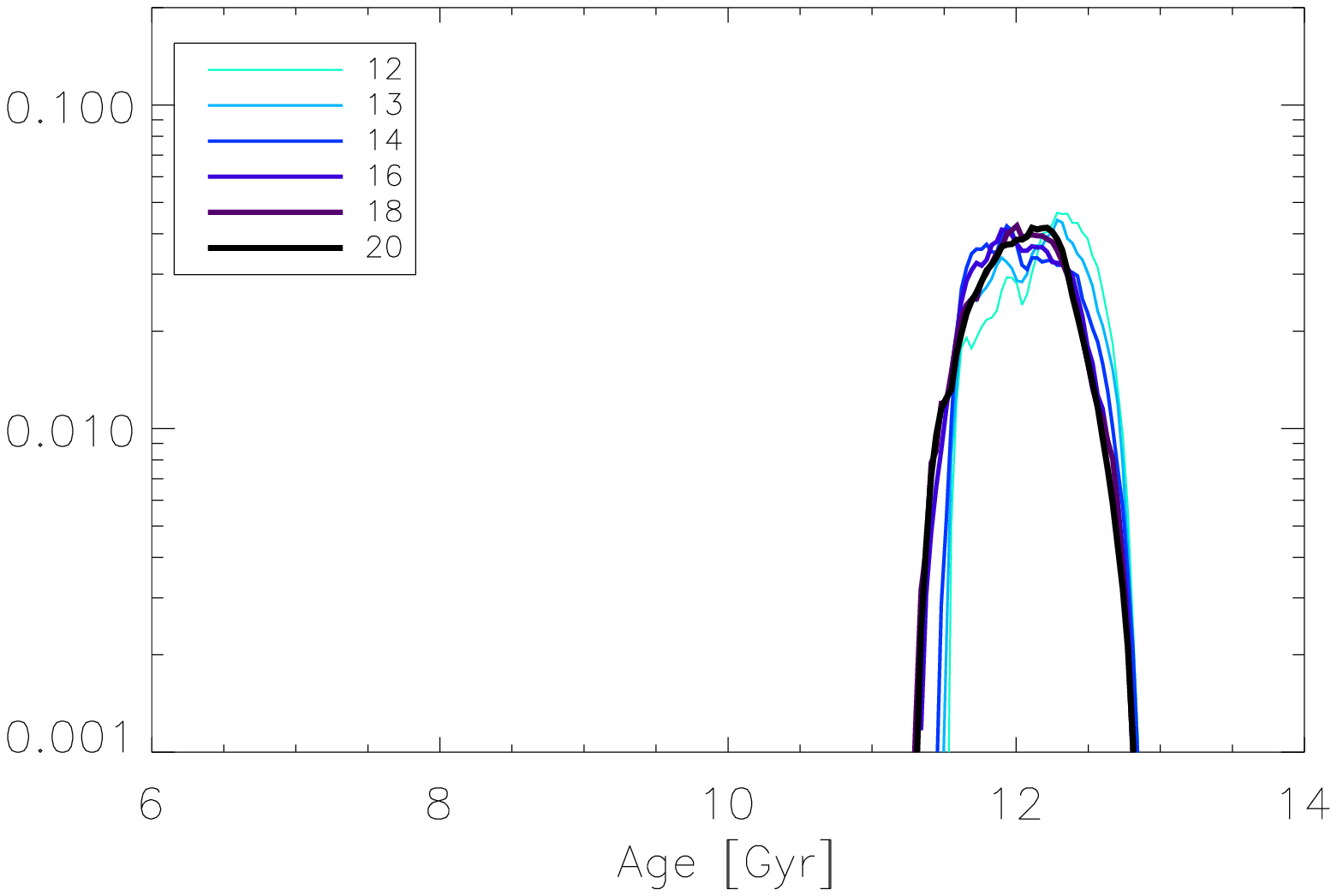}}\\
 \subfigure{
     \includegraphics*[scale = .5]{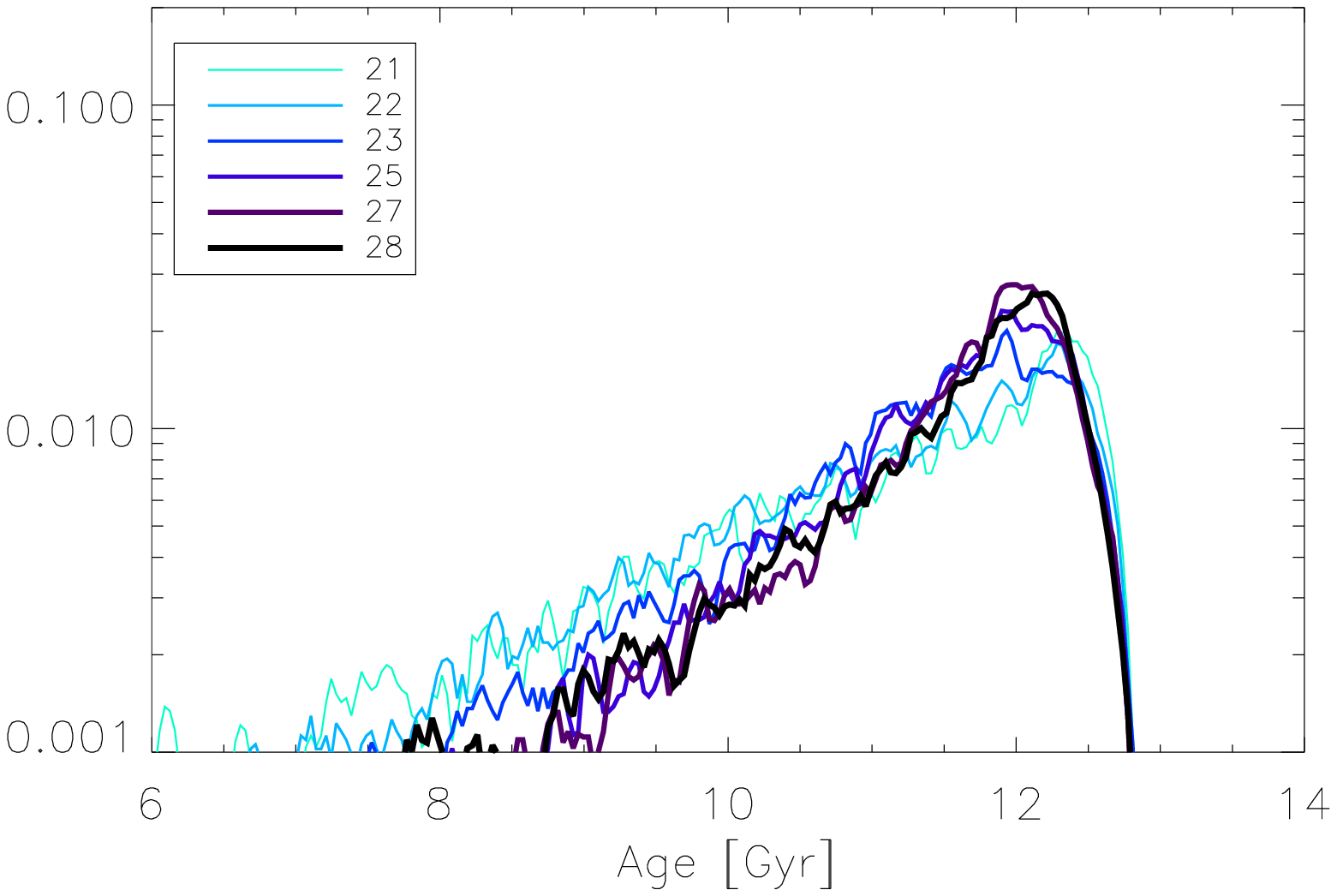}}
\caption{Distribution of stellar ages in Gyr. Simulations of the same
  halo mass are shown in the same colour in the two panels, darker
  colours and thicker lines indicate higher mass. A UV background (but
  no self shielding) is included for simulations 12-20, in the top
  panel, while the UV background is ignored in simulations 21-28, on
  the bottom. The final halo masses range from $2.3\times10^8\Ms$
  (simulations 12 and 21, lightest blue) to $1.2\times 10^9\Ms$
  (simulations 20 and 28, black). Other properties of the simulations
  are summarised in Table~1.}
\label{fig:ages-uv}
\end{figure}

For those objects we consider as the progenitors of dwarf galaxies
listed in Table~1, i.e. those massive enough to acquire dense,
star-forming gas at high redshift, we have already shown that the UV
background alone is not sufficient to shut off star formation at
$z=6$. Feedback is required in order to make the gas diffuse and to
reduce its radiative cooling efficiency. However, when this
requirement is met, the UV background radiation has a strong influence
on the star formation timescale. While the difference in total stellar
mass in a given dark matter halo varies from 30~\% for the largest
system to a factor of three for the smallest system we have studied,
this alone may not be enough to be discriminatory when comparing
mass-to-light ratios with observations. However, a substantial
difference is also found in the age and metallicity
distributions. When star formation continues beyond reionization, many
intermediate age stars with high metallicities are formed, and, as can
be seen in Table~1, this causes the median metallicity to saturate
around $[\mathrm{Fe}/\mathrm{H}] = -1.1$ when the UV background is
ignored. Moreover, while properties such as the total stellar mass and
metallicity also depend on the initial mass and (less strongly) on
other parameters of the model (see Section~\ref{sec:parameters}), the
age distribution of the stars does not. In all cases with UV
radiation, the termination of star formation at $z=6$ results in a
narrow age distribution, as can be seen in Figure~\ref{fig:ages-uv},
while in all cases without the UV background, there is a pronounced
intermediate-age tail. This is significant, because there appear to be
examples of both types of galaxies in the Local Group
\citep{Grebel-2004}. We have tested the dependence on the overall UV
intensity, and found qualitatively similar results when we decreased
it by up to a factor of 10 from the \cite{Haardt-1996}
level. Furthermore, while there may be local variations in the UV
background, particularly at higher redshift and from sources other
than quasars \citep[e.g.][]{Ciardi-2003}, the mean free path of UV
photons in the intergalactic medium is on the order of tens of Mpc
\citep{Bolton-2004}. This seems to rule out the possibility that the
observed variation in star formation histories within the Local Group
dwarf spheroidals can be attributed solely to small-scale variations
of the UV background radiation level originating from quasars at large
distances.

\subsection{The Effect of Self-Shielding} \label{sec:self-shielding}

\begin{figure*}
  \begin{center}
    \begin{tabular}{ccccc}
      \hspace{-.2in} \includegraphics*[width = .25\textwidth]{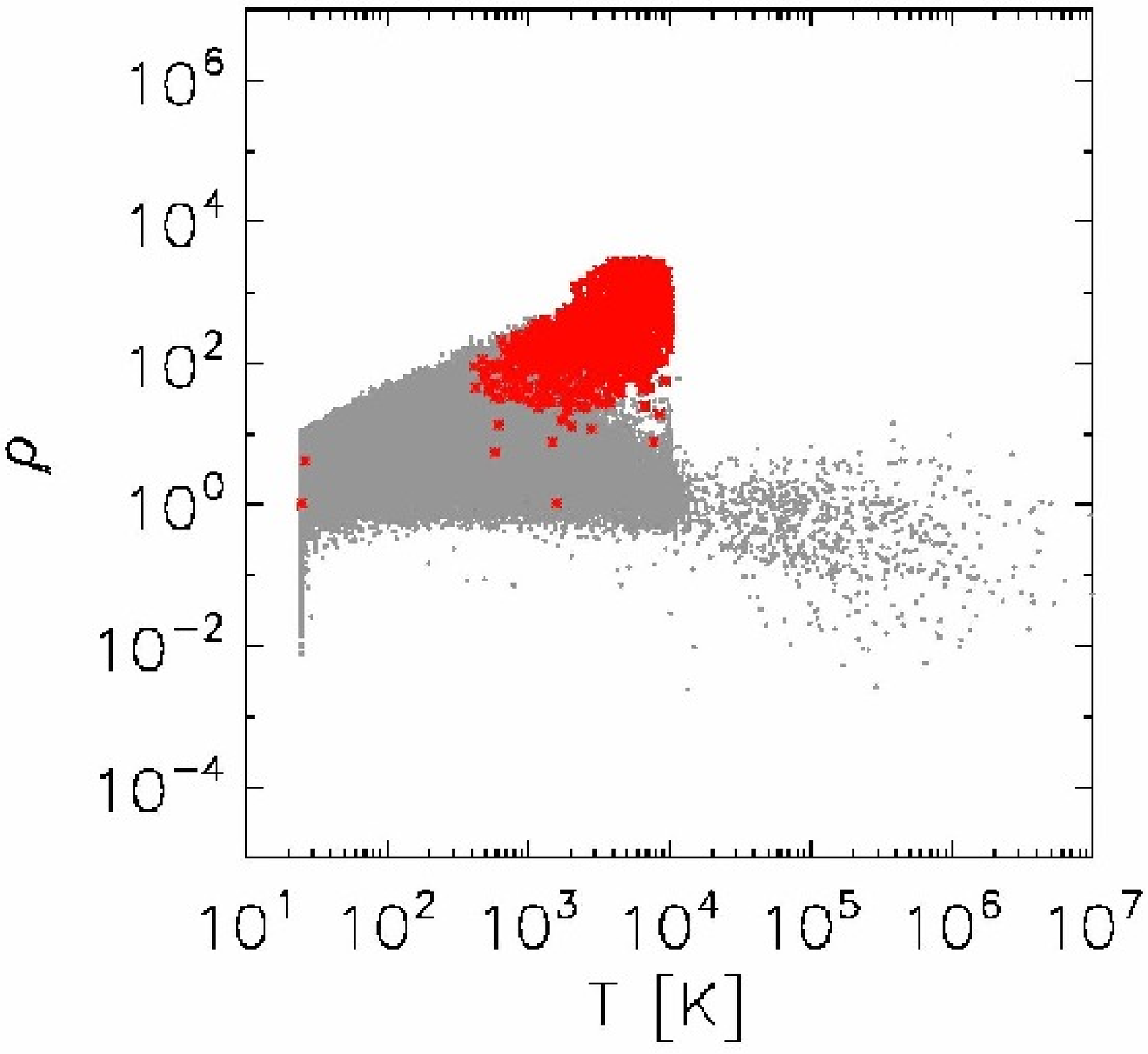} &
      \hspace{-.2in} \includegraphics*[width = .2\textwidth]{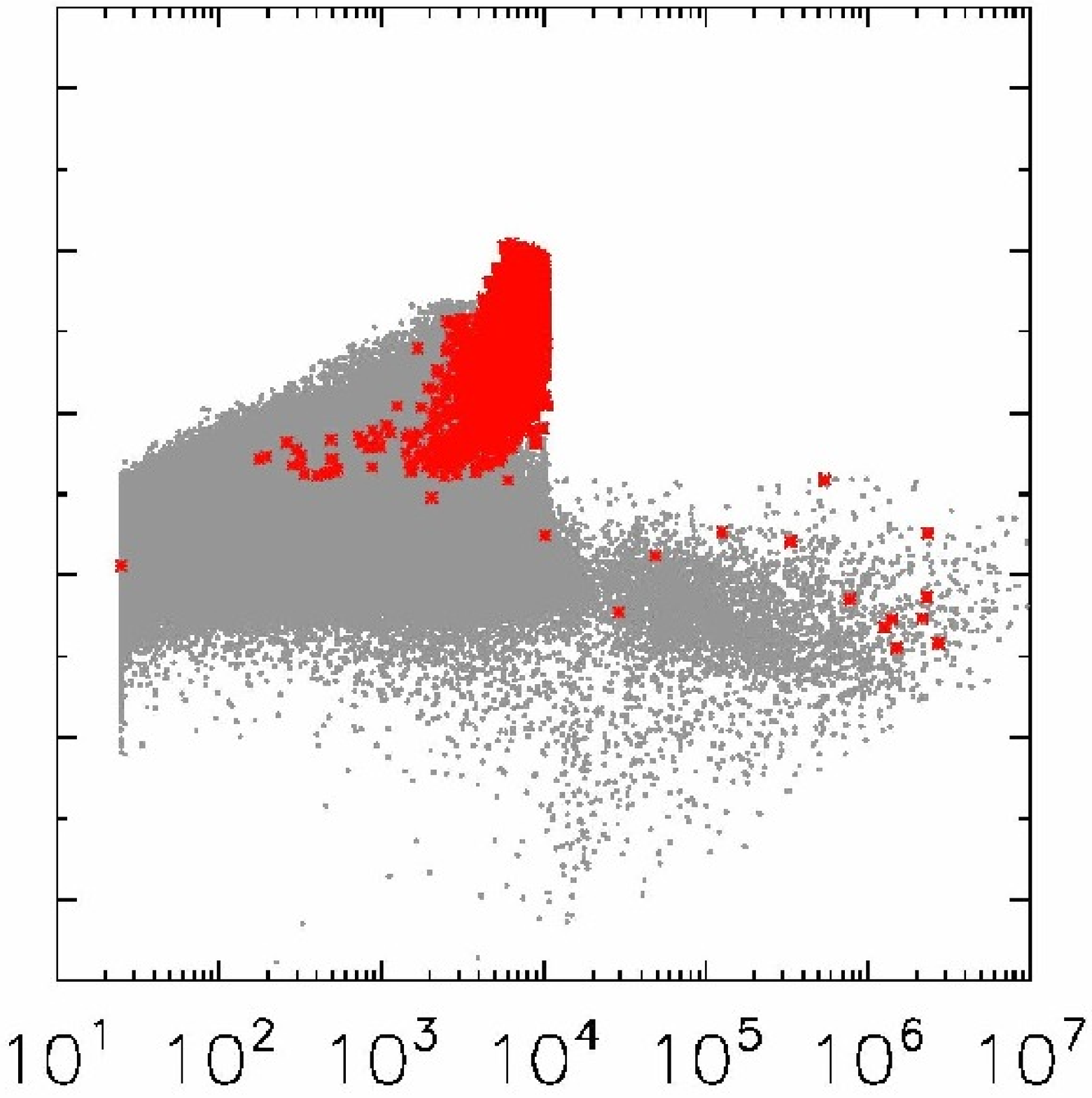} &
      \hspace{-.2in} \includegraphics*[width = .2\textwidth]{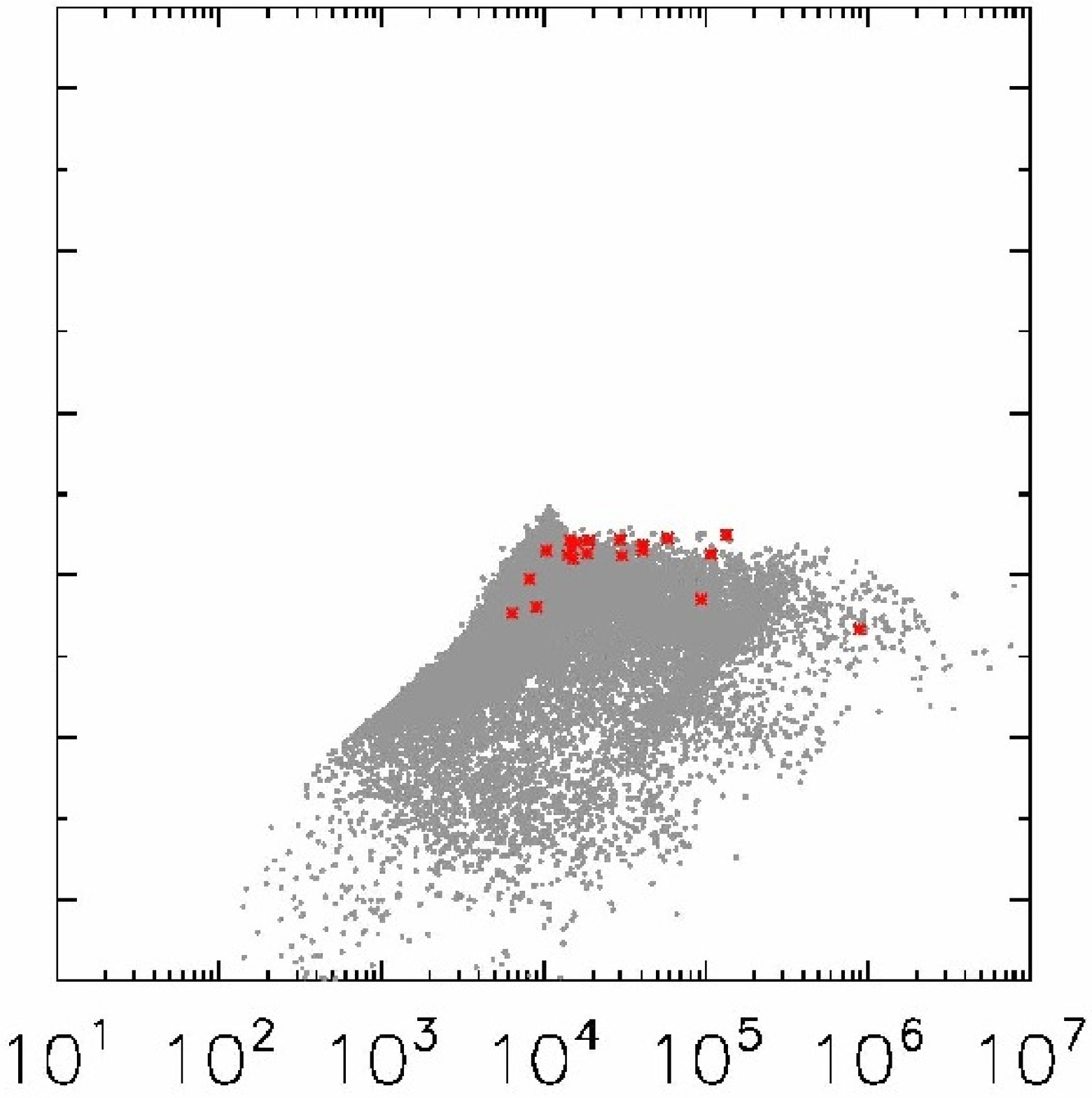} &
      \hspace{-.2in} \includegraphics*[width = .2\textwidth]{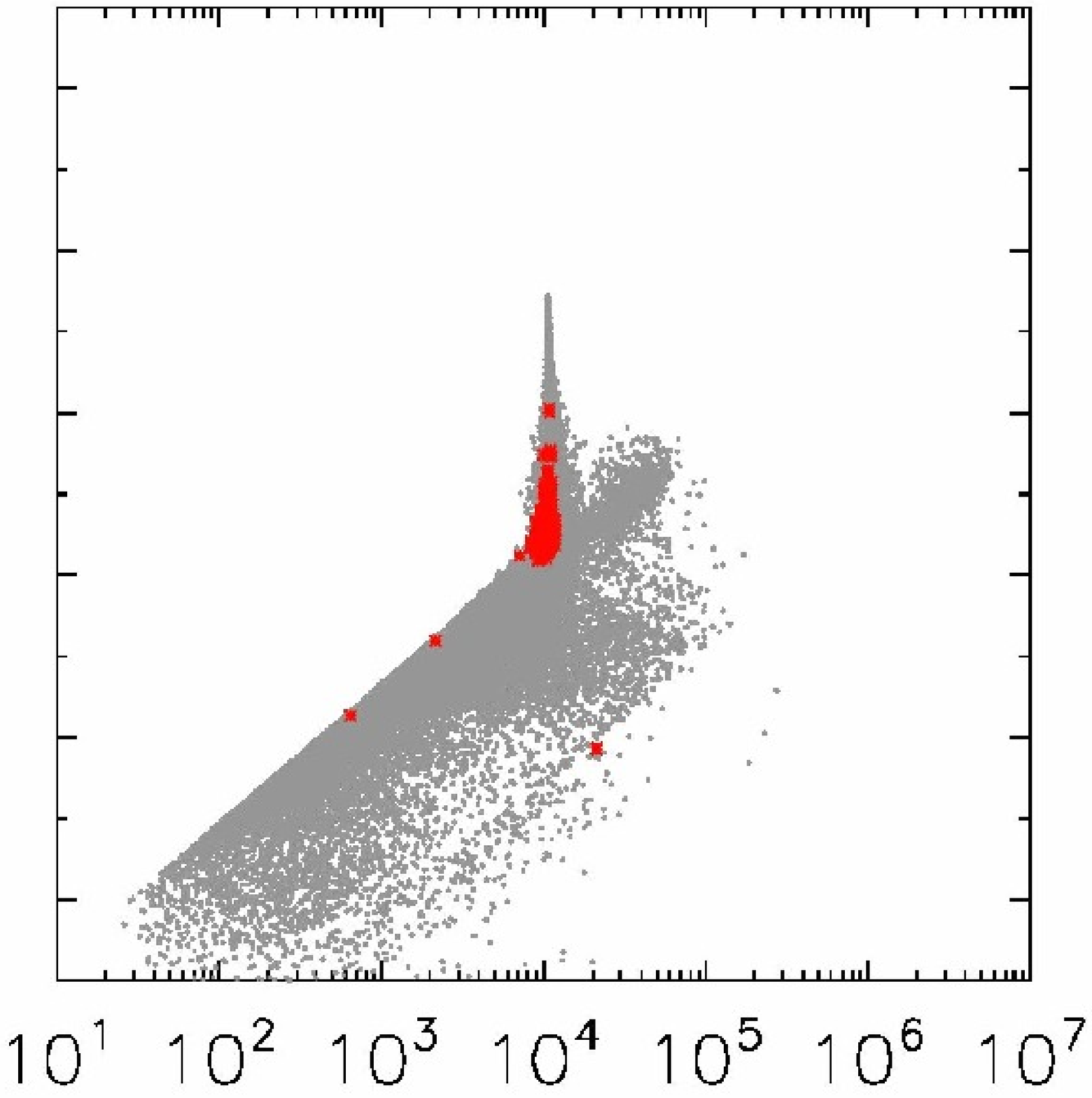} &
      \hspace{-.2in} \includegraphics*[width = .2\textwidth]{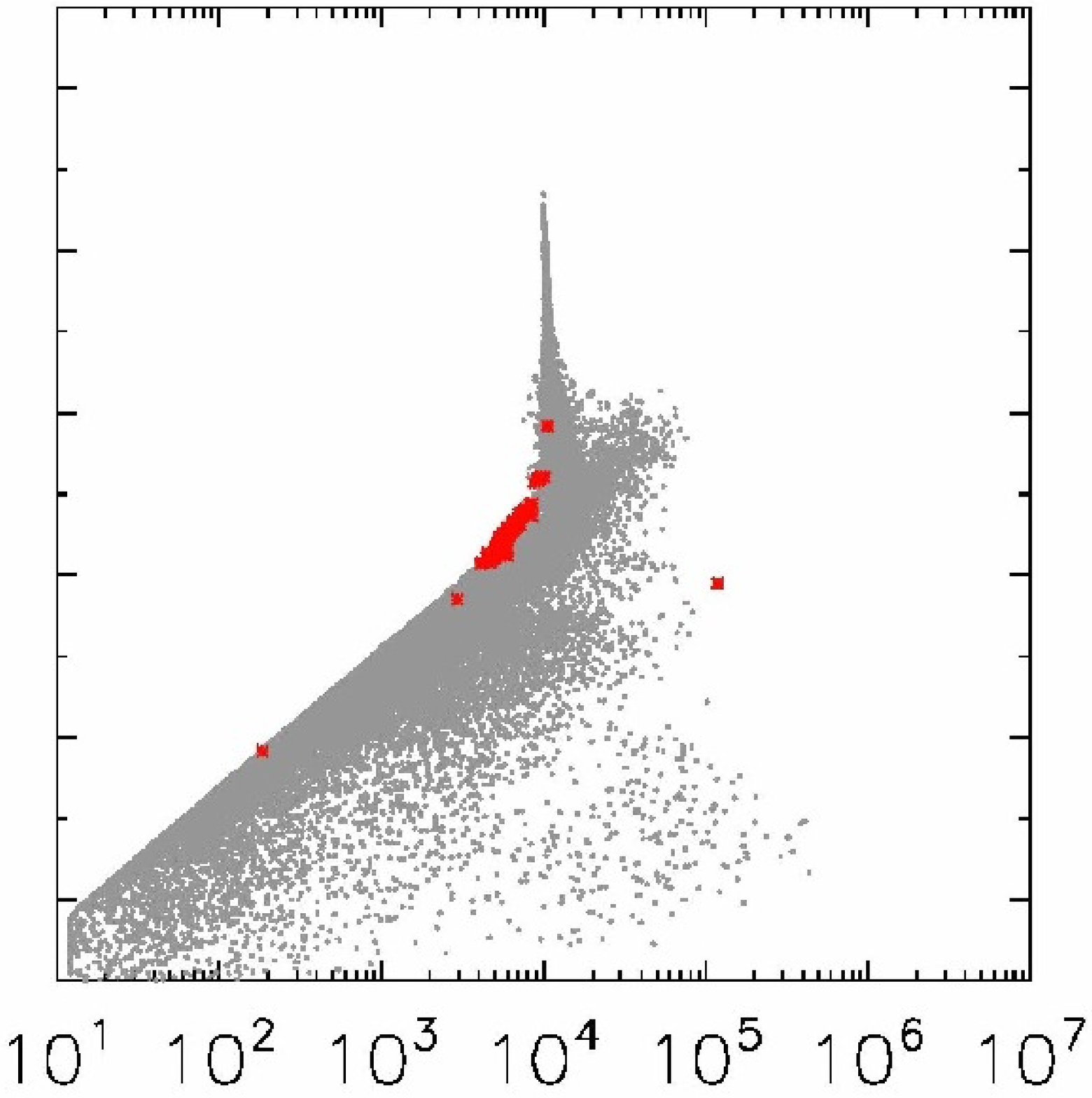} \\ 
      \hspace{-.2in} \includegraphics*[width = .25\textwidth]{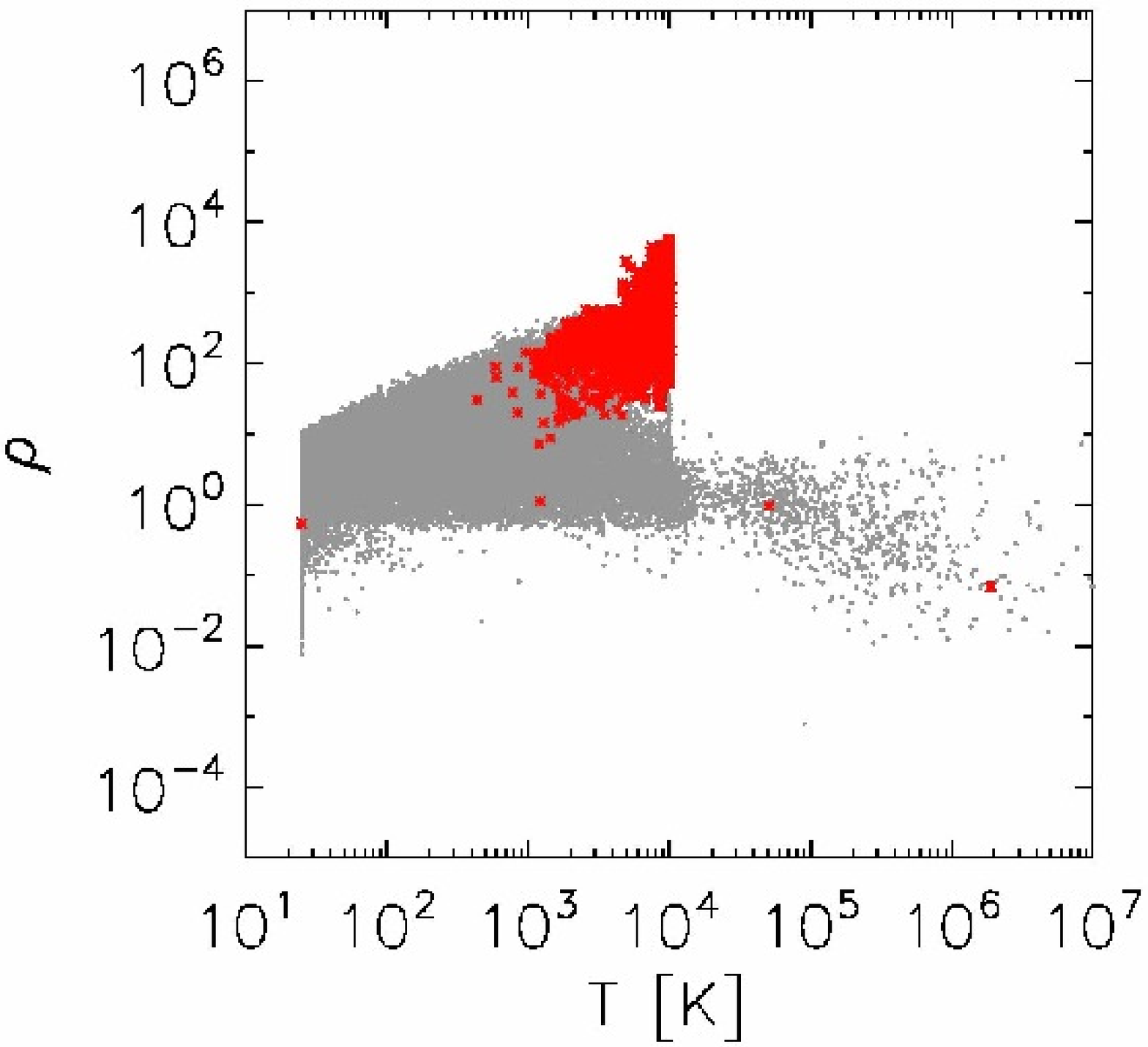} &
      \hspace{-.2in} \includegraphics*[width = .2\textwidth]{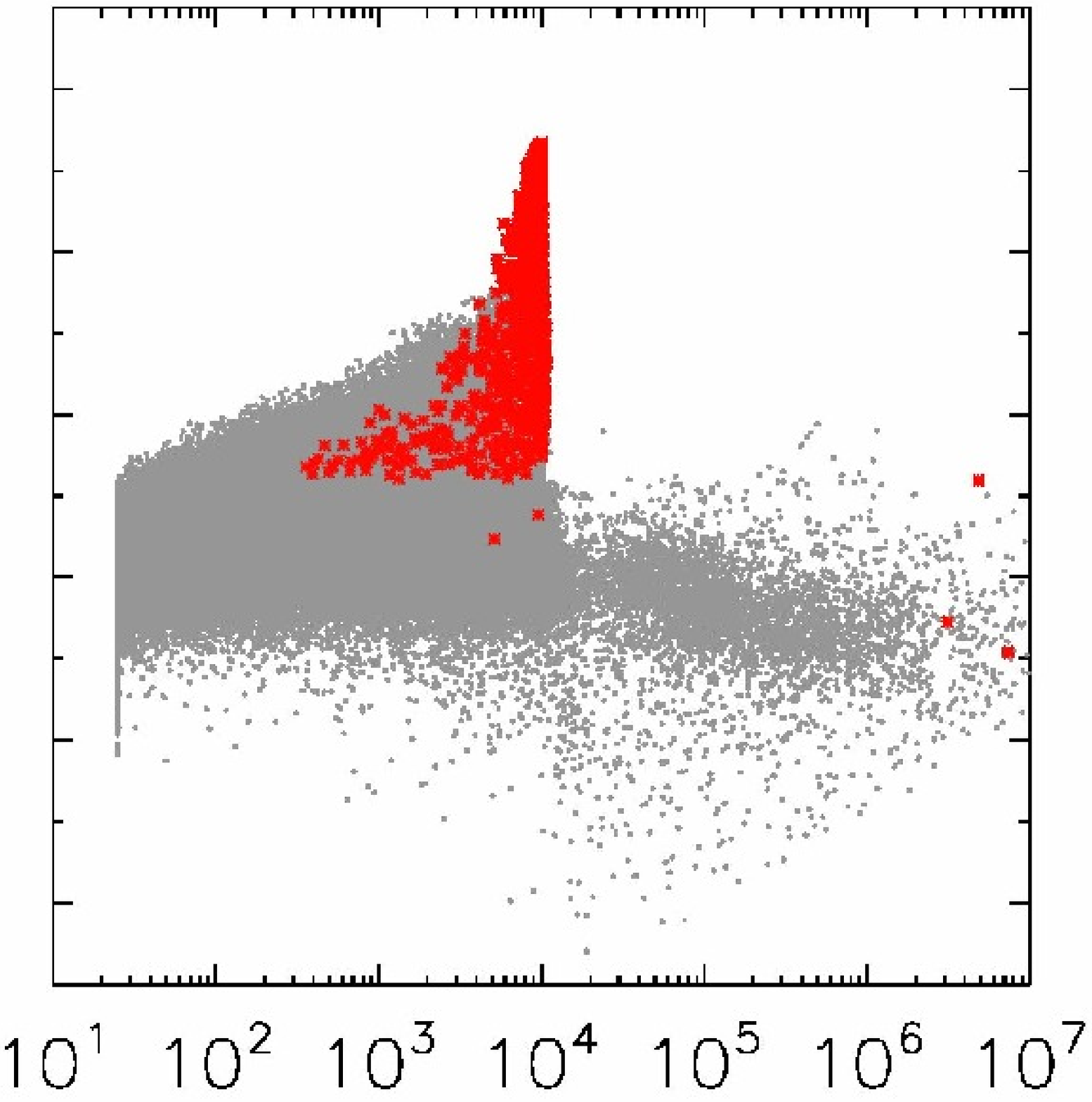} &
      \hspace{-.2in} \includegraphics*[width = .2\textwidth]{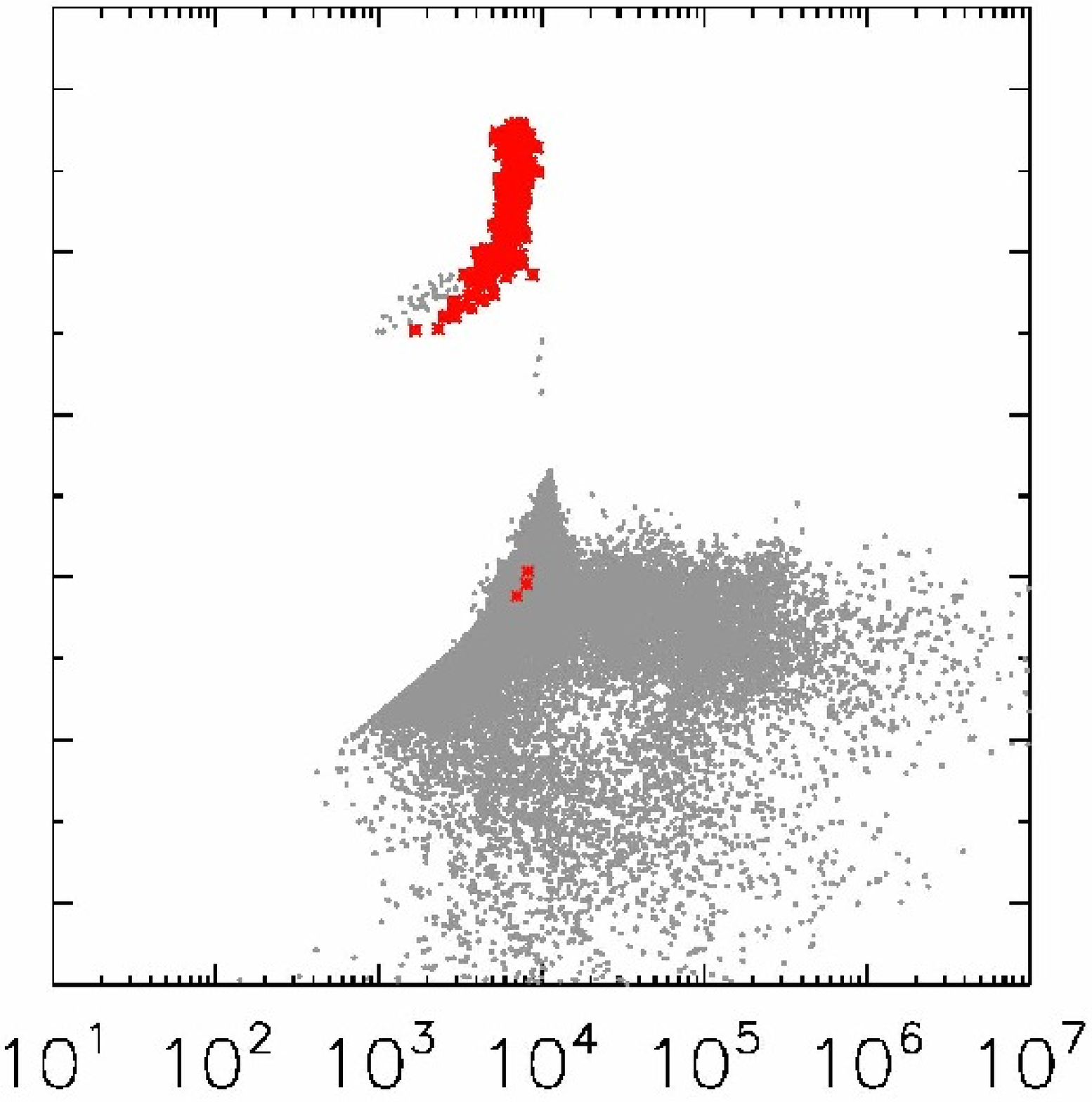} &
      \hspace{-.2in} \includegraphics*[width = .2\textwidth]{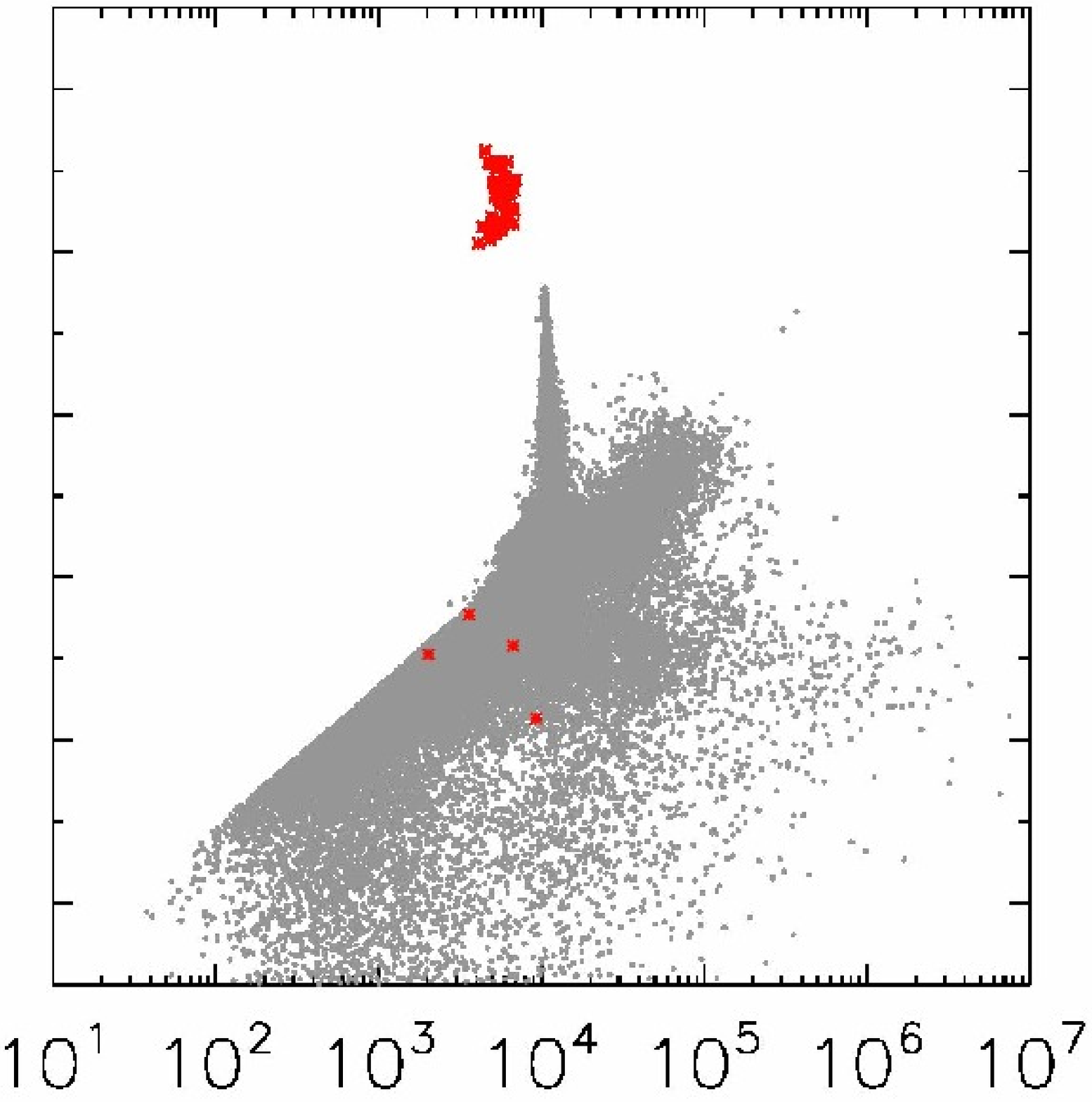} &
      \hspace{-.2in} \includegraphics*[width = .2\textwidth]{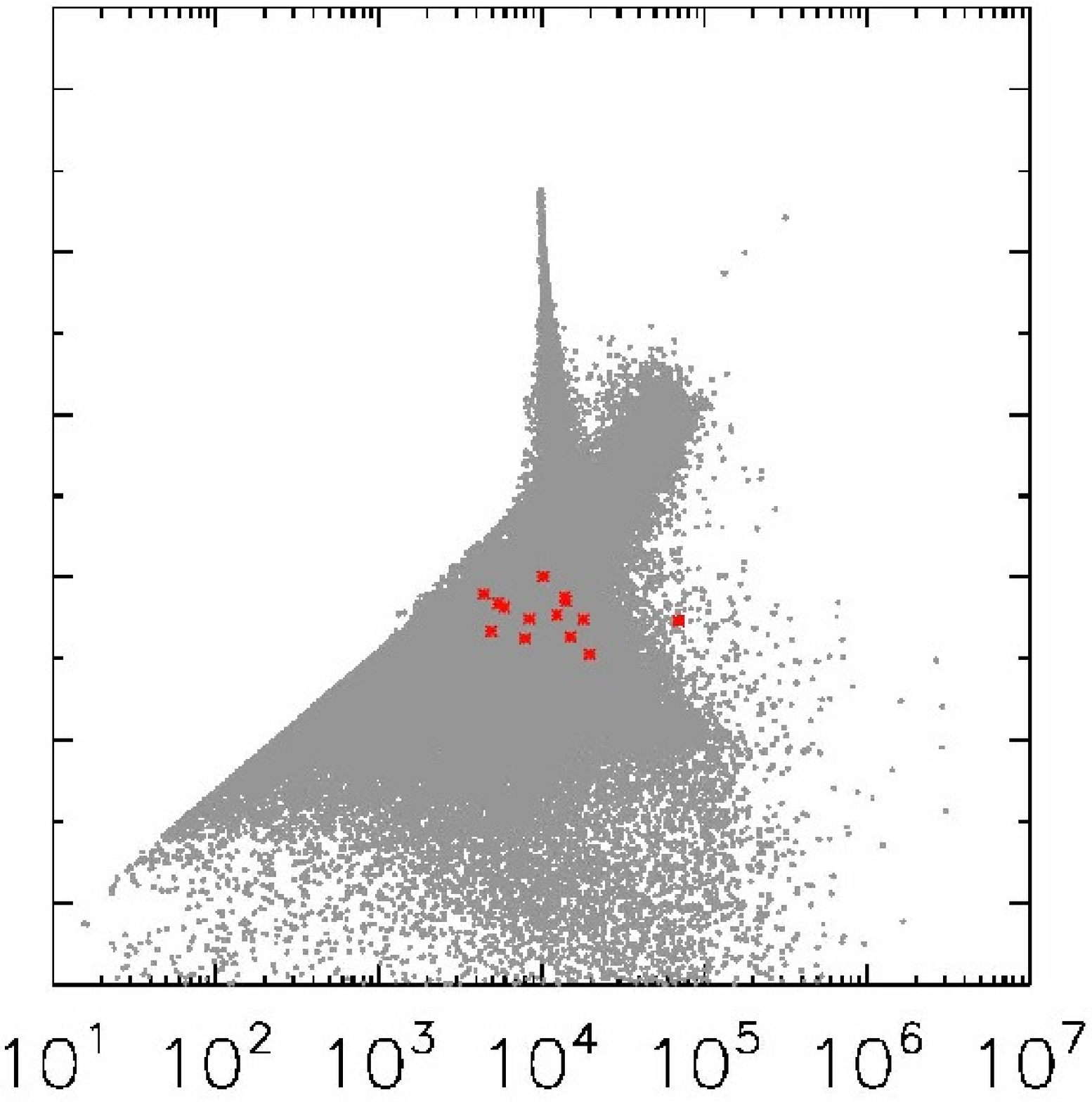}
    \end{tabular}
  \end{center}

\caption{Temperature and density of gas particles. Red dots indicate
  gas that is bound to the central halo, while grey dots are for
  particles in all other parts of the simulated volumes. Results from
  two simulations of different mass illustrate the effect of
  self-shielding. Simulation~2, shown on top, has a final halo mass of
  $3.5 \times 10^8 \Ms$, while simulation~7, shown below, reaches $9.2
  \times 10^8 \Ms$. Both simulations include supernova feedback and UV
  radiation, similar to those shown in Figure~\ref{fig:xy-begin}. At
  redshift $z=7.6$, prior to reionization, the higher mass simulation
  has been able to keep a larger amount of dense interstellar gas in
  the centre, whereas feedback has caused the gas in the lower mass
  galaxy to be more diffuse. At redshift $3.5$, the gas in the lower
  mass galaxy has been lost, while the higher mass galaxy has kept its
  high density gas, allowing it to form stars up to redshift $z=1$. In
  both cases, grey dots, associated with the IGM and smaller haloes,
  are distributed similarly to the case with UV heating but no
  shielding in Figure~\ref{fig:xy-begin}, indicating that
  self-shielding is not efficient in these low density environments.}
\label{fig:rho-temp-ss}
\end{figure*}

We have also performed simulations where we approximate the effects of
self-shielding of the dense interstellar medium against the UV
background. While we do not include radiative transfer in these
simulations, we use a threshold on the density of neutral hydrogen
(HI) gas of n$_{\mathrm{HI}} = 1.4 \times 10^{-2}$cm$^{-3}$, following
the results of \cite{Tajiri-1998}. Including this effect leads to an
interesting dichotomy. At the more massive end, as shown in the bottom
row of Figure~\ref{fig:rho-temp-ss}, the evolution in the central object
proceeds almost as in the case with no UV background, while at the low
mass end, as shown in the top row of the same figure, the evolution is
similar to the case with UV heating but no shielding.

\begin{figure}
  \includegraphics*[scale = .5]{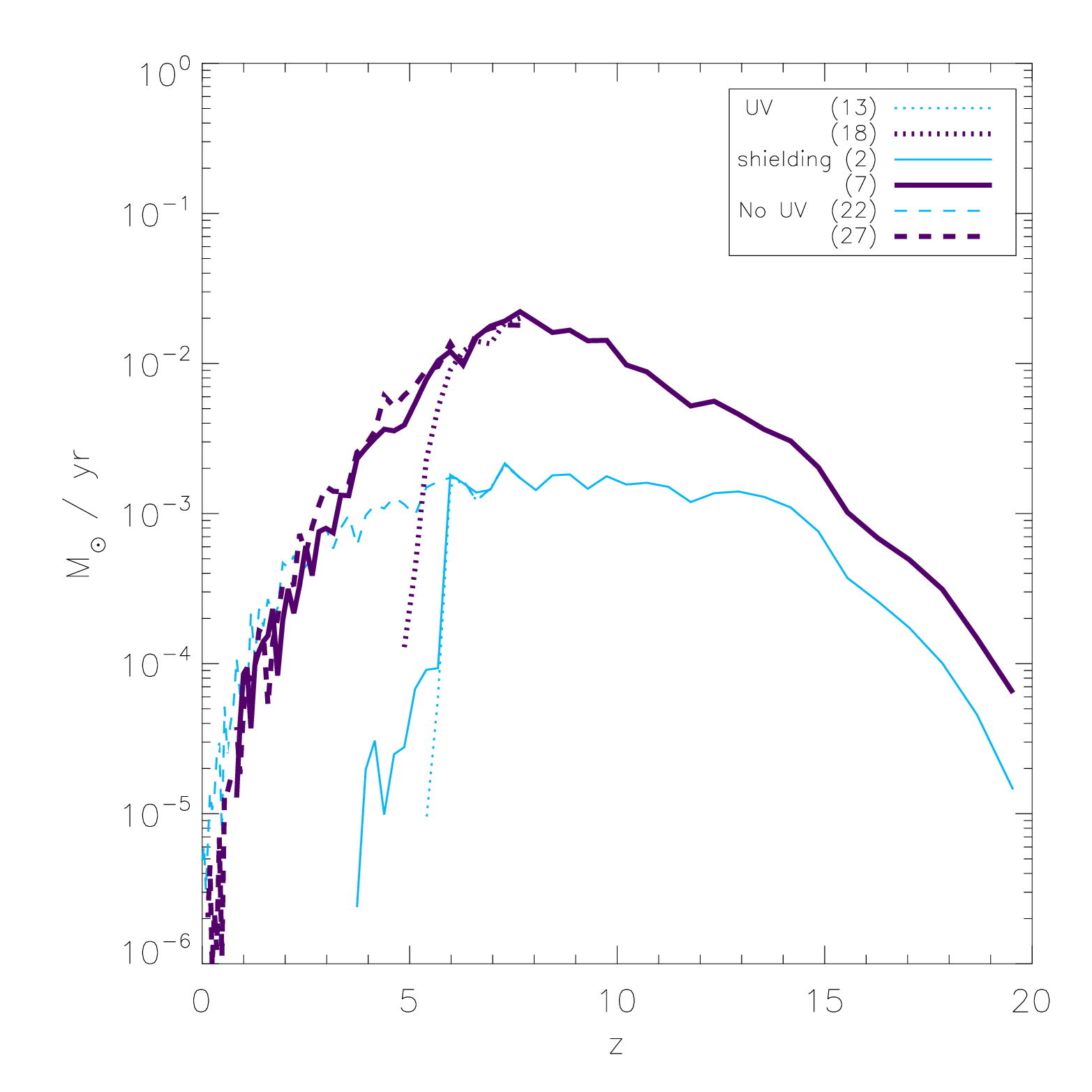}
  \caption{Star formation rate in $\Ms$ yr$^{-1}$, for a total of six
    simulations in two groups of different halo masses. Simulations 2,
    13 and 22, which have a final halo mass of $\sim 3.5 \times 10^8
    \Ms$, are plotted as thick blue lines, while simulations 7, 18 and
    27 reach $\sim 9 \times 10^8 \Ms$, and are plotted as thin purple
    lines, in correspondence to the colours used in
    Figures~\ref{fig:ages-uv} and~\ref{fig:ages-ss} for the same
    masses. All simulations include supernova feedback, they differ in
    the treatment of the UV background and/or self-shielding.
    Simulations 22 and 27 (dashed lines) include no UV
    radiation. Simulations 13 and 18 (dotted) include a UV background,
    and simulations 2 and 7 (solid) also include self-shielding. It
    can be seen that the star formation rates for all systems of a
    given mass are identical up to redshift $z=6$. After that, for
    both masses, they decline sharply in the scenarios with UV
    background and no shielding (dotted), and more gradually when the
    UV background is ignored (dashed). However, the impact of
    self-shielding (solid) is different for the two masses. In the
    high mass halo, the result with self-shielding resembles the case
    without UV radiation, whereas in the low-mass halo, the star
    formation rate drops almost as sharply with self-shielding as
    without.}
  \label{fig:sfr}
\end{figure}

\begin{figure}
     \includegraphics*[scale = .5]{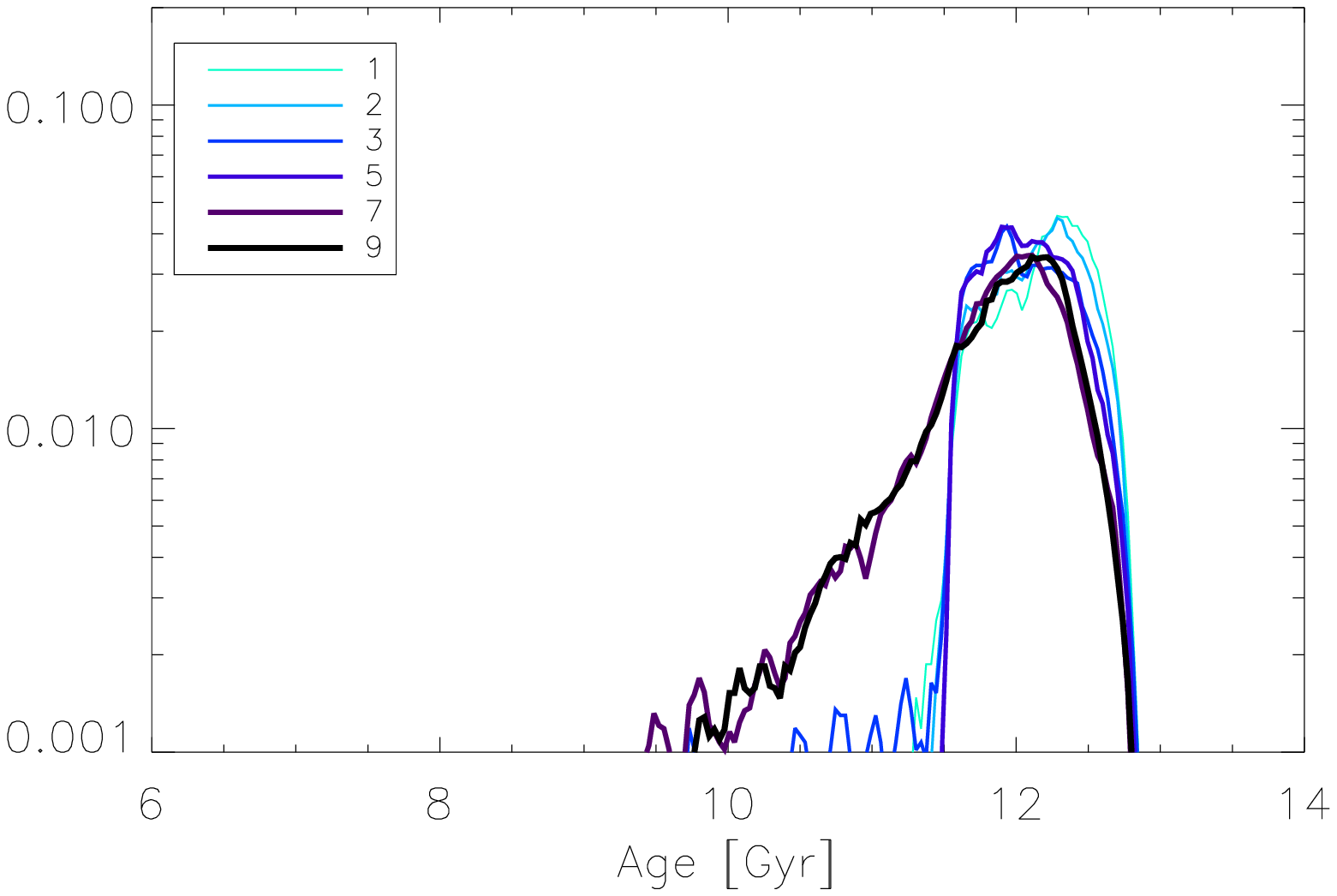}
\caption{Distribution of stellar ages in Gyr, in simulations 1-9,
  which include feedback, a UV background and self-shielding, as
  described in the text. As in Figure~\ref{fig:ages-uv}, the colours
  and line strengths indicate different halo masses, from
  $2.3\times10^8\Ms$ (simulation 1, lightest blue) to $1.2\times
  10^9\Ms$ (simulation 9, black). Other properties are listed in
  Table~1. While simulations 1-5 show only an old stellar population,
  similar to the top panel of Figure~\ref{fig:ages-uv}, a transition
  occurs around a final halo mass of $\sim8\times10^8\Ms$, above which
  self-shielding becomes effective, allowing star formation to
  continue beyond $z=6$.}
\label{fig:ages-ss}
\end{figure}

Figure~\ref{fig:sfr} shows the star formation rates over time for a
total of six simulations of two different halo masses. Simulations~2
and~7 which include shielding, as discussed above, are compared to two
sets of equal mass-mass counterparts: Simulations~13 and ~18 which
have a UV background but no shielding, and simulations~22 and 27,
which do not include UV radiation. The evolution of each triplet of a
given mass proceeds identically up to redshift $z=6$. Subsequently,
the presence of the UV background quenches star formation in both
simulations without shielding, while the two simulations without UV
radiation both see a gradual decline in their star formation rate,
solely due to feedback. However, when shielding is included, it has no
effect in the low-mass case, where the star-formation rate shows a
sharp decrease, similar to the unshielded case. In contrast, in the
high-mass case, the star formation rate with shielding closely follows
that of the corresponding simulation without UV radiation.

The age distributions shown in Figure~\ref{fig:ages-ss} reflect this
behaviour. Systems with lower mass only have small age spreads
resulting from a single burst, comparable with the results without
shielding shown in Figure~\ref{fig:ages-uv} (matching colours indicate
equal masses). Higher mass objects possess intermediate age tails,
similar to the result without a UV background.

Again, we find that while it is the response of the interstellar
medium to the UV radiation that ultimately splits the two scenarios,
it is the effect of feedback prior to reionization that is at the root
of this dichotomy. In the low-mass case, feedback dilutes the gas more
efficiently prior to reionization, and thereby prevents it from
self-shielding. In the high-mass case, the gas in the centre remains
dense enough to become self-shielding and to prevent a blow-away.

While the approximation of self-shielding is very crude, and should be
confirmed by more detailed analysis with full radiative transfer, it
is interesting that the critical gas density that determines whether
galaxies form stars after reionization appears to lie just at the
right level to allow the formation of both kinds of dwarf spheroidal
galaxies in simulations which include supernova feedback. On the basis
of these arguments, it appears that the inclusion of cosmic
reionization and a UV background with the possibility of
self-shielding is the physically correct assumption. In the subsequent
analysis, we continue to use the \cite{Haardt-1996} model, together
with the \cite{Tajiri-1998} approximation for self-shielding. We note,
however, that the environment in which our galaxies form is different
to that of the Local Group. This may play an additional role in the
star formation history. Possible environmental effects include not
only a local variation in the UV background, but also other mechanisms
for removing gas, or reducing its density in satellite galaxies. While
the removal of gas, by ram pressure stripping, for example, could halt
star formation directly, our results indicate that an indirect
mechanism might be just as efficient, if it makes the gas susceptible
to evaporation after $z=6$. That such environmental effects play a
role is supported by the observation that dwarf spheroidals close to
the centre of the Milky Way tend to have fewer intermediate age stars
than those further out, although this trend is less clear for the
M31-satellites \citep{Grebel-1997}.

\begin{figure*}
  \begin{center}
    \begin{tabular}{ccccc}
      \hspace{-.2in} \includegraphics*[width = .25\textwidth]{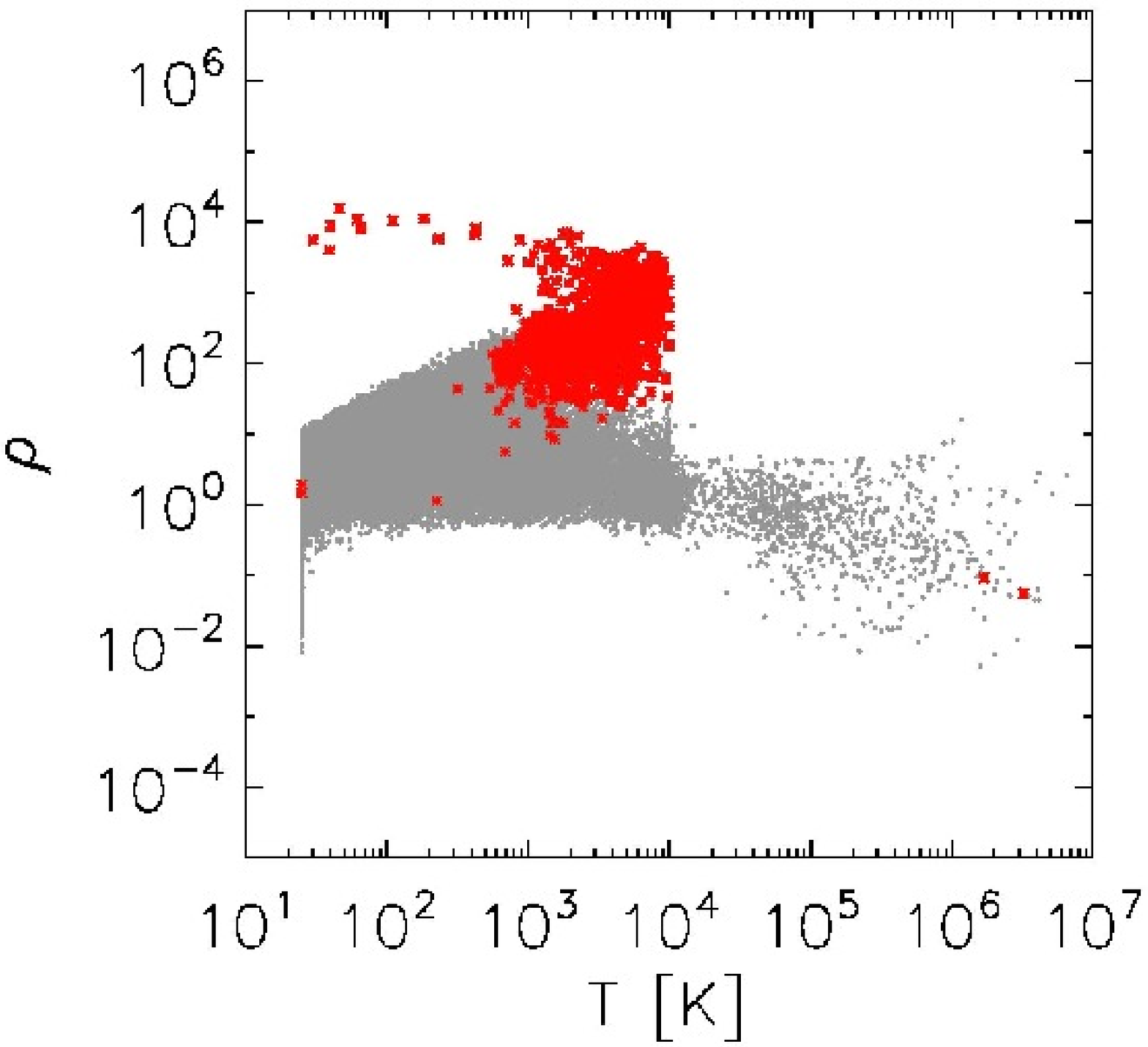} &
      \hspace{-.2in} \includegraphics*[width = .2\textwidth]{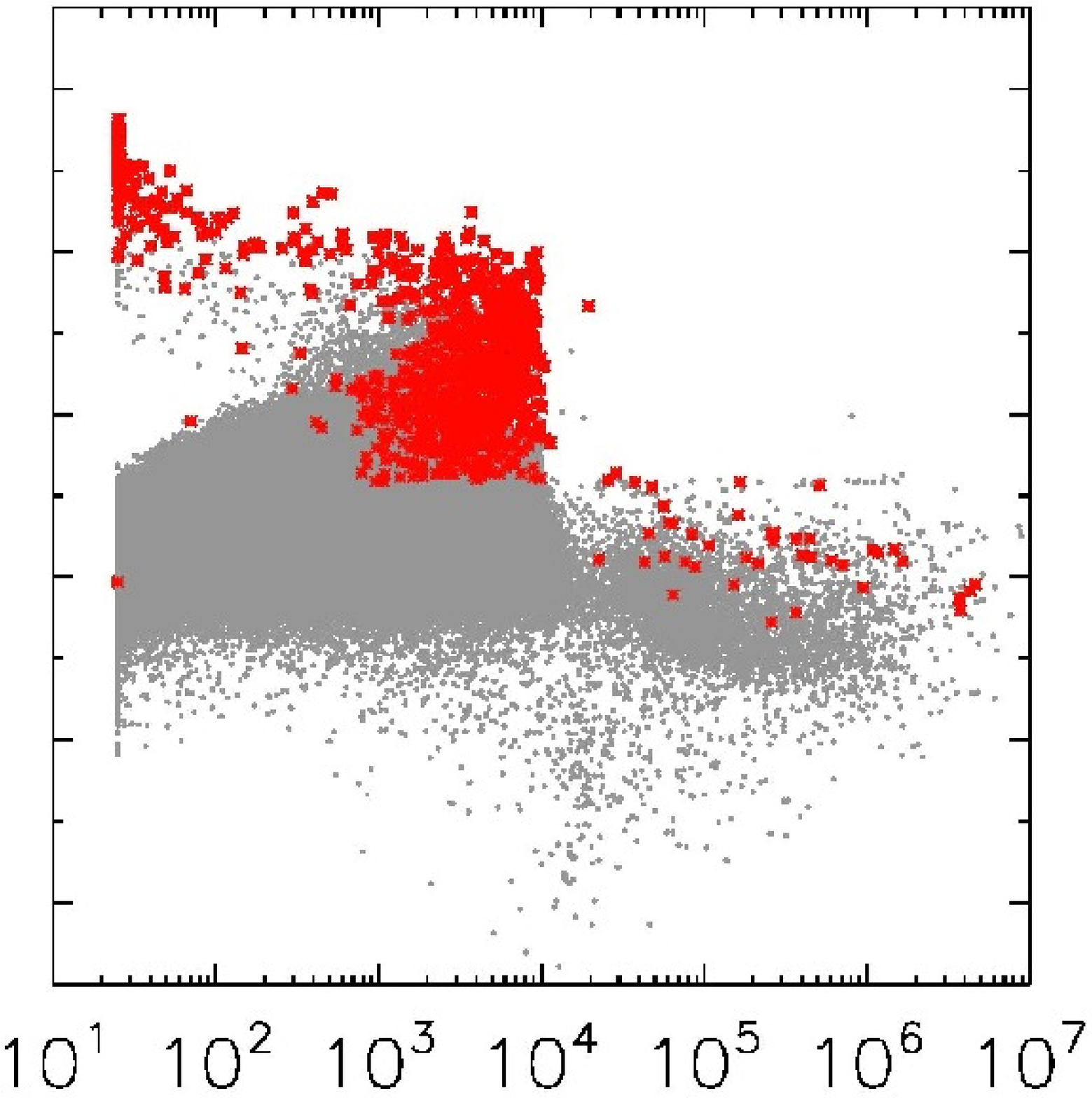} &
      \hspace{-.2in} \includegraphics*[width = .2\textwidth]{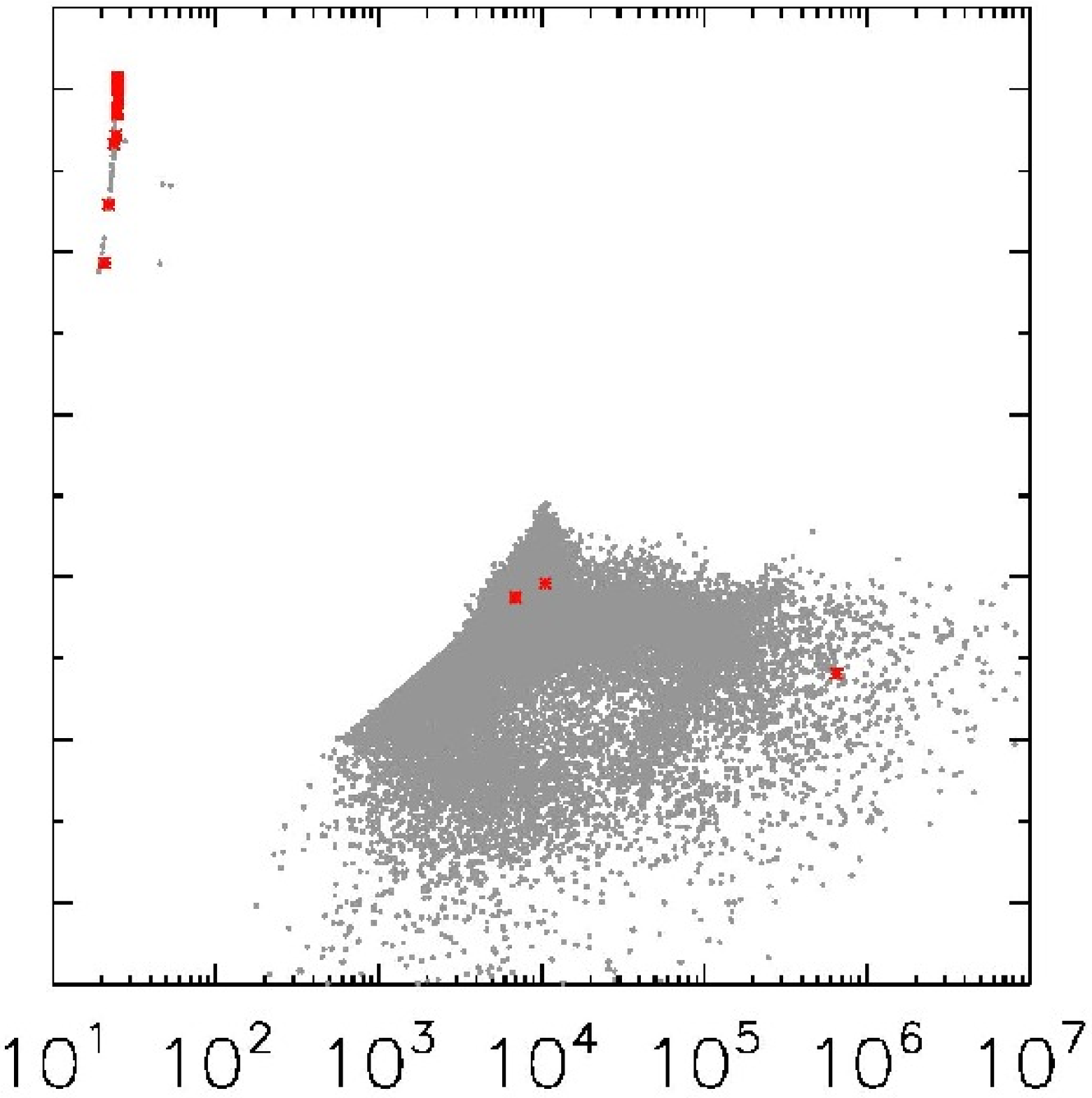} &
      \hspace{-.2in} \includegraphics*[width = .2\textwidth]{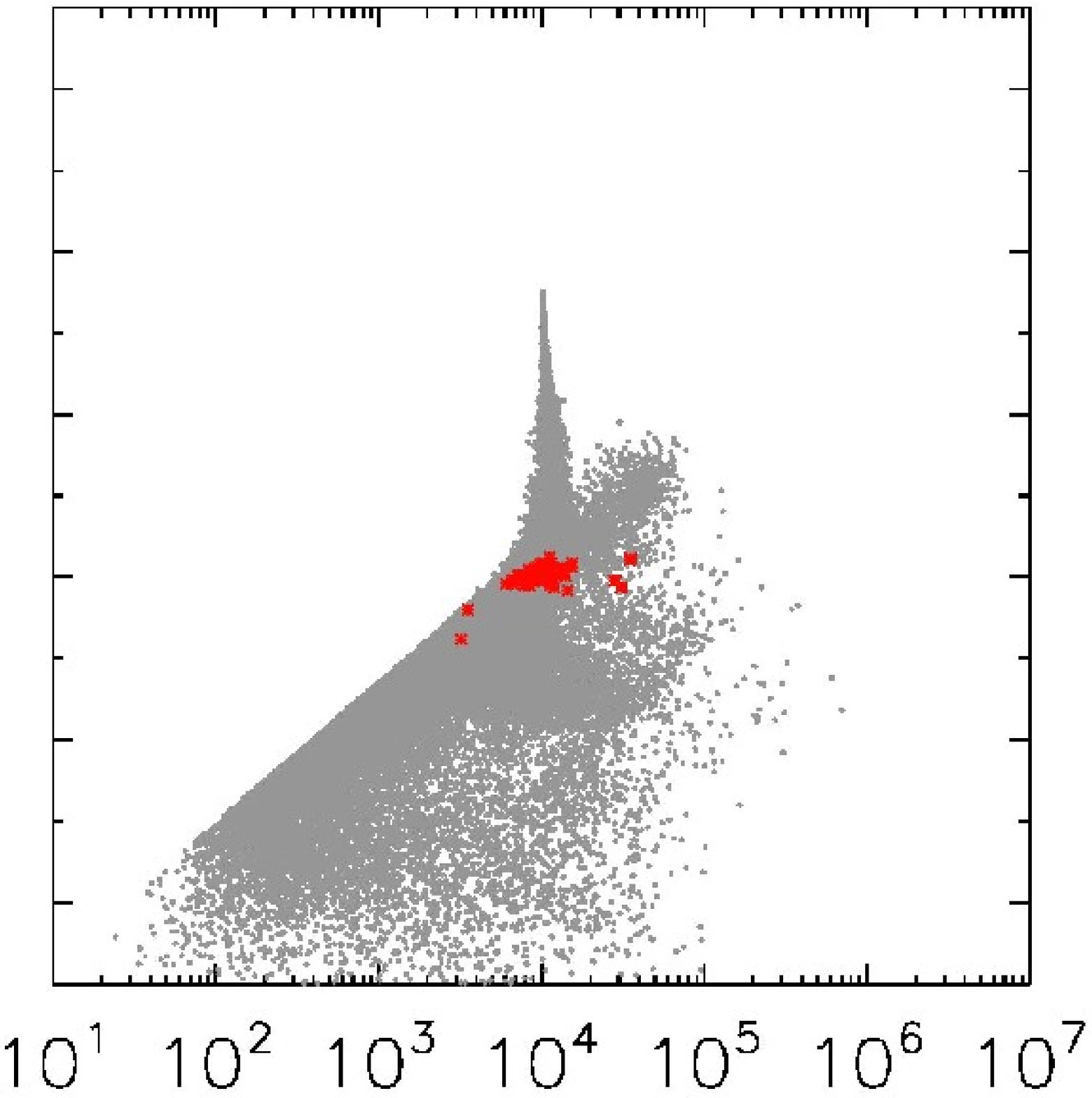} &
      \hspace{-.2in} \includegraphics*[width = .2\textwidth]{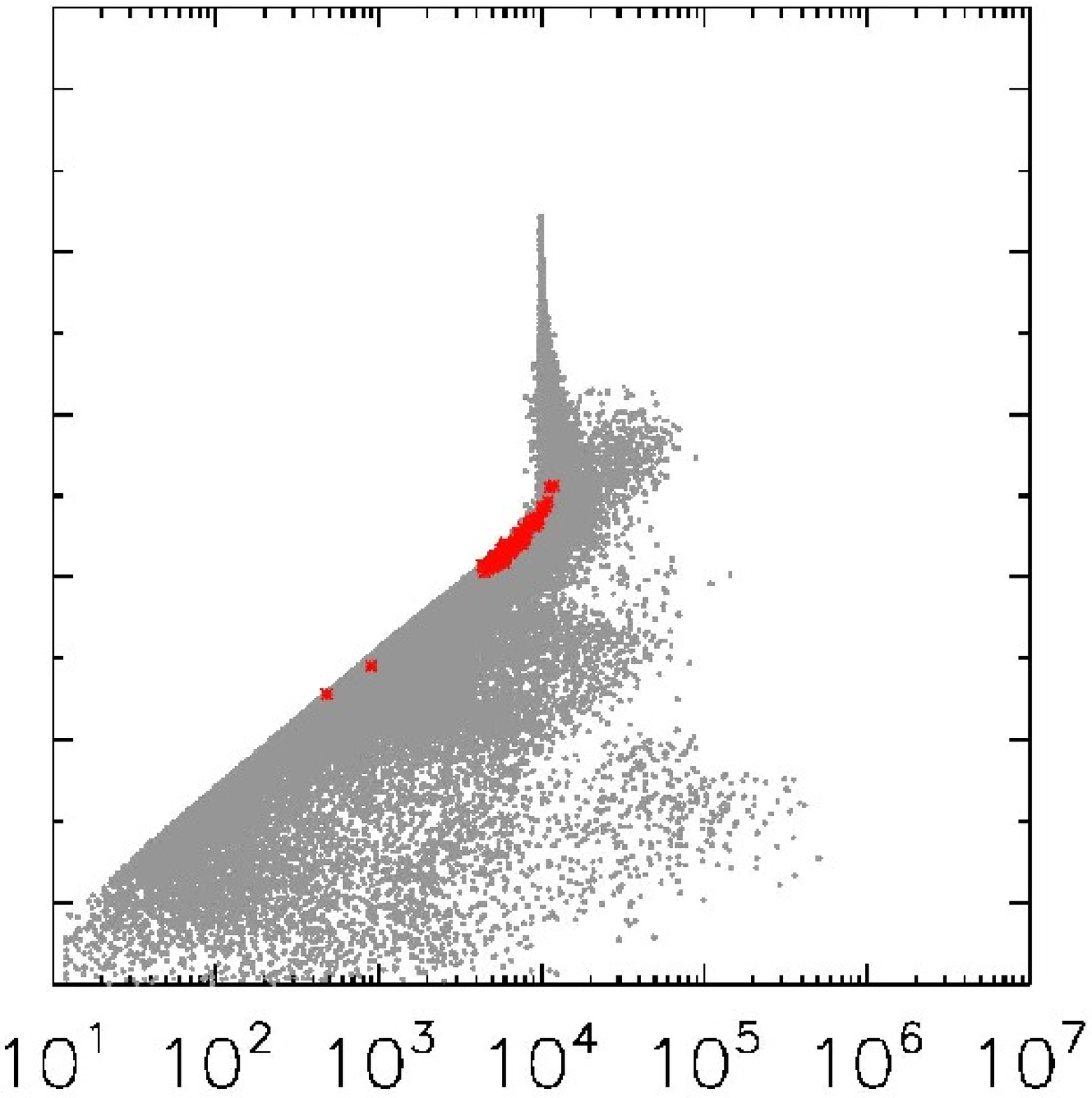} \\ 
      \hspace{-.2in} \includegraphics*[width = .25\textwidth]{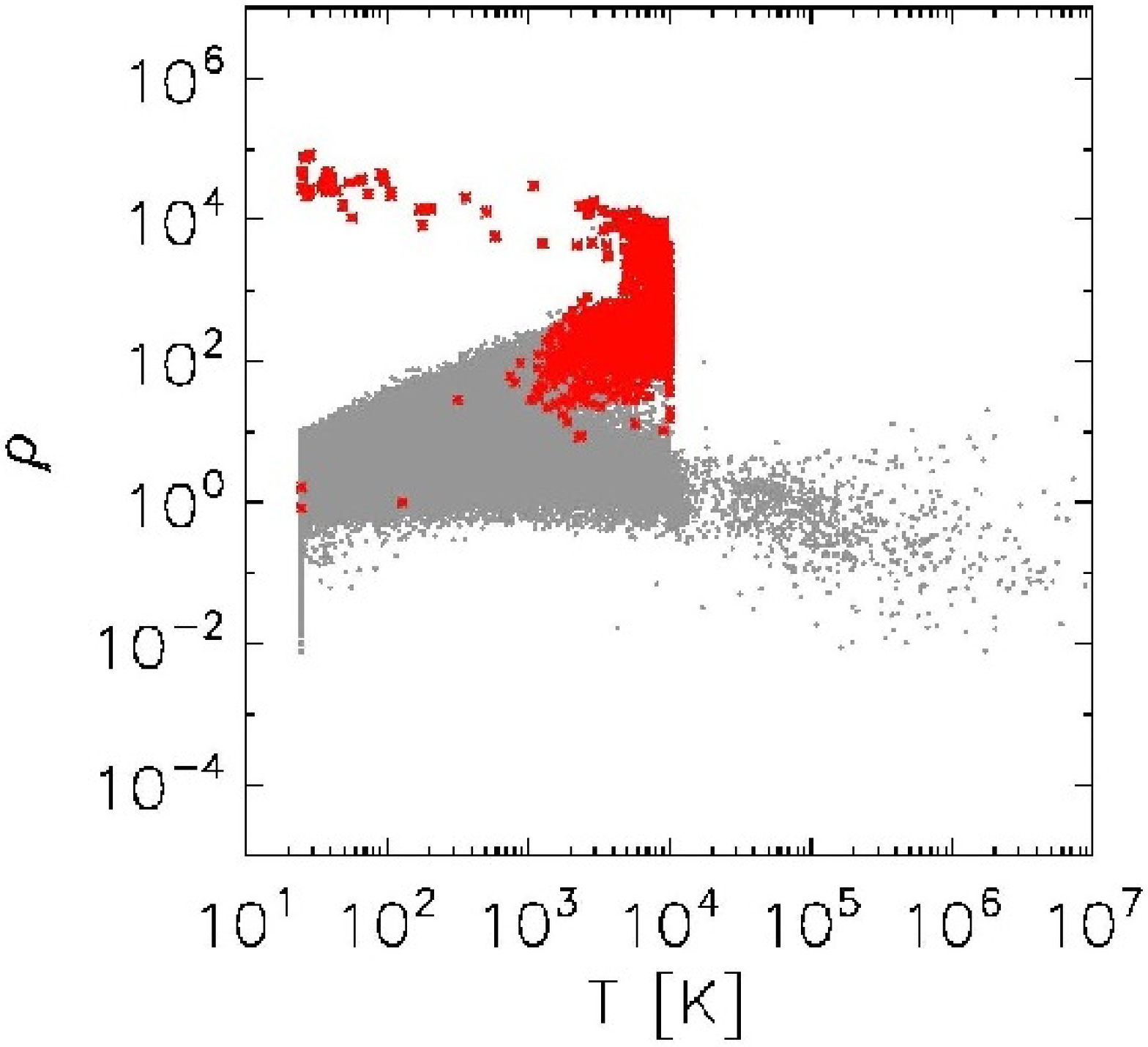} &
      \hspace{-.2in} \includegraphics*[width = .2\textwidth]{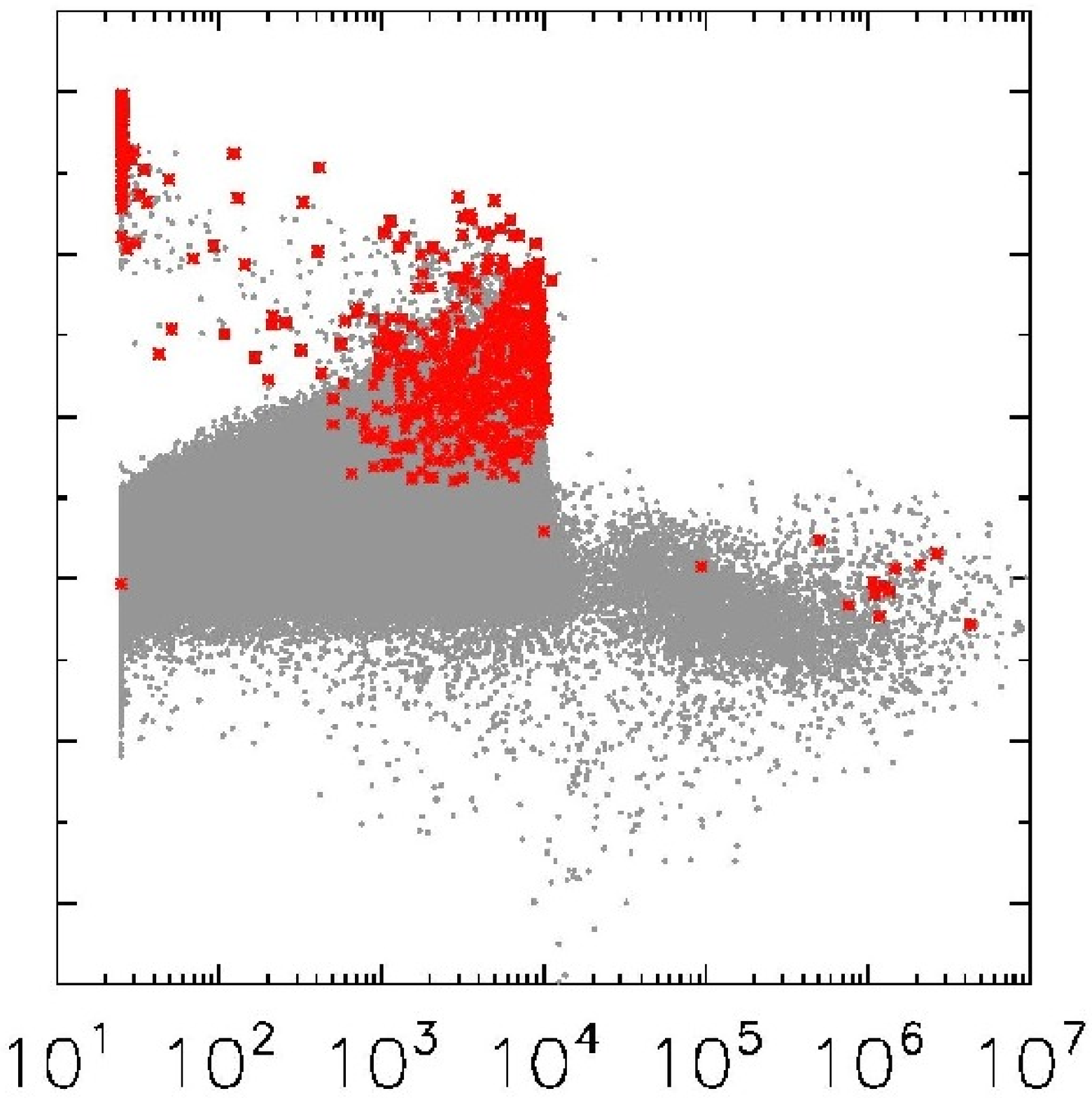} &
      \hspace{-.2in} \includegraphics*[width = .2\textwidth]{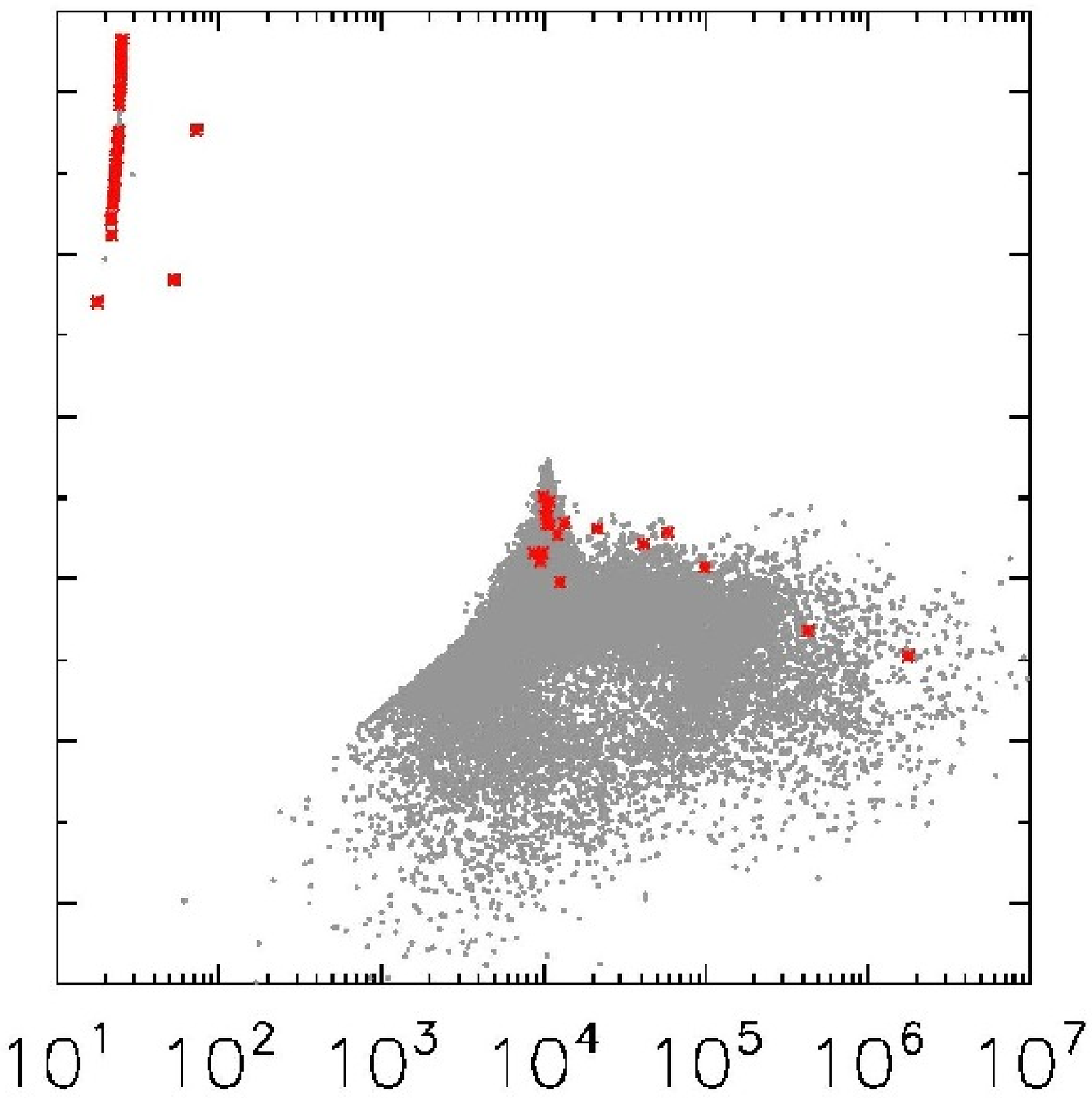} &
      \hspace{-.2in} \includegraphics*[width = .2\textwidth]{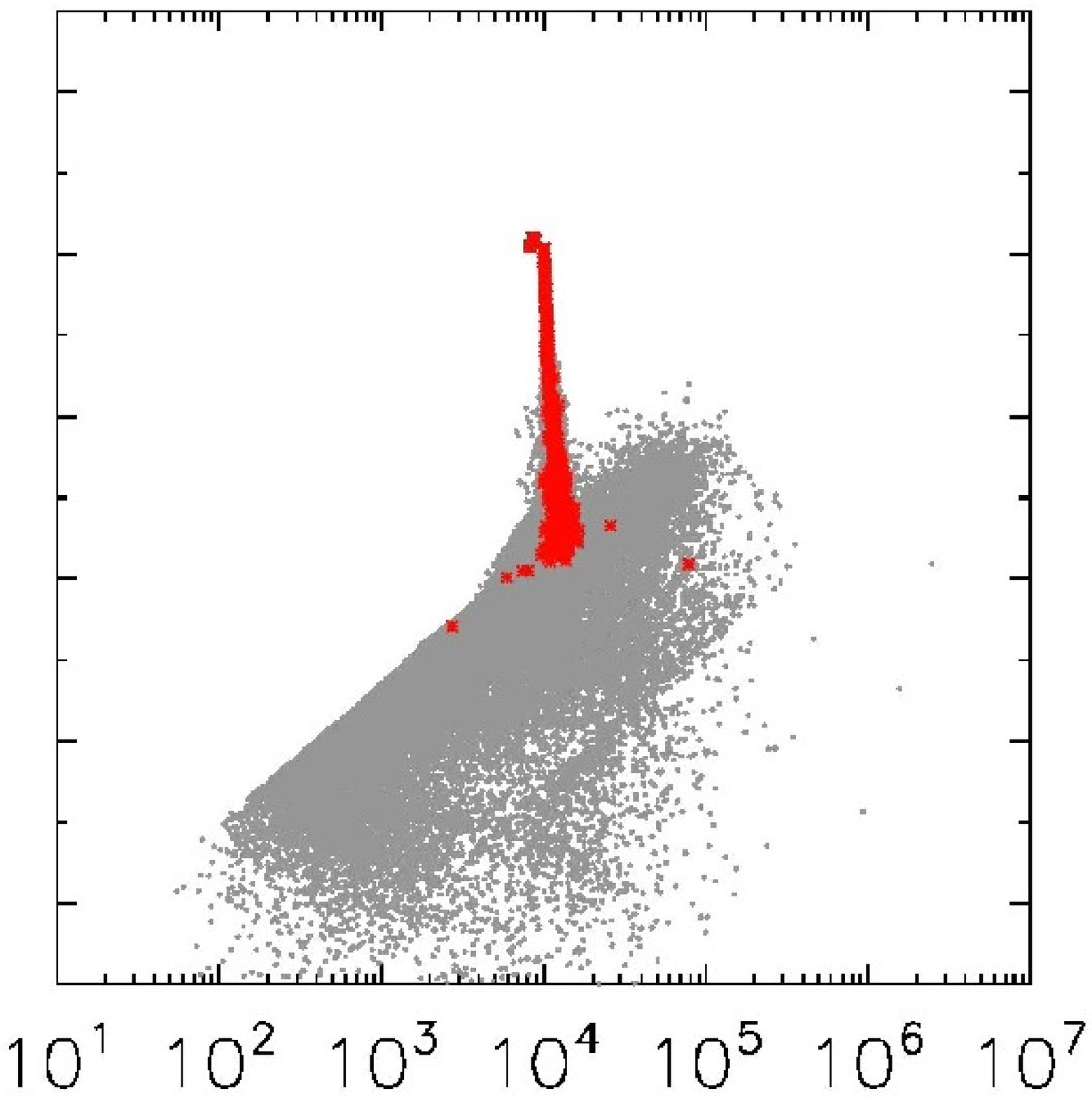} &
      \hspace{-.2in} \includegraphics*[width = .2\textwidth]{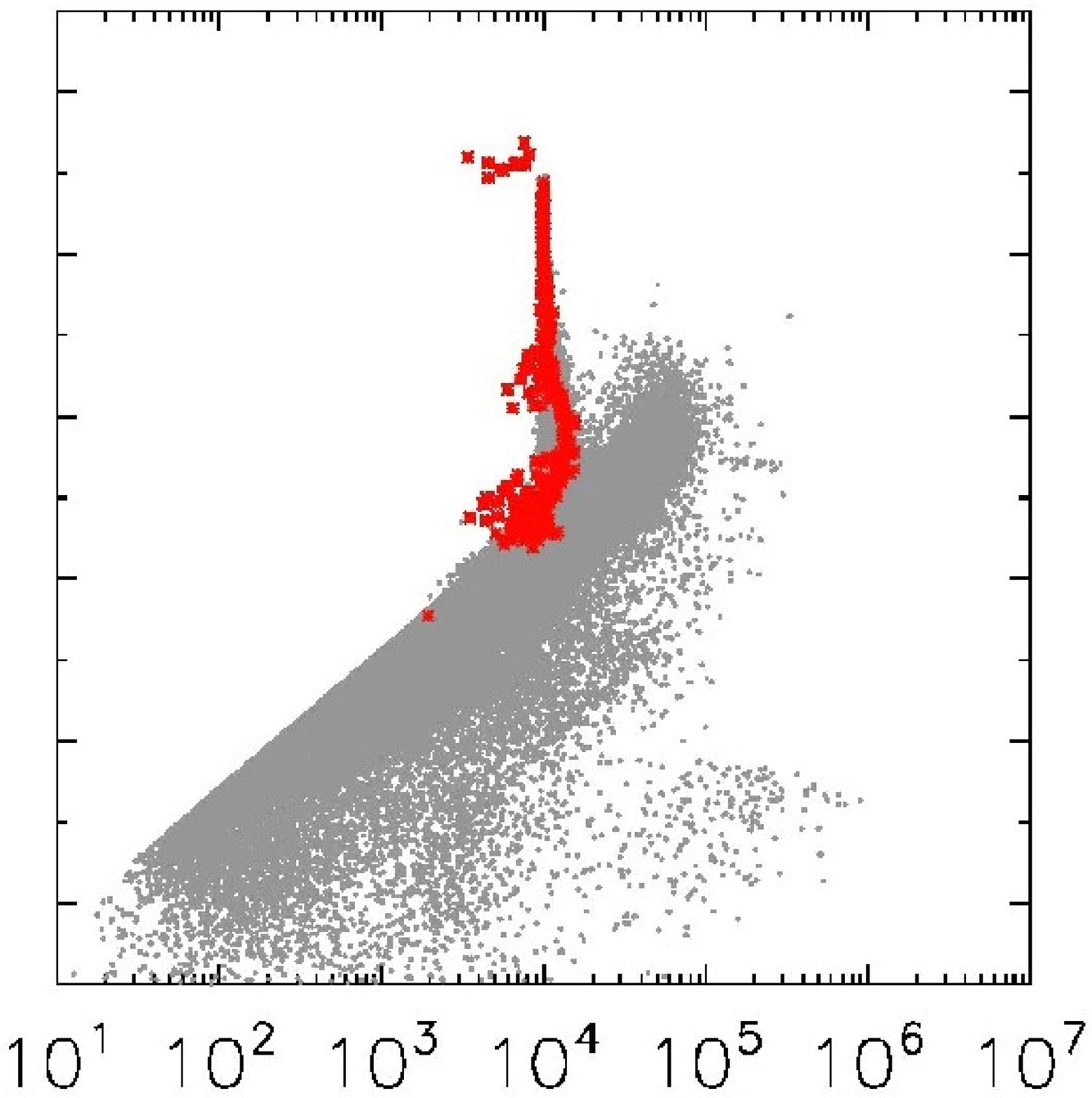}
    \end{tabular}
  \end{center}
\caption{Temperature and density of gas particles. Red dots indicate
  gas that is bound to the central halo, while grey dots are for
  particles in all other parts of the simulated volumes. Initial
  conditions for the simulations are identical to Simulations~2 and 7,
  shown in the top and bottom rows of Figure \ref{fig:rho-temp-ss},
  respectively, which include supernova feedback, UV radiation and
  self-shielding, and which reach final halo masses of $3.5 \times
  10^8$ and $\sim 9 \times 10^8 \Ms$. As a result of additional low
  temperature cooling in both haloes, early star formation and
  supernova driven outflows occur at a slightly higher rate. In the
  low mass case, the subsequent evolution is very similar to the
  situation shown in the top row of Figure \ref{fig:rho-temp-ss}
  without low-temperature cooling. As in the case without low
  temperature cooling, the residual amount of gas is again too small
  to be effectively self-shielding by the time reionization sets in at
  redshift $z=6$. In the high mass scenario, the amount of residual
  gas at $z=6$ is also reduced with respect to the case without
  low-temperature cooling, and hence the effect of self shielding is
  somewhat lower, leading to slightly less subsequent star formation.}
\label{fig:rho-temp-cool}
\end{figure*}

\subsection{The Role of Low Temperature Cooling} \label{sec:LowTCooling}
As described in Section~\ref{sec:methods:cooling}, we have performed
simulations with and without metal and molecular cooling below
$10^4$~K. \cite{Bromm-2001} found in their simulations that atomic
hydrogen cooling alone is not sufficient to form the observed dwarf
galaxies, and they as well as other authors
\citep[e.g.][]{Mashchenko-2008, Revaz-2009} have found different ways
to include low temperature cooling due to molecules and metals in
their simulations. By contrast, \cite{Mayer-2005} and others have only
considered cooling above $10^4$~K. We have repeated several
simulations with additional low temperature cooling, using the
extended cooling functions of
\cite{Maio-2007}. Figure~\ref{fig:rho-temp-cool} shows the
distribution of gas particles in the temperature-density plane at
different redshifts for two simulations with low-temperature
cooling. Both simulations have identical initial conditions to
Simulations ~2 and~7, respectively, and include the full physical model
of supernova feedback, UV radiation and self shielding. Figure
~\ref{fig:rho-temp-cool} can be compared to
Figure~\ref{fig:rho-temp-ss}, which shows the evolution without low
temperature cooling. In each case, with low temperature cooling, star
formation proceeds at slightly higher efficiency at high redshifts. As
a consequence of supernova feedback acting at a lower halo mass, the
interstellar gas mass peaks at higher redshift. For the lower mass
scenarios, the effect on the total stellar mass is minimal. In the
high mass case, however, the decreased gas mass at $z=6$ decreases the
efficiency of self-shielding against the cosmic UV background, and
hence the amount of subsequent star formation.

In general, we find that the inclusion of low-temperature cooling does
not have a very strong effect on the formation of dwarf galaxies in
the mass range of $10^8$ to $10^9 \Ms$, which we have considered in
this study, and which have all began to form stars before
reionization. It does not qualitatively alter the response of the
interstellar medium to supernova feedback or the UV background.

However, we cannot exclude the possibility that low temperature cooling
may have a significant effect on the formation and evolution of
galaxies with even lower halo masses, whose virial temperatures are
far below $10^4$~K. In all of our simulations, we have assumed the gas
to be metal-free before the first stars are formed and release
metals ab initio. As the cooling function in Figure~\ref{fig:temp-cooling}
shows, possible pre-enrichment of the intergalactic medium could
enhance the cooling efficiency, which might also play a role in this
case.

\section{Exploring the Parameter Space} \label{sec:parameters}
Even though our model is physically motivated, it also contains a
  certain degree of parameterisation in addition to the numerical
  parameters discussed in Section~\ref{sec:methods}, which cannot be
determined ab initio in our simulations. We have therefore explored a
range of variables that have a direct physical significance, some of
which we hope to constrain by comparison with observations, and others
that may simply help to explain the variation amongst the observed
systems.

\begin{figure}
  \includegraphics[scale = .5]{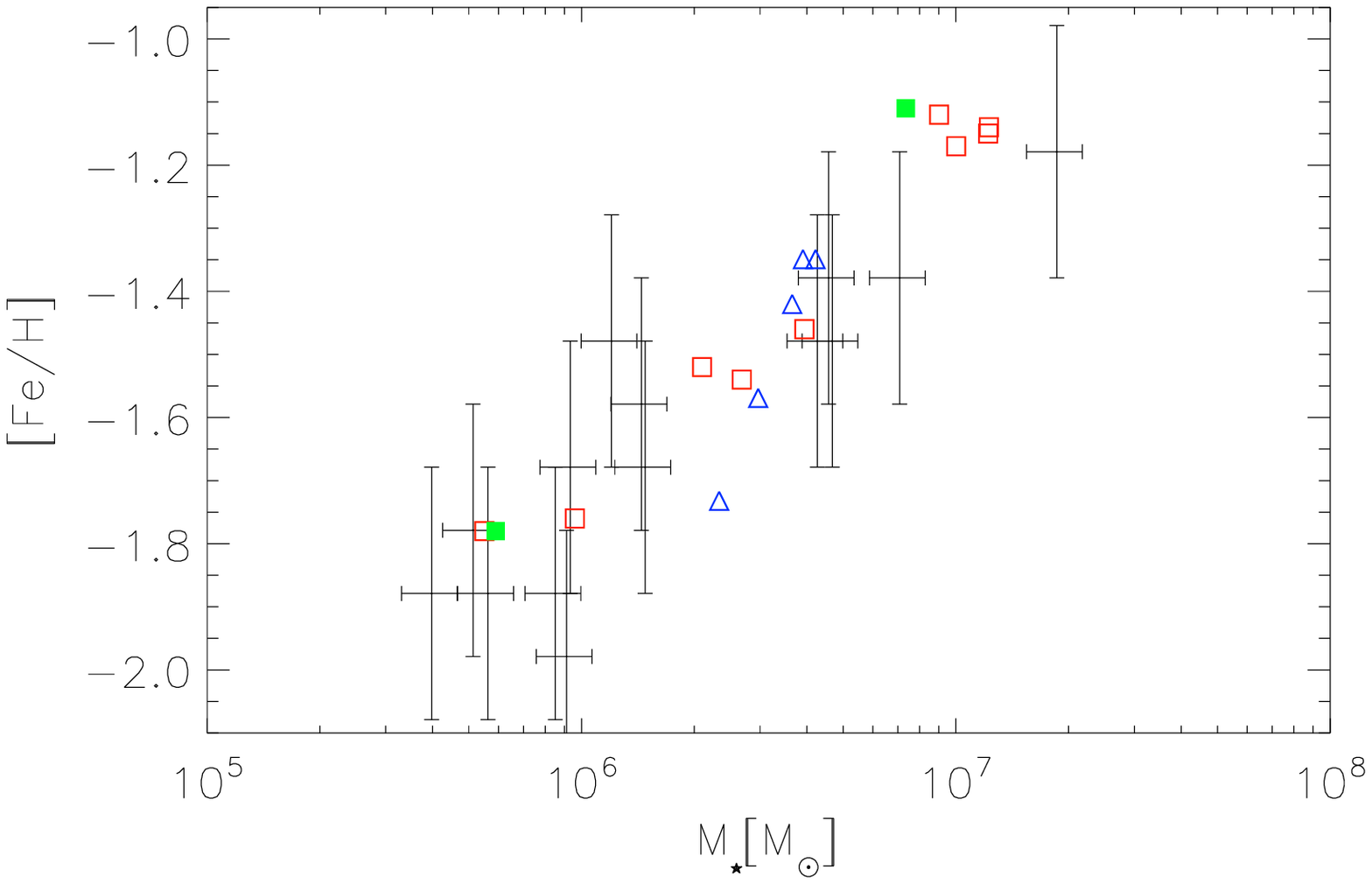}
  \caption{Mean metallicity and stellar mass for 14 observed dwarf
    spheroidals, in black with error bars, together with the results
    from our simulations. Red squares show a sequence of simulations
    (1-9) with varying initial masses, which gives a good fit to the
    observations. Also shown, with blue triangles, is a sequence of
    simulations with varying parameters of $c_\star$. While it
    intersects with the observed relation, the slope is much too steep
    compared with observations. The two filled, green squares are the
    results of two simulations with increased resolution, as discussed
    in Section~\ref{sec:resolution}. The observational uncertainties,
    as given by Woo et al., are 0.2 dex for [Fe/H] and 0.17 dex for
    stellar mass.}
\label{fig:lum-met}
\end{figure}

\begin{figure}
  \includegraphics[scale = .5]{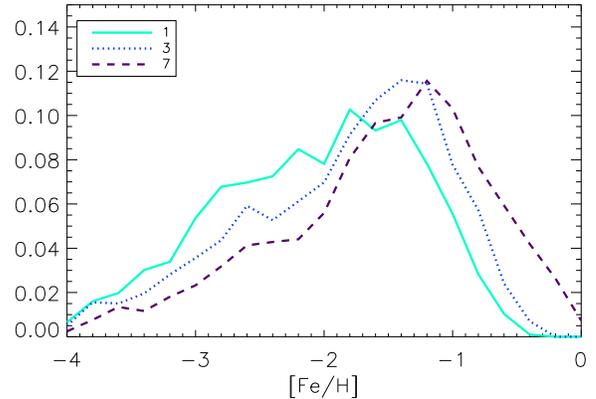}
  \caption{Relative, mass-weighted metallicity distribution of
    individual stars for the three simulations 1, 4 and 7, which vary
    in final halo mass.}
\label{fig:Fe}
\end{figure}

\begin{figure}
  \includegraphics[scale = .5]{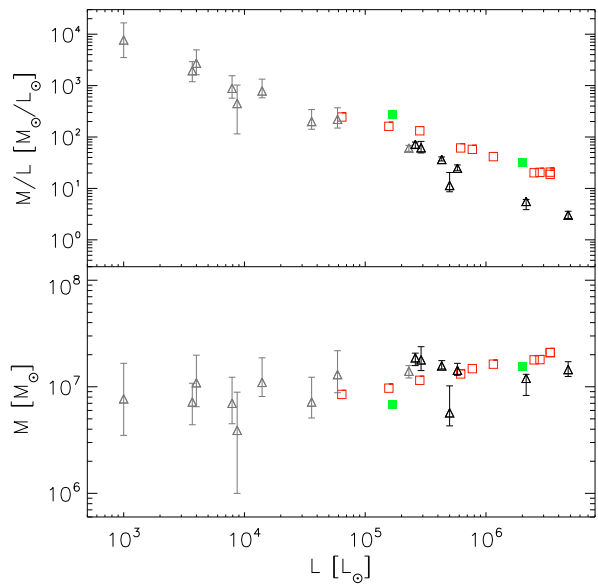}
  \caption{Mass-to-light ratio within 300 pc as a function of total
    luminosity. Simulations 1 through 9 from Table~1 and one
      additional, lower mass simulation are plotted as red squares,
      together with the 8 `classical' Milky Way satellites and the 10
      `SDSS Dwarfs' contained in the analysis of Strigari et al., in
      black and grey triangles with error bars, respectively. The two
      filled, green squares are the results of two simulations with
      increased resolution discussed in Section~\ref{sec:resolution}.}
\label{fig:l-ml}
\end{figure}

\subsection{Total Mass: Scaling Relations}
Measurements of stellar kinematics of the Local Group dwarf spheroidal
and ultra-faint dwarf galaxies have recently revealed a striking
similarity in the inferred virial mass contained within the central
300 pc. It is consistent with a common value of $10^7 \Ms$ over
several orders of magnitude in luminosity \citep{Strigari-2008}. This
suggests that all dwarf spheroidals reside in similar dark matter
haloes. Why then do they have such a large variation in stellar mass?

Some of the effects of the depth of the potential well have already
been described in Sections \ref{sec:UV} and \ref{sec:self-shielding},
reflecting the fact that no parameter can really be studied in
isolation. In this section, we look at the series of simulations that
includes all of the physical processes: cooling, star formation,
feedback, UV radiation and self-shielding, but focus on a comparison
with the observed scaling relations. As described in Section
\ref{sec:ICs}, we have scaled the initial conditions at constant
density, which results in final virial masses between 2.3~$\times
10^8$ and 1.2~$\times 10^9 \Ms$. This corresponds to masses within
300~pc between 0.9~$\times 10^7$ and 1.8~$\times 10^7 \Ms$, similar to
those obtained by \citeauthor{Strigari-2008}. We nevertheless find a
surprisingly large variation in stellar mass, luminosity, central mass
to light ratio and metallicity, as summarised in Table~1. The final
stellar masses range between $5.5 \times 10^5$ and $1.2 \times 10^7
\Ms$, whilst the median iron abundance ranges from
$[\mathrm{Fe}/\mathrm{H}] = -1.78 $ to $-1.12$.

In Figure~\ref{fig:lum-met}, we show that this is sufficient to
reproduce the well-known mass-metallicity relation
\citep[e.g.][]{Mateo-1998} of dwarf spheroidals. We compare the
results from our simulations, shown as red open squares, to those of
14 `classical' Local Group dwarf spheroidals, as given by
\cite{Woo-2008}, overplotted as black triangles. We find that there is
good agreement, both in the range of metallicities obtained, as well
as in the slope of the relation, and that this is not affected by
resolution.

We also show the distributions of metallicities of individual star
particles per galaxy in Figure~\ref{fig:Fe}. Comparing this with
observed distributions, e.g. by \cite{Helmi-2006}, we find an
overabundance of both metal-rich and metal-poor stars for a given
median metallicity. We attribute this to a lack of dissipative
metal-mixing in the interstellar medium of our simulations. As a
result, pockets of relatively metal-poor (or metal-rich) gas survive
longer, and are able to form more stars of corresponding metallicity,
which is reflected in a comparatively broad stellar metallicity
distribution, as well as steeper negative metallictity gradients
compared with observations.

In Figure~\ref{fig:l-ml} we compare the same set of simulations to
observed Milky Way satellite galaxies in terms of their luminosities
and mass-to-light ratios. The observational sample is identical to the
one used by \cite{Strigari-2008}, and comprises 8 `classical' dwarf
spheroidals, as well as 10 of the newly-discovered ultra-faint
galaxies \citep{Willman-2005,Belokurov-2007}. While the observed
galaxies span an even larger range in luminosity than the ones we have
simulated, we find a similar, tight inverse correlation between
luminosity and mass-to-light ratio. Whereas \citeauthor{Strigari-2008}
find only a very weak dependence $M_{0.3} \propto L^{0.03\pm 0.03}$,
our relation is somewhat steeper at $M_{0.3} \propto L^{0.24}$. This
is still a remarkably weak dependence, and it allows us to reproduce a
large range in luminosity with an $M_{0.3}$ mass that varies by only a
factor of two. As we discuss below, the remaining difference may point
to the fact that our model does not yet include all physical effects,
and that our assumption of an underlying mass distribution is not the
full story. \citeauthor{Strigari-2008} also note that for the most
luminous dwarf spheroidals such as Fornax, the mass-to-light ratios
relating the mass within 300 pc to the total luminosity in the
observed galaxies tend to be underestimates, since their stellar
populations are typically more extended.

\subsection{Kinematics}
As shown in Table~1, the mean one-dimensional velocity dispersions in
each galaxy resulting from our simulations are in the range of 6.5 to
9.7~kms$^{-1}$. This is comparable to the observed range of 7 to
10~kms$^{-1}$ for six of the seven `classical' Local Group dwarf
spheroidals in the sample of \cite{Walker-2007}. The one exception,
Fornax, has a velocity dispersion of about 12~kms$^{-1}$. Its stellar
population, which includes several globular clusters, is more
spatially extended, and its stellar mass is also slightly higher than
that of the most luminous galaxy in our simulations. At the faint end,
an extrapolation of our results might also be consistent with the
corresponding values of the eight ultra-faint Milky Way satellites
presented in \cite{Simon-2007}, which have velocity dispersions
between 3.3 and 7.6~kms$^{-1}$.

We also investigated the influence of supernova feedback on the shape
of the dark matter distribution. It is still an open question whether
flat cores, rather than the cusps predicted by dissipationless cold
dark-matter models exist in the central regions of dwarf galaxy
haloes. While for a stellar system with uniform mass-to-light ratio,
the shape of the gravitational potential can be uniquely determined
from the observed velocity dispersion and surface brightness profiles,
in the case of the highly dark-matter dominated dwarf spheroidal
galaxies, the unknown mass-to-light ratios result in a degeneracy
between the gravitational potential variation and the velocity
anisotropy \citep{Dejonghe-1992}. The same data, when analysed with
different anisotropy assumptions, can therefore result in different
density profiles, and as \cite{VandenBosch-2001} and \cite{Evans-2008}
have shown, the stellar kinematics of dwarf spheroidal galaxies are
generally not sufficient to distinguish between cored and cusped
profiles. Nevertheless, reports of central-density cores in dwarf
galaxies \citep[e.g.][]{Carignan-1989, deBlok-2001, Lokas-2002} have
been considered as evidence for warm dark-matter
\citep[e.g.][]{Moore-1994}. Within the framework of $\Lambda$CDM,
numerical simulations by \cite{Navarro-1996}, \cite{Read-2005},
\cite{Mashchenko-2008} and others have suggested that cores of kpc
scale may form either as a result of dynamical coupling to
supernova-induced bulk gas motions, or the rapid ejection of large
amounts of baryonic matter. Our simulations fail to fulfil these
requirements in two ways. The ejection of gas is not sufficiently
rapid (which would also be difficult to reconcile with the observed
age-spreads), and our dark matter haloes continue to evolve and grow
after star formation and supernova rates have peaked, instead of
simply settling to an equilibrium configuration. As a result, we do
not observe the formation of cores in our runs with feedback. The
final dark matter density distributions can be described by
NFW-profiles up to the resolution limit.

\subsection{Star Formation Efficiency}\label{sec:sfe}
We have also run a number of simulations where we have varied the
star-formation parameter $c_\star$, the constant of proportionality
that enters the Schmidt law for the rate at which cold gas gets turned
into stars (see Section~\ref{sec:sf}). When the star formation is
parameterized in this way in galactic chemical evolution models for
late-type galaxies, the choice of $c_\star$ has a strong influence on
the star formation rate \citep{Ferreras-2001}, and hence the stellar
age distribution, as well as on the final stellar mass. We find no
such strong influence in our simulated dwarf spheroidals, in agreement
with \cite{Katz-1996} and others. The star formation rate increases
with $c_\star$ at all redshifts, and as a result, the final stellar
mass scales as roughly $M_\star \propto c_\star^{0.25}$ over the range
of $c_\star$ between 0.01 and 0.1. In the example of initial
conditions identical to simulation~4 in Table~1, this corresponds to a
range in final stellar masses of $2.3$ to $4.2 \times 10^6 \Ms$. This
relatively weak dependence points to the fact that in dwarf galaxies,
the main factor that determines the overall star formation is not the
specific efficiency of turning cold gas into stars, but the amount of
feedback and UV heating they can sustain before star formation gets
shut down, which in our models strongly depends on the depth of the
potential well.

There is nevertheless some degeneracy between star formation
efficiency and halo mass when it comes to the amount of stars
formed. This can be broken partially by considering chemical
evolution. In Figure~\ref{fig:lum-met} we have included a simulation
sequence of differing star formation efficiency but identical initial
conditions, represented by blue diamonds, and we compare it to the
observed mass-metallicity relation, as well as to the relation
obtained from the sequence of simulations with varying total
masses. Besides the much narrower range in stellar mass of the
$c_\star$ sequence, its slope is also too steep when compared with
observations, which in contrast, are well-matched by the varying mass
sequence. While the amount of scatter prevents us from selecting a
particular value of $c_\star$, it appears that the range in
luminosities and the mass-metallicity relation cannot be explained by
a simple scaling of the star formation efficiency. For most of our
simulations, we have adopted a value of $c_\star=0.05$, in agreement
with \cite{Stinson-2007} and \cite{Mashchenko-2008}.

\begin{figure}
 \includegraphics*[scale = .48]{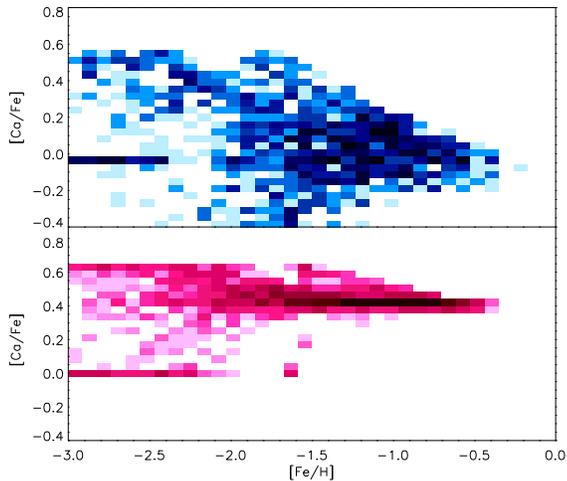}
\caption{Abundance ratios of [Ca/Fe] vs [Fe/H] of the stars, in
  simulations with a minimum supernova Ia lifetime of $10^8$ years
  (top) and $5 \times 10^8$ years (bottom). The distribution shown on
  top is for simulation~1, and shows the characteristic drop due to
  the transition from pure type II to type Ia element ratios. The
  second simulation is identical to simulation~1 in all other
  parameters, but due to the increased lifetime of the SN type Ia
  progenitors, the relative abundances remain fixed at the SN type II
  ratios.}
\label{fig:ca-fe}
\end{figure}

\subsection{Supernova-progenitor lifetimes}\label{sec:lifetimes}
Our feedback model includes both supernovae type II and type Ia. The
delay time of supernovae type II is theoretically constrained to be on
the scale of Myrs, but due to the uncertain nature of their
progenitors, that of supernovae type Ia is much more uncertain. We
find that in our simulations, the bulk of the thermal feedback
released in time to influence the star formation history is provided
by supernovae type II. However, the delay time for supernovae type Ia
influences both the total iron enrichment, and the position of the
turnover point on the [$\alpha$/Fe] / [Fe/H] diagram. In dwarf
spheroidal galaxies with very short star formation episodes, this
effect is particularly strong. In all simulations presented in this
work, we assume a uniform delay time distribution with a maximum delay
time of 1~Gyr. We have performed simulations with minimum delay times
between 100 Myrs and several Gyrs, and found that once the minimum
lifetime is increased above several hundred Myrs, the [$\alpha$/Fe]
ratios are too high, and there is no visible turnover point on the
[$\alpha$/Fe]/[Fe/H] diagram, contrary to the observed distribution of
red giants in Local Group dwarf spheroidals
\cite[e.g.][]{Shetrone-2003, Tolstoy-2003}. The two scenarios are
illustrated in Figure~\ref{fig:ca-fe}, which shows the [Ca/Fe] ratios
for minimum lifetimes of $10^8$ years (upper panel) and $5 \times
10^8$ years (lower panel) in two simulations with initial conditions
identical to simulation~2 in Table~1 that both have an age-spread of
$\sim 1.1$ Gyrs. With a more careful analysis and better constraints
on other aspects of the chemical evolution model, such as the mixing
of elements, the initial mass function and the yields, those dwarf
spheroidal galaxies which show evidence for enrichment by type Ia
supernovae despite an apparently short star-forming phase might
therefore provide an upper bound on the minimum lifetime of supernova
type Ia progenitors. On the other hand, we are satisfied that allowing
the lifetime to be a `free parameter' of the model, the best fit to
the observations is obtained with a minimum lifetime of around 100
Myrs, compatible with the value suggested by \cite{Matteucci-2001}.

\section{Summary}\label{sec:conclusion}
We have studied the formation and evolution of dwarf galaxies with
halo masses in the range of $\sim2\times 10^8$ to $10^9~\Ms$ in full
cosmological simulations including cooling, supernova feedback and UV
radiation. Our models have resulted in the formation of galaxies
similar to the Local Group dwarf spheroidals. They span a wide range
in luminosity, $6.4~\times~10^4$ to $3.4\times10^6~\Ls$ and median
metallicity, from [Fe/H]$ = -1.83$ to $-1.12$. The variation in total
mass, 2.3 to 11.8~$\times 10^8~\Ms$, is surprisingly small, but it is
comparable to the values inferred from observations in the Local
Group. The range of velocity dispersions, 6.5 to 9.7~kms$^{-1}$, is
also in good agreement with the observed range. Our simulations have
resulted in two kinds of age distributions, either a single burst of
star formation lasting around 1 Gyr, or a burst followed by a tail
extending over several Gyrs. Both of these have Local Group
analogues. However, in some sense the sample of dwarf spheroidal
galaxies in the Local Group is even more diverse. Our limited set of
initial conditions did not produce a system as luminous and extended,
or with such a large age-spread as Fornax, nor were we able to resolve
systems as faint as some of the ultra-faint dwarf galaxies.

We have shown that in our simulations, feedback from supernovae and
the cosmic UV background shape both the dynamical and the chemical
evolution of dwarf spheroidal galaxies. As a result, these are
inseparably linked, which is reflected in the scaling laws such as the
mass-metallicity relation. Feedback is essential for the evolution of
all galaxies in our models, while additional UV radiation is required
to reproduce the full range of observed galaxies, particularly those
which only have a short burst of star formation. We have demonstrated
that, with a sensible choice of parameters, the formation of systems
comparable to Local Group dwarf spheroidal galaxies is possible. While
we stress that we do not suggest that the dwarf spheroidal population
in the Local Group reflects merely a variation in halo mass, we
conclude that it is possible to reproduce the wide range of observed
stellar masses and metallicities, as well as differing star formation
histories within a single evolutionary scenario, and for a narrow
range of dynamical masses, as observed.

Our simulations put the Local Group dwarf galaxies in a unique
position where the gravitational potential is at a critical value with
respect to the combined effects of supernova feedback and UV heating.
We find that both are necessary to reproduce the observations. In this
scenario, a number of dynamical and stellar evolution effects conspire
to reproduce the observed scaling relations. More massive galaxies
start off with a proportionally higher initial gas mass, and in
addition turn a larger fraction of it into stars, because their deeper
potential wells moderate supernova-driven outflows and also allow the
gas to self-shield against the UV background. On top of that, more
efficient recycling of the gas leads to higher metallicities, and a
more extended star formation history results in a younger stellar
population with higher specific luminosities. While all these effects
undoubtedly play a role in the real systems, differing assembly and
accretion histories, and differing environments are likely also to
influence dwarf spheroidal structure.

With regard to the `laboratory' characteristics of dwarf spheroidals
mentioned in our introduction and underlined by our results, as well
as in view of the enormous difference in scale compared with the disk
galaxies studied by \cite{Scannapieco-2008d}, the addition of
low-temperature cooling and self-shielding has relatively minor
effects. It is a noteworthy and reassuring result that the same
numerical model effectively works for both kinds of galaxies. Our
models do not yet include the mixing of elements in the interstellar
medium, or cooling outside the collisional excitation equilibrium. In
addition, their environment differs from the Local Group, where
environmental effects are clearly reflected in relations such as the
apparent dependence of star-formation timescale on Galactic distance
\citep{VDB-1994, Grebel-1997}. Nevertheless, it remains to be seen how
important they are compared to feedback and the UV background. Our
choice of initial mass function has produced a residual population of
metal-free stars in the simulations, which are not observed. The
inclusion of early enrichment by Population III stars might remove
this discrepancy. Simulations by \cite{Wise-2008a} have recently
studied how massive Population III stars, and the resulting
photoionization at redshifts $\sim$~30 may affect the interstellar
medium in dwarf galaxies. This would be an interesting addition to our
models. Addressing these issues should lead to a more complete
understanding of the evolution of dwarf galaxies, and should also
allow us to make more detailed comparisons with individual objects,
exploiting the high quality data that have become available in recent
years.

\section*{Acknowledgements}
We would like to thank Volker Springel for his support with the
numerical methods that made this work possible, and Adrian Jenkins for
his advice and for preparing the initial conditions. We also thank
Pascale Jablonka for her advice, which was essential to this and to
earlier work. Finally, we would like to thank our anonymous referee
for his comments, which have improved the quality of this work. The
simulations were carried out at the computing centre of the Max-Planck
Society in Garching.

{} 

\label{lastpage}

\end{document}